\documentclass[12pt,preprint]{aastex}

\usepackage{rotating}

\newcommand{\msol}{M{$_{\odot}$}}

\newcommand{\kms}{km~s{$^{-1}$}}

\newcommand{\Msol}{M$_{\odot}$}

\slugcomment{DRAFT: \today}
\shorttitle{NGC7000 H$_2$}
\shortauthors{Bally et al.}

\begin{document}

\title{Outflows,  Dusty Cores, and a Burst of  Star Formation
         in the North America and Pelican Nebulae  \\}

\author{ John Bally\altaffilmark{1,7},               
               Adam Ginsburg\altaffilmark{2},
               Ron Probst\altaffilmark{3,7},  
               Bo Reipurth\altaffilmark{4}, \\
               Yancy L. Shirley\altaffilmark{5}, 
               and 
               Guy S. Stringfellow\altaffilmark{6,7} 
	    }
	    
\affil{{$^1$}{\it{
      Department of Astrophysical and Planetary Sciences,\\
      University of Colorado, UCB 389, 
      Boulder, CO 80309}      
      \email{John.Bally@colorado.edu}}  }
\affil{{$^2$}{\it{
      European Southern Observatory, \\
      Karl-Schwarzschild-Str. 2, 
      85748 Garching bei Munchen, Germany}      
      \email{aginsburg@eso.org}}  }
\affil{{$^3$}{\it{
      National Optical Astronomy Observatory, \\
       950 North Cherry Avenue, 
       Tucson, AZ 85719 }
       \email{probst@noao.edu }}}
\affil{{$^4$}{\it {
      Institute for Astronomy, University of Hawaii at Manoa, \\
      640 N. A'ohoku Place, Hilo, HI 96720}
      \email{ reipurth@ifa.hawaii.edu}}  }
\affil{{$^5$}{\it {
      Steward Observatory, University of Arizona, \\
      933 North Cherry Ave., \\
      Tucson, AZ, 85721}
       \email{ yshirley@as.arizona.edu}}  }
\affil{{$^6$}{\it{Center for Astrophysics and Space Astronomy, \\
      University of Colorado, UCB 389,
      Boulder, CO 80309}
      \email{Guy.Stringfellow@colorado.edu}}  }  

\footnotetext[ 7]
{Visiting Astronomer, Kitt Peal National Observatory, National Optical 
Astronomy Observatories, which is operated by the Association of Universities
for Research in Astronomy (AURA), Inc. under cooperative agreement with
the National Science Foundation.}

\begin{abstract}

We present observations of near-infrared  2.12 $\mu$m molecular hydrogen
outflows emerging from  1.1 mm dust  continuum clumps in the North America and 
Pelican Nebula (NAP) complex selected from the Bolocam Galactic 
Plane Survey (BGPS).        Hundreds of individual shocks powered by over 50 outflows
from young stars are identified, indicating that the dusty molecular clumps surrounding the
NGC 7000 / IC 5070 / W80 HII region are among the most active sites of on-going
star formation in the Solar vicinity.     A  spectacular X-shaped outflow, MHO 3400,  emerges 
from a  young star system embedded in a dense clump more than a parsec from the 
ionization  front associated with the Pelican Nebula (IC 5070).   Suspected to be a binary,
the source drives a pair of outflows  with orientations  differing  by  80$\arcdeg$.  
Each flow exhibits  S-shaped symmetry and multiple shocks  indicating a pulsed and  
precessing  jet.      The `Gulf of Mexico'  located south of the North  America Nebula 
(NGC~7000),  contains a  dense cluster of  molecular hydrogen objects (MHOs), 
Herbig-Haro (HH)  objects,   and  over 300 YSOs, indicating  a recent  burst of star formation.  
The largest outflow detected thus far in the North America and  Pelican Nebula complex, 
the 1.6 parsec long  MHO 3417 flow,  emerges from a  500 \msol\  BGPS clump and   
may be powered by a forming massive star.    Several prominent outflows such as 
MHO 3427  appear to be powered by highly embedded YSOs only visible at 
$\lambda  > 70 ~  \mu$m.       An `activity index'  formed by dividing 
the number of shocks by the mass of the cloud containing their source stars is used to
estimate the relative evolutionary states of Bolocam clumps.   Outflows can be used as 
indicators of the evolutionary state of  clumps  detected  in mm and sub-mm dust continuum 
surveys. 

\end{abstract}

\keywords{
ISM: Herbig-Haro objects  --
ISM:  individual (NGC 7000, IC5070, LDN 935)  --
ISM:  jets and outflows --
stars: formation  \\
}


\section{Introduction}

Recent millimeter and sub-millimeter wavelength surveys of the Galactic plane
have found many thousands of dense, dusty clumps  embedded within molecular
clouds, many of which are destined to form groups, associations, or clusters of stars.  
The Bolocam Galactic Plane Survey (BGPS) used the Caltech Sub-millimeter Observatory
(CSO) to survey the  northern Galactic  plane at a wavelength of 1.1 mm with 
an angular resolution $\sim$  33\arcsec\ , finding about $10^4$ clumps with masses 
ranging from a few to over $10^4$ M$_{\odot}$  \citep{Aguirre2010,Ginsburg2013} .   
A key observational challenge is to determine the evolutionary
state of dense, self-gravitating clumps to determine which ones are pre-stellar,  or actively
forming stars, or near the end of  star-formation  and in the process of being
disrupted. 

The amount of star formation, and hence the evolutionary state of a clump, can be
estimated from its content of young stellar objects (YSOs).  Embedded YSOs can be
identified  by their infrared excess emission or protostellar outflows. 
In this study, we investigate the distribution of protostellar outflows as traced by near
infrared (NIR)  emission in the 1.644 $\mu$m [Fe II] and 2.122 $\mu$m
H$_2$  lines which trace shocks in lightly-obscured  (A$_V$ less than about 
15 magnitudes) molecular gas in the North America and Pelican Nebulae.  
Shocks detected in visual-wavelength  H$\alpha$ and [S II] emission (Herbig-Haro
objects) which can be seen when $A_V$ is less than a few magnitudes  and 
{\it Spitzer} Space Telescope detections of YSOs are combined with our
data to trace  recent and on-going star formation in these clumps.

The North America and Pelican (hereafter NAP) Nebulae (NGC 7000 and IC 5070)  
are parts of a single, 3\arcdeg\ diameter HII region in Cygnus called W80 which is 
located at the  high-Galactic-longitude side of the Cygnus-X complex of  star 
formation regions in a  spiral arc or spur in our  Galaxy \citep{Reipurth2008,Guieu2009}. 
The distance to  this complex has been estimated to be between 500 pc to 
over 1 kpc  \citep{Reipurth2008}.  A  distance of  
550 pc is adopted here for the NAP complex \citep{Laugalys2006}, making the NAP one of
the nearest regions  that formed massive stars in the recent past.

\begin{figure}
\epsscale{1.0}
\center{\includegraphics[width=1.0\textwidth,angle=0]
   {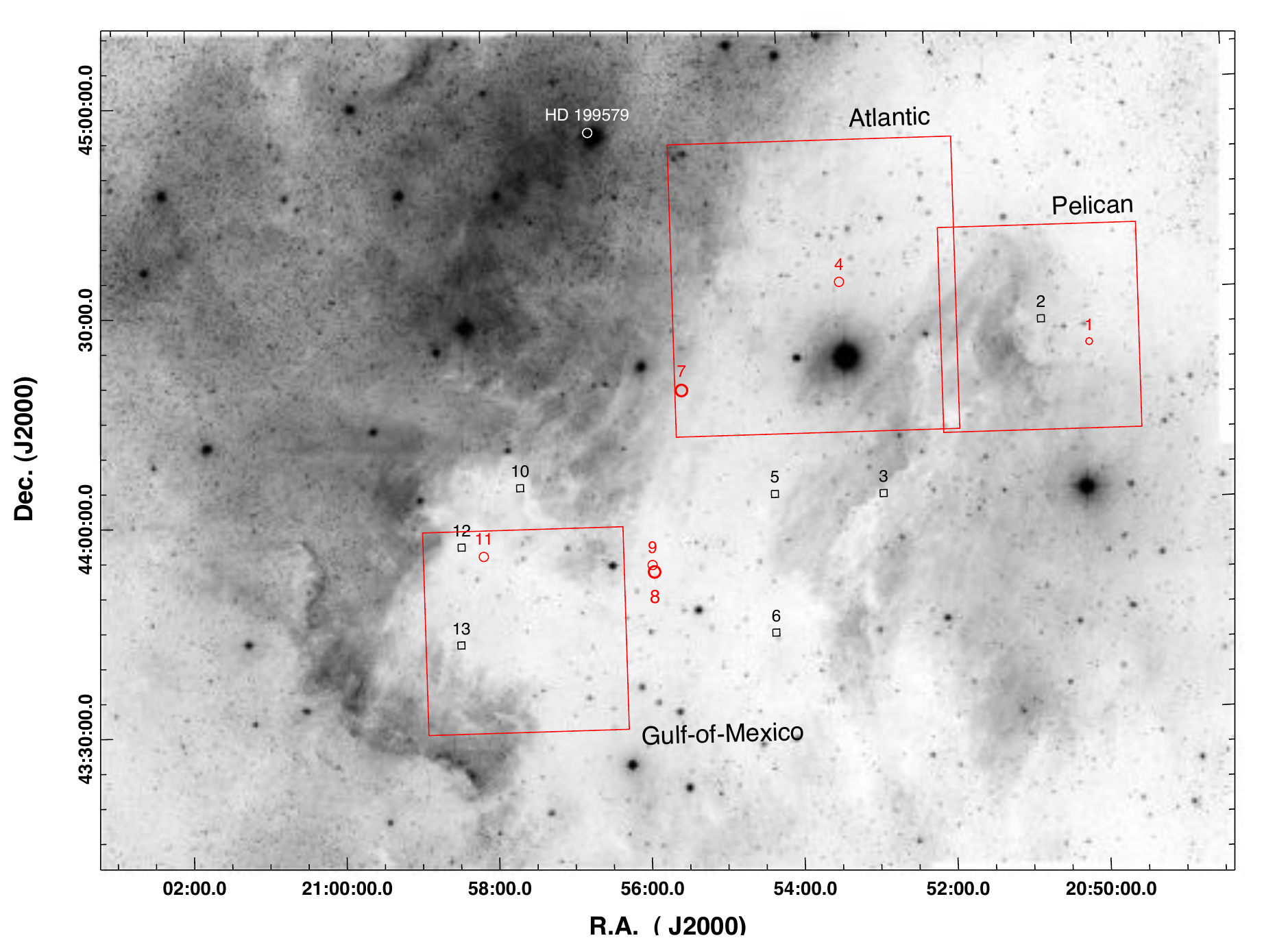}}
\caption{A wide-field image of the NGC 7000 region showing
the regions observed in this study taken from the digitized 
red sensitive (E) plates of the Palomar Sky Survey. The locations 
of stars identified by 
Straizys \& Laugalys (2008) are indicated:  Red circles mark stars
they classified as spectral type OB.   The two  circles with larger radii
(\#7 and \#8) indicate the two most massive O5 stars. 
Squares indicate massive
stars thought to be AGB stars based on photometry.   The large
red boxes indicate the fields imaged with NEWFIRM.  }
\label{fig1}
\end{figure}

W80 is a mature HII region which has swept-up and accelerated considerable amount
of molecular gas \citep{BallyScoville1980} and  the molecular clouds at its periphery 
exhibit on-going star formation activity.  The two nebulae, NGC 7000 and IC 5070,  
are separated by a large dust cloud called L935.    The highly opaque 
southern portion of L935  resembles  the  `Gulf of Mexico' (hereafter Gulf) while 
the less opaque central portion  looks like the `Atlantic Ocean',  giving the North 
America Nebula its name (Figure \ref{fig1}).  
L935 is the foreground part of a giant molecular cloud 
associated with W80.  Molecular clouds associated with the NAP  have
a  total mass of about $3 \times 10^4$ M$_{\odot}$ 
\citep{BallyScoville1980} to $5 \times 10^4$ M$_{\odot}$ \citep{FeldtWendker1993}.     
The North America and Pelican Nebulae each contain an association of 
T Tauri stars \citep{HerbigBell1988}.   The clouds northwest of the Pelican and 
those in the `Gulf' region each contain a large number of protostellar
outflows traced by visual-wavelength shocks - Herbig-Haro Objects - 
\citep{BallyReipurth2003,Armond2011}.

The radial velocity field in CO indicates that the  expansion of  the W80 HII region 
may have swept-up, compressed, and accelerated  the clouds surrounding W80,  
resulting in  3 to 5 km~s$^{-1}$ blueshift of L935 (the `Gulf' and the `Atlantic Ocean') 
with respect to the clouds lying  along the projected edge of the HII region
\citep{BallyScoville1980}.      Most of the massive  stars responsible for ionization of 
W80 are hidden behind L935.  \citet{StraivzysLaugalys2008}  identified a number
of candidate massive stars in this region.  In particular, their star \#7 
(2MASS J20552516+4418144), located behind the northeastern edge of the
``Atlantic Ocean", and \#8 (the  Comer{\'o}n \& Pasquali star; 2MASS J20555125+4352246), 
located northwest of the `Gulf'  have  spectral types O5V \citep{ComeronPasquali2005}.   

The molecular clouds associated with the W80 complex contain several dozen dense dust 
clumps recently mapped at a wavelength of 1.1 mm by the BGPS 
described in \citet{Aguirre2010},   \citet{Rosolowsky2010},  and \citet{Ginsburg2013}.   The 
Gulf of Mexico is seen as an infrared dark cloud (IRDC) against the background emission
at wavelengths between 3.6 and 24 $\mu$m.   This cloud, G084.81--01.09,  is associated
with  the largest and brightest 1.1 mm clumps in the NAP  and has been mapped 
recently in several transitions of NH$_3$, $^{12}$CO, $^{13}$CO, C$^{18}$O, and HCO$^+$,
providing measurements of central velocities, line-widths, densities, and gas temperatures
\citep{Zhang2011}.   These authors found six ammonia clumps with masses ranging from
60 to 250 \msol ,  gas temperatures of order 12 K, and molecular hydrogen densities of 
order $n(H_2) \sim 10^5$ cm$\rm{^{-3}}$.   \citet{Guieu2009} presented 
{\it Spitzer} Space Telescope IRAC observations and \citet{Rebull2011} used  {\it Spitzer}  MIPS  
data to identify about 2,000 candidate YSOs in the NAP of which over 300 are 
concentrated in the Gulf.     Thus, the NAP contains one of the largest concentrations 
of YSOs in the Solar vicinity.

The observations discussed here are described in Section 2.   The observational 
results are described in Section 3 starting with the Pelican region (Section 3.1), the
Atlantic region (Section 3.2), and the Gulf of Mexico region (Section 3.3).  Section  4
discusses the properties of the BGPS clumps and a group of small cometary clouds
seen in the IR images.     Section 5 presents a discussion of the results.  
Subsection 5.1 discusses the evolutionary states of the NAP clumps, subsection
5.2 looks at triggering, and subsection 5.3 reviews the star formation efficiency in the 
NAP.    Section 6 presents the conclusions.

\section{Observations}

In this paper,  near-infrared,  narrow-band images of 2.122 $\mu$m emission from 
shock-excited and  fluorescent H$_2$  are presented.    The dominant physical 
mechanisms in our sources are inferred from morphology, association with optical 
emission lines, and comparison to other sources studied with spectroscopy. 
For example, in the Orion Nebula region, spectra and narrow-band imaging of the  
2.12 $\mu$m and 2.25 $\mu$m H$_2$  transitions  show that the background 
molecular clouds exhibit both shock-excited  and fluorescent  H$_2$ emission.    
The shocked component exhibits small-scale, clumpy structure
with large intensity variations \citep{Bally2011} while the fluorescent 
component tends consist of large-scale filaments with a relatively constant 
surface-brightness e.g. \citep{Usuda1996}.   

\begin{figure}
\epsscale{1.0}
\center{\includegraphics[width=1.0\textwidth,angle=0]
   {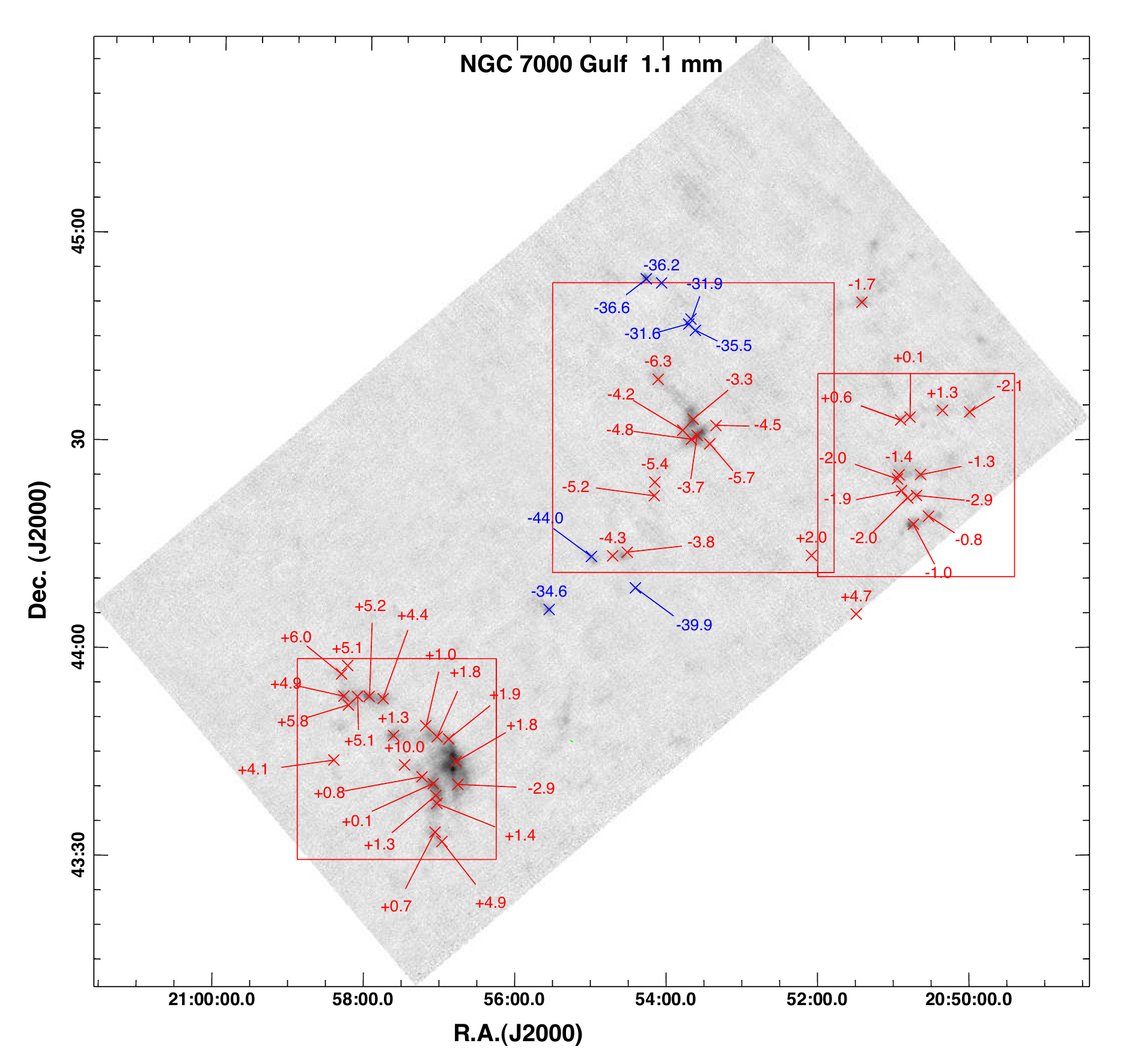}}
\caption{A 1.1 mm dust continuum image showing clumps
in the NGC 7000 region taken from the Bolocam Galactic 
Plane Survey (BGPS) in Galactic coordinates.   
Locations where the J = 3-2 transition of HCO$^+$ were
detected are marked with an X.  The numbers next to each X indicate the 
LSR velocity determined from a Gaussian fit to the spectrum.
Values with $-10.0 < V_{LSR} < +10.0$ are shown in red;
values with $V_{LSR} < -10.0$ are shown in blue.
The large red boxes indicate the fields imaged with NEWFIRM as in Figure 1. 
}
\label{fig2}
\end{figure}

Figure \ref{fig1} shows the locations of the three near-IR fields observed in the NAP complex.  
These data reveal a large collection of protostellar  jets and outflows emerging from various 
dense clumps in the NAP cloud complex.     Figure \ref{fig2} shows the 1.1 mm BGPS 
map with their radial velocities.   The column densities  of clumps associated with 
NGC~7000 were  determined from the BGPS 1.1 mm fluxes using an assumed dust 
temperature of 20 K and  the formulae given in \citet{Aguirre2010} and \citet{Bally2010}.
The areas in which column densities were estimated are shown in Figure \ref{fig3}.    
Clump  radial velocities  and line-widths were measured with the 10-meter  
Sub-Millimeter Telescope on Mt. Graham using the J = 3-2 transition of HCO$^+$.    
The abundance of  shock-excited H$_2$ 
features and Herbig-Haro objects place constraints on the 
amount of recent star formation  in each 1.1 mm clump.    While some BGPS clumps 
are relatively inactive, others  burst with outflow activity; the former objects may be in a 
less evolved evolutionary  state than the latter.    Furthermore, the locations of these 
outflows and associated  source stars relative to the W80 ionization fronts provide 
clues that can differentiate between  triggered and pre-existing star formation that was 
uncovered by the expanding HII region.  
Figure \ref{fig4} shows the fields of view observed in the near-infrared superimposed on 
{\it Spitzer}  Space Telescope images obtained at 24 $\mu$m (red), 8 $\mu$m (green), 
and 4.5 $\mu$m (blue) with BGPS contours superimposed and suspected massive stars
marked.  All coordinates discussed in this papers are given as J2000.

\subsection{NEWFIRM Near-Infrared Images}

The new observations reported here were obtained during 30 October to
5 November 2009 using the NEWFIRM near-IR imaging camera \citep{Probst2008} 
on the 4-m Mayall telescope at Kitt Peak National Observatory.     Each observation  
of the ``Pelican''  and the ``Gulf-of-Mexico" fields  (Figure 1) consisted of a dithered 
set of 11  separate exposures with offsets of up to 100\arcsec\  from the nominal pointing 
center to  fill-in the gaps in NEWFIRM's four  2048 by 2048 pixel focal plane arrays.    
Each reduced image has a field of view 29\arcmin\ $\times$ 29\arcmin\ 
with a scale of 0.39\arcsec\ per pixel.       For observations of the ``Atlantic"
field, larger offsets were use and resulted in a roughly 41\arcmin\  $\times$ 
42\arcmin\  field being imaged.

Observations were obtained 
through  1\% narrow-band filters centered on the 2.122 $\mu$m S(1) line in the 
ro-vibrational  spectrum of H$_2$ and the 1.644 $\mu$m line of [Fe II] using 
exposure times of 180 seconds per frame to give a total exposure
time per pixel of 33 minutes.    Broad-band images for continuum subtraction
were taken through standard 
broad-band J, H, and K$_s$ filters  using exposure times of 30 seconds per frame and
the same dither pattern used for H$_2$ imaging to provide a total exposure time of
5.5 minutes per pixel.   However,  during these exposures, the weather conditions 
were variable with occasional clouds. As a result, only cosmetically poor K$_s$ images 
were obtained in the Gulf of Mexico region.  Although of sufficient  quality to distinguish 
line emission from reflection nebulae,  these broad-band images did not  enable
continuum subtraction of the  narrow-band images in this field.  Continuum subtracted
images were generated only in the Pelican and  Atlantic regions.

The images presented here were reduced using the NEWFIRM 
Quick Reduce Pipeline \citep{Daly2008}.  The pipeline extracts a sky image
by median combination of an unregistered stack of images and then subtracts this
sky frame from each image.    Dome flats were used to flat-field individual images 
which were then registered  and median combined.    Stars 
in the 2MASS catalog were identified automatically on the images and used to 
remove optical distortions inherent in the re-imaging optics of NEWFIRM and to
establish a World Coordinate System (WCS) for each combined frame.
The mean flux of all identified 2MASS stars in the field are used for 
flux calibration.

The K$_s$ images were registered to the narrow-band H$_2$ images with
sub-pixel precision.   The K$_s$ intensity
scale was adjusted  by eye to match the average fluxes of stars in the corresponding
narrow-band image and subtracted from it to produce a continuum subtracted narrow-band
image.   Variations in the point spread function between broad and narrow-band
images sometimes resulted in imperfectly subtracted stars.   In order to preserve the
H$_2$ morphology, we did not alter the image PSFs to address this issue.

\subsection{Bolocam 1.1 mm Continuum from Cool Dust Clumps}

The observations reported here were obtained with Bolocam\footnote
{http://www.cso.caltech.edu/bolocam} between  June 2005 and July 2007
on the Caltech Submillimeter Observatory (CSO) 10-meter diameter telescope.
Bolocam \citep{glenn03} is a 144-element bolometer array with between
110 to 115 detectors working during the observations reported here.
The instrument consists of a monolithic wafer of silicon nitride
micromesh AC-biased bolometers cooled to 260 mK.  All data were
obtained with a 45 GHz bandwidth filter centered at 268 GHz ($\lambda$
= 1.1 mm) which excludes the bright 230 GHz J=2-1 CO line.  However, as
discussed in \citet{Aguirre2010},  the effective central
frequency is 271.1 GHz.   The individual bolometers are arranged on a
uniform hexagonal grid.  Each bolometer has an effective Gaussian
beam with a FWHM diameter of 31\arcsec , but the maps presented here
have an effective resolution of 33\arcsec\  due to beam-smearing by the
scanning strategy, data sampling rate, and data reduction as discussed
by \citet{Aguirre2010}.  Full details of the Bolocam Galactic Plane Survey
(BGPS) are presented in \citep{Aguirre2010}.      In this paper, data from
the BGPS Second Data Release (V2.0) are used \citet{Ginsburg2013}.
The resulting data are
available to the public from IPAC/IRSA at  
http://irsa.ipac.caltech.edu/data/BOLOCAM$\_$GPS.

The BGPS data can be converted into estimates of dust and gas column
density and mass using the methods outlined in \citet{Bally2010}.

\subsection{Sub-mm Spectroscopy from the Mt. Graham 10 meter 
Sub-Millimeter Telescope}

Observations of the J = 3-2 transitions of N$_2$H$^+$ and HCO$^+$
were conducted with the Heinrich Hertz Submillimeter Telescope on 
Mount Graham, Arizona as part of a survey of BGPS clumps
\citep{Schlingman2011, Shirley2013}. The data were taken over the course of 44 
nights beginning in February 2009 and ending in May 2011 with the ALMA 
Band-6 dual-polarization sideband-separating prototype receiver in a 
4-IF setup. 
With this setup, both the 
upper and lower sidebands (USB and LSB, respectively) are observed 
simultaneously in horizontal polarization (H$_{pol}$) and vertical polarization 
(V$_{pol}$) 
using two different linearly polarized feeds on the receiver.  The receiver was 
tuned to place the HCO$^+$ J = 3-2 (267.5576259 GHz) line in the center 
of the LSB. The IF was set to 6 GHz, which offsets the N$_2$H$^+$+ J = 3-2 
line (279.5118379 GHz) in the USB by +47.47 km s$^{-1}$.  The signals 
were recorded by the 1 GHz filterbanks  (1 MHz per channel, 512 MHz 
bandwidth in 4-IF mode; LSB velocity resolution $\Delta V_{ch}$ = 1.12 km s$^{-1}$
and  USB velocity resolution $\Delta V_{ch}$  = 1.07 km s$^{-1}$) in each 
polarization and  sideband pair 
(V$_{pol}$ LSB, V$_{pol}$ USB, H$_{pol}$ LSB, H$_{pol}$ USB).
133 positions were observed in the NGC 7000 NAP complex.    
Figure \ref{fig2} 
shows the 1.1 mm dust continuum image from BGPS with the radial velocities 
of  a representative sub-sample of clumps marked.  While most clumps
(those marked in red in the electronic versions of the figures) 
have clump radial velocities between $-10$ and +10  km s$^{-1}$
seen in HCO$^+$ in the NAP complex,  others (those marked in blue 
in the electronic version of the figures) have LSR velocities outside 
of this range and are  presumed to be background clumps.

\section{Results}

 \subsection{The Pelican Region: A Quadrupolar jet and Fluorescent Edges}
  
Figures \ref{fig5} and \ref{fig6}  show the 27\arcmin\ field containing the 
Pelican Nebula (IC5070)  ionization front and the molecular cloud that lies to 
its northwest.  Contours of dust  continuum emission at a wavelength of 1.1 mm 
are superimposed in Figure \ref{fig6}.    As discussed in 
\citet{Aguirre2010, Ginsburg2013},  the 1.1 mm BGPS data are high-pass filtered 
and do not trace  structure on scales larger than about 3\arcmin\ to 5\arcmin .   
Thus, these data emphasize  dense clumps and cores.   The locations of the 
Herbig-Haro (HH) objects found by \citet{BallyReipurth2003} are marked by 
squares in Figure \ref{fig6}.     Suspected shock-excited sources
of H$_2$ emission are shown with circles and numbers which correspond to the entries
in Table~1.   Several clusters of H$_2$ shocks in the Pelican region 
can be clearly associated with coherent
outflows;  these are  labeled as Molecular Hydrogen emission-line Objects 
(MHOs) and given numbers 3400 through 3412 (C. J. Davis; private communication.  
See http://www.astro.ljmu.ac.uk/MHCat/ for the catalog and list of criteria used
to define MHOs).      

The most striking feature of Figure \ref{fig6}  is the presence of 
extended filamentary  H$_2$ emission along the southeast-facing edges of the 
molecular cloud with  a relatively constant intensity.   We infer that these glowing 
cloud edges trace UV-excited fluorescent H$_2$ emission and indicate the direction of
UV illumination.  Several prominent protrusions (``elephant trunks" or ``pillars") extend from
the main cloud towards the southeast and provide additional indicators 
for the location of the exciting stars of the W80 HII region.    
The irradiated jet HH 555 \citep{BallyReipurth2003} is located  at the
eastern tip of the longest protrusion near the center left of the image.   
\citet{Guieu2009} found a source at 70 $\mu$m in {\it Spitzer}  Space Telescope images
at the base of the jet.  The associated clump shadows the elephant trunk which points
at \citet{StraivzysLaugalys2008} star number \#7 (Figure \ref{fig1}),  one of the
two massive stars with spectral type O5, located at a projected separation of about 50\arcmin\
($\sim$ 8 pc) from the Pelican ionization-front.  While the overall orientation of the protrusion
near HH 568 along the southeast corner of the Pelican point south of this stars's location,
individual parts of this tongue, along with a number of isolated cometary clouds at the
bottom of Figure \ref{fig6} clearly point at star number \#7.  Thus, this O5 star is the most
likely source of illumination of the Pelican Nebula. 

Several 1.1 mm clumps and cores in the projected interior of the Pelican Nebula 
dust cloud northwest of the fluorescent H$_2$ emission associated with its ionization fronts 
also exhibit elongated  and cometary morphologies, suggesting that they too have been impacted
by illumination from the massive stars number \#7 and \#8, the Comer{\'o}n  \& Pasquali star
\citep{StraivzysLaugalys2008}.    These clumps  may be situated behind the 
Pelican cloud where they are  exposed to ionizing radiation.  However, the lack of
obvious fluorescent H$_2$ edges as seen near most of the Pelican Nebula ionization
fronts  raises the possibility that they are insulated 
from ionizing radiation.   These 1.1 mm clumps  may have 
been shaped by non-ionizing, softer UV or visual radiation fields that can penetrate
several magnitudes into the cloud.  
That some of these clouds (e.g. the cloud containing MHO 3401) point
towards star \#8 and not star \#7 implies that star \#8 may be older, and has influenced
the structure of W80 for a longer time.   The overall elongation of the protrusion near
HH 568 in the southeast is consistent with this interpretation.

Figure  \ref{fig6} shows three groups of MHOs;  
three MHOs are located in an east-west chain
near the top of Figure \ref{fig6} (MHO 3400, 3401,  and 3402).  A second group is below 
the center of the image just north of HH 567.    Finally, the third group is associated with the
1.1 mm cometary clumps near the bottom of Figure  \ref{fig6}. 
Figures \ref{fig7} and \ref{fig8} shows details of the MHOs located at the top of Figure \ref{fig6},
MHOs 3400, 3401,  and 3402.  Coordinates
of the various features shown in the figure are tabulated in Table \ref{table1}.

{\it MHO 3400, A Point-Symmetric Pair of Jets Emerging at Nearly Right Angles:}
The two most spectacular flows in the Pelican Nebula region, collectively MHO 3400, emerge
from an infrared source embedded within a compact 1.1 mm core (Figure \ref{fig7}). 
An east-west jet  emerges at PA $\sim$ 100\arcdeg\ from this star and a north-south 
jet  comes out
at PA $\sim$ 25\arcdeg .   Both knotty jets exhibit S-shaped, point-symmetry about the 
source, indicating a possible counter-clockwise precession of the jet axes.   
Each lobe of each jet contain approximately four bright knots within 1\arcmin\ of
the source.   The mean projected spacing of 15\arcsec\ corresponds to a physical
separation of $\rm 1.6 \times 10^{17}$ cm (0.05 pc).  For a flow speed of
100 km~s$^{-1}$ typical of the inner portions of similar outflows, the time between 
subsequent ejections is about 500 years.  \citet{Reipurth2000} suggested that knots
in HH jets may be produced by periastron passages of eccentric binaries.  
If  500 years is the orbital period of a pair
of 1 M$_{\odot}$ stars, their typical separation would be about $\rm 10^{15}$ cm
(70 AU), corresponding to an angular separation of 0.09\arcsec .

A pair of distant, faint knots, W4 and E4 may mark the most distant shocks in the east-west
jet in MHO 3400.       N5 may trace the most distant shocks powered by the north-south jet.   
These knots are located at  projected distances
of 2 to 3\arcmin\ (0.3 to 0.5 pc) from the source.   Assuming an ejection velocity of 
30 km~s$^{-1}$, a typical flow-speed for H$_2$ shocks and HH objects located at 
such distances from their sources, the dynamic ages of these shocks are about 
$1.0$ to $1.5 \times 10^4$ years.   

The MHO 3400 jet  join the relatively rare class of twin jets from (presumably) 
multiple  protostar systems in which each member powers a jet  emerging at large 
projected angles.   Other examples of  mis-aligned 
flows include the HH 111/ HH121 system in Orion \citep{GredelReipurth1994}  
and the flows in L723   \citep{Carrasco-Gonzlez2008}.      Outflows emerging
at large angles from a multiple star system indicate that the spin-axes of  
circumstellar disks are misaligned.    The S-shaped  symmetry of the MHO 3400 
jets  likely indicate that the circumstellar disks surrounding each protostar 
experience forced precession, causing their bipolar jets  to change orientation.    
Non-coplanar disks can result from  dynamical interactions in non-hierarchical
multiple star systems or from formation of the multiple star by capture
\citep{MoeckelBally2007a,MoeckelBally2007b}.   Such multiple star system 
N-body interactions  may be  common among young stars \citep{Reipurth2010}.

Source 3 in Figure \ref{fig7}, IRAS 20485+4423, has fluxes of 1.9, 4.3,  46, and 316 Jy at 
12, 25, 60, and 100 $\mu$m while  Akari \citep{Akari2010}
reports a 160 $\mu$m flux of about 103 Jy.
The star, visible in the near-IR and embedded in BGPS clump G084.713+00.336
is one of the half dozen most luminous YSOs in the Pelican region and is likely
destined to be a moderate-mass star.

{\it MHO 3401 in the cometary cloud IRDC G84.656+0.383:}  
Three  H$_2$ knots, E1, W1, and W2, emerge at position angle 
(PA) $\approx$ 130\arcdeg\  from a prominent 1.1 mm BGPS clump (G084.662+00.388; 
the millimeter wave counterpart of  IRDC G84.656+0.383) 
which exhibits a cometary tail towards the northwest.     The three knots are not co-linear;  
their locations suggest that the flow suffers a mild  deflection towards the north or precession.  
The large  circle (\#10 in Table 1) below MHO 3400 marks the  location of a possible 
bow shock  associated with MHO 3401.  High-resolution spectroscopy is needed to 
determine if this  feature is indeed a shock or a fluorescent cloud edge.

MHO 3401 is embedded in a cometary clump and infrared dark cloud (IRDC) 
in the Pelican field,  G84.656+0.383,  seen in silhouette at 24 $\mu$m and shorter 
wavelengths.   It contains two highly embedded protostars  marked by the black 
circles  and numbered  1 and 2 in Figure \ref{fig7}.     YSO1, located below the 
axis defined by MHO 3401  E1, W1, and W2 in Figure \ref{fig7},  is visible in the 
{\it Spitzer}  IRAC 4.5 to 8 $\mu$m bands but
invisible at 24 $\mu$m.  YSO2, located in the core of the IRDC between
MHO 3401 E1 and W1 in Figure \ref{fig7} is the more likely candidate source
of this outflow.    It is seen as a faint, conical reflection nebula opening 
towards the northwest and MHO 3401 components W1 and W2
in the {\it Spitzer} 4.5 $\mu$m image, a dim point source much fainter than 
YSO1 in the {\it Spitzer } 8 $\mu$m image, and the dominant point source in this
IRDC at 24 $\mu$m.    YSO2 is visible at 70 $\mu$m in both the {\it Spitzer} and Herschel
Hi-GAL 70 $\mu$m images and peaks in the 160 $\mu$m band.   
From these colors,   we infer that YSO2 is a low luminosity Class 0/I
protostar while YSO1 is probably a Class I/II object which exhibits no 
evidence for outflow activity in the near-IR.

{\it MHO 3402:}  
This complex of bright, compact shocks is located near the Pelican 
ionization-shock front  (Figure \ref{fig8}). 
MHO 3402  consists of a sub-arcminute chain of knots emerging from a
tight aggregate of stars and a bright far-IR to sub-millimeter protostar embedded 
within a   cometary 1.1 mm BGPS clump,  G084.754+00.258, 
that points towards star \#7.   The IRAS catalog lists a source which is not point-like
(IRAS R2049+4421) with a 100 $\mu$m flux of 127 Jy.   The {\it Spitzer} images reveal a small
cluster of about a  half-dozen stars visible from 3.5 to 8 $\mu$m that correspond to the 
brightest 2 $\mu$m  stars.    However at  24 $\mu$m, the brightest source by far  is embedded in 
the clump near the location of H$_2$ knot MHO 3402 W3 (black circle in Figure \ref{fig8}
and labeled 'IRS 4'). 
This is the only YSO in this region visible in the Herschel Hi-GAL 70 $\mu$m images.   It is listed in the
WISE satellite all-sky data release (Cutri et al. 2012) 
as J205046.59+443323.1.  The NEWFIRM
K$_s$  image shows a compact 8\arcsec\ diameter bipolar reflection nebula surrounding this  
YSO with a position angle of 45$\arcdeg$,  similar to the orientation of the flow defined 
by a line connecting MHO 3402 W1, W2, and W3. 
The images suggest that this source drives a jet consisting of H$_2$ knots 3402 W1, W2, and W3.    
The chain consisting of MHO 3402  E1 through E4 may trace an additional flow from this small aggregate
of YSOs.   Knot MHO 3402 E1, a southwest-facing bow shock,  is by far the brightest.  
There is no {\it Spitzer}  source near its location.   Thus, it is unclear which YSO in this aggregate 
is the driving source.   A 2 $\mu$m star located about 5\arcsec\ northeast of MHO 3402 E1 is 
a potential source.   Knots E2 and E3 may trace another
flow from the star located about 5\arcsec\ southeast of knot E1.  This star has been identified
as a YSO by \citet{Guieu2009} based on {\it Spitzer} photometry.

\subsubsection{MHOs  in  the southeast portion of the  Pelican Nebula:} 

The extensive network of fluorescent filaments along cloud edges make the identification
of shock excited emission difficult in this region.  Two criteria were used to select and
mark candidate MHOs.  Features which stand out because they have higher surface brightness
than nearby filaments, or those which have very different morphologies (e.g. bow-shapes, or
isolated knots along axes defined by other candidate MHOs) are marked as MHOs.   Individual
clusters of knots are grouped and given MHO numbers. Admittedly, these criteria are highly
subjective.    These criteria result in the identification of several groups of compact knots of 
H$_2$ emission in the southeast  corner of the Pelican Nebula in the cloud interior 
(Figure \ref{fig9}).    Spectroscopy is required to confirm the shock nature of these objects. 

{\it MHO 3403:}
HH~567, the brightest HH object in the Pelican Nebula Region 
\citep{BallyReipurth2003}, is associated  with object  21 and 22 in 
Table \ref{table1}.   As shown in  Figure \ref{fig9}, these two H$_2$ knots are located 
close to and point to a bright {\it Spitzer}  source visible at 3.5 to 24 $\mu$m at 
20:50:36.8, +44:21:39 (IRS 7a) which is likely the driver of HH~567 and the 
associated H$_2$ shocks.      Several 
additional filaments and knots of H$_2$ are located southwest of object 21.  However
their nature (shocks or fluorescent edges) remain unclear.

{\it MHO 3404:}
Knots  23, 24,  25, and 26 form a chain extending about 2\arcmin\ northwest of a faint
24 $\mu$m  {\it Spitzer}  source IRS 7c  (Figure \ref{fig9}) at J2000 = 20:50:34.8, +44:22:23. 
This chain is designated MHO 3404.

{\it MHO 3405:}  
MHO 3405 is the molecular hydrogen counterpart of the irradiated jet HH~555 
\citep{BallyReipurth2003} emerging from a prominent pillar extending east of the 
Pelican ionization front.    The north-south ridge of H$_2$ emission at the
eastern tip of the pillar (see Figures \ref{fig6} and \ref{fig9})  traces the surface of the jet.
The surface layers of the pillar are also aglow with slightly fainter H$_2$ emission.   
It  is unclear weather the H$_2$ emission associated with the HH 555 
jet surface is  fluorescent or traces shocks.   However,  the H$_2$ surface brightness 
is similar to the Pelican cloud edges, suggesting that the
fluorescent mechanism dominates. 

{\it MHO 3406:} 
{\it Spitzer}  source  IRS 8 (Figure \ref{fig9}), corresponding to IRAS 20489+4410,
is the brightest YSO in this portion of the Pelican cloud at 24 $\mu$m.  Knot 27 
located about  20\arcsec\  south and elongated towards this YSO is located
about 45\arcsec\ east of HH~567. 
Knots 28 and 29  extend about one arcminute  north of the IRAS
source.  These three knots constitute MHO 3406. 

{\it MHO 3407:}
Knots 30, 31, and 32 form a southeast to northwest chain whose axis passes through
IRS 14, a dim IR source visible at wavelengths of 70 $\mu$m and longer in the 
Herschel data.   Each of these three knots is elongated along the line defined by
their positions. This chain constitutes MHO 3407.  

{\it MHO 3408:}
A pair of knots,  objects 33 and 34 
about 50\arcsec\  and 120\arcsec\ southwest  of IRS 13 at 
20:50:27.5, +44:23:29 forms MHO 3408.    A faint knot in between
but closer to object 33 may also be shock excited.
Knots 24 and 32,  grouped with MHO 3404 and MHO 3407 respectively, 
also lie along a line defined by MHO 3408, making groupings of knots 
into flows in this region highly uncertain.  

{\it MHO 3409:}
MHO 3409 consist of a pair of knots, 35 and 36 located north of IRS 6 at the top of 
Figure \ref{fig9}.

{\it MHO 3410:}
A chain of irradiated HH objects, HH~563 to 565 extends south from IRAS 20489+4406
(IRS 11) in Figures \ref{fig6} and  \ref{fig9} \citep{BallyReipurth2003}.   
This infrared source is the brightest 70 $\mu$m YSO in the Pelican region.  Objects 38, 37, and 
36 are located along the axis expected for the counterflow associated with HH~563 to 565. 
Object 39 is located south of IRS 11 half way between HH 565 and the infrared source.
This is MHO 3410.  
The {\it Spitzer} 3.5 $\mu$m images shows a reflection nebula extending north-south from this
YSO, providing support for the suggested connection between this IRAS source, 
HH~563 to 565 (see Figure \ref{fig6}) and H$_2$ knots.  

{\it MHO 3411:}
A bright, compact H$_2$ knot located east of the axis defined by MHO 3410 and
the HH 563-565 chain is MHO 3411 (object 40 in Table 1).

{\it MHO 3412:}
Finally, MHO 3412 (object 41 in Table~1) is a small bow-shaped H$_2$ knot located 
on the axis of a small 2 $\mu$m reflection  nebula embedded within the tip of a small 
cometary globule located in the southwest corner of Figures \ref{fig5} and \ref{fig6}.

\subsection{Low Level Star Formation Activity in the Atlantic}

Figure 4 shows that  the ``Atlantic''  field about a degree due 
east of the Pelican Nebula contains two bright infrared nebulae  in 
the portion of the L935 dark cloud which separates the Pelican Nebula from
the North America Nebula.    Comparison of  Figures 1 and 4 indicates
that these IR-nebulae are completely obscured at visual wavelengths.
The negative radial velocities of molecular gas associated with the northern
object in Figure 2  ($-$31 to $-$37 \kms )  indicate that it
lies far behind  the NAP complex.   However, clumps in the
vicinity of the IR-nebula near  the center of  the``Atlantic"  field and 
located close to star number 4 in Figure 1 and labeled ``Atlantic main"
in the 1.1 mm BGPS image (Figure 3) have 
radial velocities indicating a possible association with the NAP.
This IR-nebula  contains  a 6\arcmin\ diameter HII region,
G085.06$-$00.16,   with a 5 GHz flux of about 1 Jy along with  
a compact  cluster of stars  seen in the infrared images.  
The Planck Early Release Compact Source Catalogue 
\citep{PlanckCollaboration2011} includes a 5\arcmin\ diameter
source with a 100 GHz flux of about 5 Jy.

The IRAS point source catalog does not list any entries near the
``Atlantic main'' IR-nebula G085.06-00.16,.  However, a search with 
Visier finds two bright 60 and 100 $\mu$m sources listed in the IRAS
serendipitous survey catalog \citep{Kleinmann1986};  
IRAS 20518+4420 and IRAS 20519+4420 with 60
$\mu$m fluxes of about 750 Jy.     The 2MASS survey
lists many near-IR sources, with the brightest one identified 
as 2MASS 20534117+44311378.   This source is marked in
Figures \ref{fig10} and \ref{fig11}.
  
The Bolocam source catalog Bolocat lists five  1.1 mm  clumps in the
``Atlantic main" region
within a 5\arcmin\ radius of 20:53:40.9 +44:31:34.  As shown in Figure 2, all 
have radial velocities between V$_{LSR}$ = -3.3 and -5.7 \kms .   The 
NEWFIRM near IR images (Figure 10) show that the associated cloud is 
cometary with fluorescent  edges facing towards the southeast, 
indicating illumination from this direction.       

The 8 and 24 $\mu$m {\it Spitzer}  images reveal cometary 
features along the rim of the HII region that 
point toward the star cluster contained in the IR-nebula.   Thus, the OB star (or 
stars)  responsible for the  ionization of the HII region and the bright mid- to
far-IR emission from its immediate surroundings is suspected to be located
within a few arc seconds of  20:53:42.4, +44:31:54.

Figure \ref{fig10} shows a non-continuum subtracted,  narrow-band  H$_2$ image 
of the region.  Several compact isolated H$_2$ knots 
are evident in Figure \ref{fig10}  and may trace shocks from outflows  associated with
young stars in this compact cluster (Table~2).

{\it MHO 3413 and MHO 3415:}
The brightest H$_2$ feature, MHO 3413 in Table 2,  is located 
near the south rim of the cometary cloud.   It  is unusual in that it exhibits a prominent chain 
of knots in the [FeII] line at 1.644 $\mu$m (Figure \ref{fig11}) consisting of knots 1 though
6 in Table~2.   Only knot 1 is visible in H$_2$, the rest are only seen in [FeII] (Figure
\ref{fig11}).   However, the relatively isolated H$_2$ feature, MHO 3415 (knot 11 in
Table~2 and Figures \ref{fig10} and \ref{fig11})  near the upper-left corner of
Figure \ref{fig10} is located within 1 degree of the axis of this
[FeII] jet.   MHO 3415 may therefore trace an outlying shock in this flow, in which case 
it  is a candidate  parsec-scale outflow.   
The source may be a {\it Spitzer}-detected 3.6 to 8 and 24 $\mu$m YSO at 
20:53:30.9, +44:30:03     located midway between [FeII] knots 1 and 2.  Although 
this YSO is visible in the H and K$_s$ images, its flux increases with increasing wavelength
up to at least 24 $\mu$m.   The [FeII] image shows a jet extending from this source to
knot  1.    

{\it MHO 3414:}
This MHO contains four knots  (7,8,9 and 10 in Table~2 and in 
Figures \ref{fig10} and \ref{fig11}) which form a chain which may
trace a single outflow.  A 24 $\mu$m-bright {\it Spitzer} YSO located  at 
20:53:36.5, +44:34:33 is located a few arcseconds southeast of knot 8 and
may be the driving source of this southeast-northwest oriented flow.   A faint  
2.2 $\mu$m reflection nebula connects this knot to the {\it Spitzer} YSO, 
supporting the interpretation  that this source drives this outflow.   

{\it MHO 3416:}
Finally, a faint pair of H$_2$ and [FeII] dominated knots are located about 1.5\arcmin\ west
of the star 4 \citep{StraivzysLaugalys2008} which they list as a candidate massive star
with spectral type O9 to B0.    The axis defined by the two knots also points back
towards the 24 $\mu$m {\it Spitzer} YSO marked by the `X'  North of [Fe II] knot 
2 in Figure 10 potentially associated with MHO 3414.  Thus, the source is unclear.

\subsection{The Gulf of Mexico Region}

The Gulf region shown in Figures \ref{fig12} through  \ref{fig19c} is  the most 
active site of on-going star formation in the NAP region. 
This region contains  many MHOs, HH objects, 
IRAC-identified Class I, II , and III YSOs,  and highly embedded Class 0  or I YSOs 
seen mostly at 24 and 70 $\mu$m that could be drivers for the various shocks. 
Early surveys for H$\alpha$ emission line and flare stars \citep{Herbig1958,Welin1973}
identified over 60 young stars.   The recent study by \citet{Armond2011} identified 35 
HH objects (HH 636 through 663 and HH 952 through 958), 5 possible HH objects, and
30 new H$\alpha$ emission line stars.    
The {\it Spitzer}  IRAC and MIPS images, combined with
2MASS photometry  \citep{Guieu2009, Rebull2011} led to the identification of about 
2,076 YSOs in the NAP complex of which 375 are located in  the `Gulf of Mexico'.
Table~3 lists over a hundred individual shocks many of which have complex 
substructure.   

Figure \ref{fig12} shows the full field of view H$_2$ image of the Gulf region which can be 
subdivided into three sub-regions:   The region  `Gulf SW2' (Figure \ref{fig13}) 
consists  of the collection of  bright BGPS clumps on the right-side of 
Figure \ref{fig12}.   In Figure \ref{fig3}, these clumps are further divided into
`SW2 main',  `SW2 SE',  and `SW2 S' for determination of column densities and
masses.    The region dubbed  `Gulf SW1'
(Figure \ref{fig14})  is associated with the compact BGPS clump and an IRDC
visible in silhouette at wavelengths of 24 $\mu$m and shorter  (G84.963-1.173) 
located just  above the center of  Figure \ref{fig12}.   The northeast portion of the Gulf 
located in the upper-left portion of Figure \ref{fig12}  is dubbed `Gulf core'
(Figures \ref{fig15} through  \ref{fig18}).   Grouping HH objects and H$_2$ shocks into 
coherent outflows is  difficult owing to the large number of individual features and 
the high density of YSOs.     Below, the discussion of MHOs  proceeds roughly from  
southwest to northeast. 

\subsubsection{A Giant Outflow from the Most Massive Clump in the Gulf:  `SW2 main'}

 The brightest and largest 1.1 mm BGPS clump in the NAP complex (`SW 2 main'; Figure
 \ref{fig3})  is located  about 15\arcmin\ southwest of the very active region of star formation 
 in the `Gulf core' (consisting of `core E' and `core W' in Figure \ref{fig3}).  
 Although this large and structured 1.1 mm clump exhibits  three to 
 five times  more total 1.1 mm emission,  and is therefore likely to be more massive by the same
 factor than the highly active complex at the northeast end of the Gulf,   it exhibits a much 
 lower level of outflow  activity.   The `Gulf SW2' region contains several reflection nebulae, 
 several dozen {\it Spitzer} / IRAC YSOs, 24, and 70 $\mu$m sources, and about a dozen 
 H$_2$ shocks  and contains the largest outflow detected in the NAP complex, 
 MHO 3417 (Table~3).  
 
 The two brightest H$_2$ features in the 
 `SW2 main'  clump consist of oppositely facing bow shocks 
 located in the upper-right and 
 lower-left in Figure \ref{fig13}.   The northern bow (H$_2$ feature 1 in Table~3) was
 detected by \citet{Armond2011} at visual wavelengths and designated HH~953.  The
 southern bow (H$_2$ feature 5), HH~954, is much fainter at visual wavelengths
 than the northern shock.   These two  shocks  are symmetrically placed about the 1.1 mm 
 peak  of  `Gulf  SW2'.      Several faint H$_2$ knots (objects 2,3,4, 6, and 7) lie on or near the 
 axis defined by this flow.   This giant flow is designated MHO 3417  in Table~3.
 The projected  separation between HH 953 and 954   is about 10\arcmin\ in projection,
 or about 1.6 parsecs at the distance to the NAP complex.    

 The two brightest 24 and 70 $\mu$m sources in `Gulf SW2'  in the {\it Spitzer} data are 
 located on the flow axis.
 The northern source, IRS 1 (Table~3),  is brighter at 70 $\mu$m and more obscured.  H$_2$
 knot 2 (Table~3) is located about 30\arcsec\ northwest of the location of IRS~1.  The
 southern source, IRS 2 (Table~3), is the brightest source at 8 $\mu$m in the {\it Spitzer}  images.
 A 2 $\mu$m source and faint  reflection  nebula (R1 in Table~3)  on the north side of this clump
 is evident in the NEWFIRM data.   It is unclear which IR source is the driver of the MHO 3417
 HH 953/954 outflow.  \citet{Zhang2011} suggests that a massive star is forming in the 
 `SW2 main'  clump.   But, it is not clear which of the mid-IR sources may be the
 suspected massive star.  
 
\subsubsection{Other MHOs  in Gulf  `SW2 main'}
 
MHO 3418 (Figure \ref{fig13}) consists of faint H$_2$ knots  (8 and 9 in Table~3) 
located near and southeast of a 24 and 70 $\mu$m source embedded in the 
southern BGPS peak in `SW2 main'.  This far-IR source is designated IRS 3 in Table~3. 
A faint H$_2$ jet extends from this source to reflection nebula
R3.   It is possible that knot 8 traces a small outflow from a deeply embedded
source in R3 while knot 9 is powered by IRS 3.   There is a faint H$_2$ knot about 15\arcsec\
east-northeast of the 70 $\mu$m source.  Thus, MHO 3418 may consist of several small
flows from highly embedded low luminosity YSOs.

MHO 3419 (Figure \ref{fig13})  consists of a bright H$_2$ arc  (number 10 in Table~3) 
located about 50\arcsec\ due north of reflection nebula R2 and the IR source IRS 4 and about 
and 30\arcsec\ southwest of IRS~5.  IRS~5 is much brighter at 24 $\mu$m (it is the second
brightest source in `SW2 main' at 8 $\mu$m) than IRS~4.   The source of this shock is unclear.
  
Figure \ref{fig13b} shows the BGPS clumps SW2 SE and SW2 S and MHOs 3420
through 3424.
MHO 3420 is a compact, east-facing  bow shock (number 11 in Table~3)
located in between the two dust filaments
SW2 main and SW2 SE (Figure \ref{fig3}) in the Gulf SW2 clump.

MHO 3421 is a bright bow-tie shaped compact H$_2$ feature (knot 12) and a filament 
(knots 13 and 14) associated with a compact aggregate of a half dozen YSOs
embedded in the north-end of BGPS clump `SW2S' (Figure \ref{fig3}).   The
H$_2$ features are visible in the {\it Spitzer}  IRAC 4.5 $\mu$m images and thus
are `extended green  objects' (EGOs).  Faint [SII] emission is also seen from
knot 12 in the images presented by \citet{Armond2011} and it is here designated as HH 1087.

MHO 3422 (Figure \ref{fig13b}) consists of H$_2$ knots 15 and 16 located southeast 
of reflection  nebulae R5 and R6 in the `SW2 SE' clump.   A bright 70 $\mu$m source, 
IRS 7, is located at the southern tip of this clump about 30\arcsec\ north of the 
K-band reflection nebula R5.   R6 is associated with a 24 $\mu$m source.

MHO 3423  consists of a pair  of faint H$_2$ knots (17 and 18  in Table 3) located south 
of the north-end of BGPS filament `SW2 SE'  which contains several embedded IR sources. 

MHO 3424 (Figure \ref{fig13b}) consists of a chain of faint H$_2$ knots (19, 20, 21 in Table 3) 
aligned along a east-southeast to west-northwest axis located in the space between 
BGPS filament `SW2 SE' and `SW2 main' and north of MHO 3423.  
It is possible that the knots in  MHO 3423 and
3424 trace flows from the IR sources in the `SW2 SE' filament.

MHO 3425 (Figure \ref{fig13}) consists of a diffuse H$_2$ feature located about 
3\arcmin\  east of the  `SW2 main' clump.  Its morphology may indicate a south-facing shock.

Given the relatively large number of {\it Spitzer}  detected YSOs, and the presence 
of at least one large outflow, the relative paucity of H$_2$ features in Gulf `SW2 main' 
compared to the Gulf `core E'  and `core W'  regions  may be due to its relatively large
column density and  foreground obscuration which may hide some shocks.  
As discussed above, the 3.6 Jy
peak emission in a 33\arcsec\  beam at $\lambda$ = 1.1 mm implies an H$_2$ column density 
of $N (H_2) = 7.2 \times 10^{22}$~cm$^{-2}$ assuming a dust temperature of 20 K,
corresponding to a visual extinction, $A_V \approx 36$ magnitudes, or larger 
if the dust temperature is lower.

 \subsubsection{Outflows from the Gulf `SW1'  Region}
 
Gulf `SW1'  is an  opaque IRDC,  G84.963-1.117,  located about 250\arcsec\ south 
of the bright infrared reflection nebula R10 (Figure \ref{fig15}) embedded in the 
Gulf of Mexico `Gulf core' region.  It is seen in silhouette at 24 $\mu$m and shorter 
wavelengths. 

A 70\arcsec\ long string of 5 knots (23 through 27)  emerge from a small cloud seen 
as an IRDC at the northwestern end of the Gulf SW1 region; this is MHO 3426.  An 
infrared source visible at  8, 24,  and 70 $\mu$m  (IRS8 in Table~3)  is located 
at the position of knot 24.    Though not  listed in the YSO compilation of 
Guieu et al. (2009),  its location at the base of  a compact 
10\arcsec\ long H$_2$  jet makes it likely to be another highly embedded Class 0 YSO.    
The jet is pointed at knots 25, 26, and 27 that together form a bow-shock facing east.
A faint shock, knot 23 is seen on the west  side of the source.  This morphology 
suggests that the east side is the blue-shifted, approaching side of this outflow.

MHO 3427 is a   330\arcsec\ long chain containing a  dozen H$_2$ shocks 
(listed as 28 through 40 in Table~3) that 
emerge from the most opaque portion and peak of the 1.1 mm emission located
near the east-end  of the Gulf SW1 clump.   Guieu et al. (2009) list nearly two dozen 
YSOs  in the vicinity of  this dark cloud, but none near the 1.1 mm  peak.       
The  brightest H$_2$ features (knots 36 and 37) appear to
form an east-facing bow shock;   knots  30 and 31 trace a  west-facing shock.  

Inspection of the {\it Spitzer} 70 $\mu$m image shows
a dim but compact source at the location noted as IRS 10 in Figure  \ref{fig14}
and Table 3.    Although this source is not seen at 24 $\mu$m, 
the symmetric placement of the two oppositely oriented bow shocks suggests
that the 70~$\mu$m source is a highly embedded Class 0 protostar.  Future far-infrared
or sensitive radio continuum  searches are needed to confirm or deny the reality of
this source.   

A prominent 2 $\mu$m reflection nebula (R9 in Table~3)  is associated with the 
brightest 4.5 to 24 $\mu$m source in this region.   Also detected at 70 $\mu$m, 
this object is listed as IRS 9 in Table~3.   At 2 $\mu$m, the nebula appears 
bipolar with a bisecting dust lane separating the brighter southern lobe from the
dimmer northern lobe.  The morphology may indicate the presence of a disk 
shadow produced by a disk with an axis at position angle of 210$\arcdeg$.   
There are no H$_2$ shocks or HH objects along the suspected axis of this disk.

 \subsubsection{A Burst of Outflows from the `Gulf core'  Region}
  
The `Gulf core' region consisting of BGPS clumps `core E' and `core W' 
in Figure \ref{fig3} contains the highest concentration of  outflows
in the NAP complex.     The high density of HH objects and H$_2$ shocks makes
the identification of individual outflows difficult as there may be multiple overlapping
outflows along many lines-of-sight.  
Figures \ref{fig15} through \ref{fig18} show our H$_2$ image and  optical wavelength 
H$\alpha$  and [SII]  images of the `Gulf core' region from \citet{Armond2011}.
Comparison with the  H$\alpha$ and [SII] images 
indicates that the several dozen H$_2$ features shown in 
Figure \ref{fig15} are generally not the same as the shocks traced at visual wavelengths.  
Most of the H$_2$ features trace portions of outflows that are not visible in  
H$\alpha$ and [SII], and therefore provide complementary data.  While the HH objects
mark  shocks in regions suffering relatively low-extinction 
($A_V$ less than a few magnitudes), the MHOs can be seen through 
an order of magnitude larger extinction.   Thus, it is likely that the visual-wavelength emission
lines mostly trace the approaching, or blue-shifted portions of outflows, while the MHOs 
trace both the approaching lobes and their counter-flows.   The absence of MHOs
at the locations of some HH objects may indicate either the complete dissociation of H$_2$
by shocks propagating through the cloud, or an absence of H$_2$ which indicates that
the shocks are  impacting atomic gas outside the molecular
cloud.   

Clusters of shocks in close proximity that may trace outflows are given MHO 
numbers.  However, future observations may show that some of these groupings 
represent overlapping flows that are unrelated.    The grouping into outflows 
is especially ambiguous in the `Gulf core' region due to crowding and confusion.   

MHO 3428,  (near the northeast edge of Figure \ref{fig14}) 
located about 2.5\arcmin\ northeast of Gulf `SW1',   consists of a pair
of faint H$_2$ knots that are molecular counterparts to HH 639.   
Knot 41 in Table~3 is associated with the brightest part of this HH object which 
appears to trace a flow propagating at PA $\sim$ 130 to 40$\arcdeg$.
MHO 3429  is an isolated knot associated with HH 638 and may be part of
the same flow as MHO 3428/HH 639.

MHO 3430 consists of two H$_2$ knots  (knot 44 and 45 in Table~3) which are 
the near-IR counterparts of HH 636 located along the eastern rim of the large
near-IR reflection nebula R10.   The southern H$_2$ knot (44)  is a 30\arcsec\ 
diameter  H$_2$ feature  resembling a south-moving bow shock.  The northern
knot (45) is located along the eastern rim of HH 636 and coincides with
a bright [SII] feature.    One of the brightest 8, 24, and 70 $\mu$m sources,
IRS 11, located at the northern end of HH 636 / MHO 3430, is associated
with the bright 2 $\mu$m reflection nebula R13.  
 
MHO 3431, located  directly north of R13 and IRS 11 contains knots  46 to 48
and may be a counterflow and near-IR counterpart to HH~637.

MHO 3432, located 30\arcsec\ due west of IRS 11 is a faint collection of H$_2$ knots;
it may be a visually obscured counterflow to HH 640 / MHO 3434.

MHO 3433 is comprised of three H$_2$ knots (51, 52, and 53) located 
170\arcsec\ to 270\arcsec\ northwest of 
IRS 11.  Knot 52 is associated with an HH object not
listed in \citet{Armond2011};  it is designated HH 1088. 

MHO 3434 is an east-west  collection of knots associated with the
western part of HH 640.    Knot 50 is located 6\arcsec\  west  of IRS 12, 
a mid-IR source that is redder but fainter than IRS 11 and associated with
2 $\mu$m reflection nebula R14. A fainter H$_2$ knot is located about
7\arcsec\ east of R14/IRS 12 which is likely to be the driving source.
 
MHO 3435 consists of a jet-like chain of H$_2$ knots 54 through 60 associated 
with the eastern part of HH 640.    It is unclear if this MHO is a part of the same
flow as MHO 3434 or a separate flow, possibly powered by IRS 11.   The nearly continuous
H$_2$ emission bends south towards knots 61 and 62, east-facing bow shock 63
(associated with HH 643), and filament 65.    It is also unclear if knots 61 through 65
are part of the same flow or trace shocks in other flows in the region.   
A very dim H$_2$ feature that resembles a bow shock  (74 in Table~3) may also 
be part of this chain since there is a faint trail of H$_2$ emission extending back to 
just south of knots 64 and 65.

MHO 3436 consists of knots 66 through 71 which form a 70\arcsec\ long linear chain at 
PA $\sim$ 34$\arcdeg$.   Knot 70 is the near-IR counterpart of HH~642.  A near-IR star
at 20:57:52.91, +43:53:28.4 associated with reflection nebula R17 is located
near the middle of this chain.  The southwestern knots 66 and 67 are much brighter
than the northeastern knots, suggesting that the southwestern side of the flow is
approaching.

MHO 3437 is a compact knot (72) located 17\arcsec\ north of R15 associated 
with mid-IR source IRS 13. The H$_2$ knot is the western, compact component of HH 641.

MHO 3438 is a bright H$_2$ shock about 5\arcsec\ southwest of a star and also
25\arcsec\ southwest of IRS 14.

MHO 3439 is a 2.8\arcmin\ long chain consisting of knots 75 to 78.  Knots 75 and
76 are bright while 77 and 78  consist of faint filaments and knots aligned with the
overall southeast-northwest orientation of the flow at PA $\sim$ 130$\arcdeg$.  Knot
75 is associated with a dim [SII] counterpart which appears to be an HH object (HH 1089).

MHO 3440 consists of three knots, 80 to 82 in Table~3 which are also seen
in [SII] as HH 648.  A faint HH object candidate is located along the axis 
defined by HH~648 at 20:58:01.84, +43:52:17.4 in the Subaru 
[SII] image indicating that MHO 3440 is a collimated flow.

MHO 3441 consists of three knots (83, 84, and 84e) that are near-IR counterparts to
HH 649.  The knots are connected by a faint H$\alpha$ and [SII] bridge in the
Subaru images.

MHO 3442 is an H$_2$ filament about 10\arcsec\ north of  MHO  3440 knot  82
and labeled as knot 82n in Table~3 and as MHO 3442 in Figure~16.

MHO 3443 is a collection of H$_2$ features around IRS 15.  The southeastern
feature (85 in Table 3) looks like a bow shock facing away from IRS 15.   The
western rim of the 40\arcsec\ diameter H$_2$ feature 88 coincides with HH~647.
Like MHO 3438, MHO 3443 is  also in close proximity to IRS 14.

MHO 3444 consists of two H$_2$ knots about 1.7\arcmin\ northwest of  the 
ÒHerbig clusterÓ, a dense aggregate of YSOs and H$\alpha$ emission line stars 
discovered by George Herbig \citet{Herbig1958}
embedded in the large reflection nebula complex R19 at the eastern end of the 
Gulf core region.

MHO 3445 consists of an H$_2$ filament and knot (91 and 92) 
about 0.7\arcmin\ northwest of the  Herbig cluster.

MHO 3446 is a prominent outflow extending southeast of the Herbig cluster.   
The FU Ori candidate HBC722 (V2493 Cyg, LkH$\alpha$ 188-G4)
that experienced a 4 to 5 
magnitude visual-wavelength flare in 2010 \citep{Semkov2010, Miller2011, Semkov2012}
is located in the Herbig cluster near the likely point of origin of MHO 3446.  
This region was observed with the SMA in the 1.3 mm-wavelength 
transitions of CO  and the adjacent continuum by \citet{Dunham2012} who found
a cluster of continuum sources, MMS1 through MMS7.   Only MMS2 and MMS3
are associated with stars in our H$_2$ image;  the rest have no near-IR counterparts.
This indicates that  the Herbig cluster contains some highly embedded sources
not yet visible in the near-IR \citep{Armond2011}.   

A bipolar CO outflow was detected by \citet{Dunham2012} at position angle
PA $\sim$ 135\arcdeg\ centered on the mm-source MMS3, located about 2\arcsec\ from
2MASS20581617+4353310 
(m$_K$ = 13.1,  m[3.6] = 11.75, m[8.0] = 8.8, m[24.0] = 3.78 mag.).  The blueshifted
lobe of this flow is closely aligned with objects 97 through 100; the redshifted lobe
appears to be associated with object 96.    However, no obvious H$_2$ features 
appear to be associated with the FU Ori candidate, HBC722.

Knots 99 and 100 in MHO 3446 are closely associated with bright knots in HH 657.
However, the northern part of this HH object lies east of and has a different orientation
than the flow orientation defined by knots 97 and 98.  Knot 101 does not
belong to either the chain of HH objects or MHOs.  Knot 96, located northwest of
the Herbig cluster may be a counterflow to the MHO flow axis  defined by knots
97 and 98.  It is likely that the various components of MHO 3446 and HH 657 trace
parts of multiple overlapping flows emerging from the Herbig cluster.

MHO 3447 is a northeast facing bow and knot located about 1\arcmin\ north of
the Herbig cluster along the axis defined by the filaments MHO 3445. 

MHO 3448 is a compact knot associated with HH 658 east of the Herbig cluster.
MHO 3449 is a pair of  diffuse H$_2$ knots southeast of HH 662.

MHO 3450 and 3451 are chains of H$_2$ knots at the northeast periphery
of the Gulf region.    MHO 3450 is centered on mid-IR source  IRS 16  located 
a few arc seconds east of knot 109 at the location of reflection nebula R21.
This is almost certainly the driving source.
MHO 3451 is a north-south flow associated with HH 958 whose driving source is
not obvious.

\section{Clump properties}

Table~4 lists the properties of the major clumps and largest cores in the NAP complex
as determined from the 1.1 mm BGPS Version 2.0 maps.  
Figure \ref{fig3} shows the regions inside which the fluxes were summed and divided by
the number of pixels per CSO Bolocam beam area.   In the standard Bolocam pipeline
processed images, the pixels are 7\arcsec .2 in diameter and there are 23.8 pixels in
each gaussian  beam with a full-width-half-maximum diameter of 33\arcsec\  
(equivalent to a 40\arcsec\ diameter top-hat beam).
Table~4 lists the coordinates of the centers of each of the ovals shown in Figure \ref{fig3}, 
the total flux in each oval in Janskys, the mass enclosed in the ovals, and the major and 
minor diameters of the regions in arcseconds. 
The mass estimates assumed a uniform dust temperature of 20 Kelvin, a
gas-to-dust ratio of 100, and that all clumps in the NAP complex are located at a 
common distance of 550 pc.   A detailed description of the column density and 
mass-estimation  method used for Bolocam BGPS data is given in Bally et al. (2010).

In the Gulf core region, the BGPS-based mass estimates can be compared to the 
published 1.3 cm ammonia-based estimates \citep{Toujima2011,Zhang2011}  . 
The `Gulf core' region corresponds to \citet{Toujima2011}  `sub-clumps'  A1, A2 and A3 while
our `Gulf SW2' corresponds to  \citet{Toujima2011} clump B.   Below, their  mass
estimates are scaled to a distance of 550 pc (they used a distance of 600 pc).    

\citet{Toujima2011}  core A1 (ammonia mass of
10 \Msol ) corresponds to our `core E' with a Bolocam-based mass of 32 \Msol .    
Their A2 + A3 (ammonia mass of 24 \Msol  )  corresponds to our `core W'  for which
we estimate a mass of 64 \Msol .   The total mass of   the `Gulf core' region (their region A)
has an ammonia-based  LTE mass of 80 \Msol\  and a virial mass of 105 \Msol \  that can 
be compared to a Bolocam-based mass of  96 \Msol . Thus,   
although \citet{Toujima2011} find smaller masses for their `sub-clumps',  perhaps because
they associate a smaller region with each core, the overall
mass estimates for our `Gulf core' region are in excellent agreement with our Bolocam-based
estimate.   

The  `SW1'  core is seen as a minor lump in the \citet{Toujima2011} map and no mass 
estimate was provided in their paper. 
 \citet{Toujima2011} region B and our region `SW2 main' + `SW2 SE') have 
ammonia based  masses  of 294 \Msol\  to  380 \Msol .   For the corresponding region, 
the Bolocam-based mass is 596 \Msol .   As discussed by \citet{Zhang2011}, Gulf SW2
is also known as  MSX dark cloud G084.81-01.09.    This object may potentially form a 
massive star.  \citet{Zhang2011} also mapped ammonia and compared the results  to 
Bolocam BGPS V1.0 images.   \citet{Zhang2011} P1 corresponds to our `SW2 main' 
clump with an ammonia mass of about 200 \msol\  and BGPS mass of  
477 \msol .   \citet{Zhang2011} P2 with a mass 120 to 255 \msol\
corresponds to our `SW2 SE'  with 119 \msol .  \citet{Zhang2011} P3, P4, and P5 with a
mass of $\approx$ 300 \msol\  corresponds to our `Gulf SW2 S'  with a BGPS mass of 
only 26 \msol .   
Summing all of the \citet{Zhang2011} regions gives a molecular mass estimate of
about  830 \msol , somewhat larger than our total mass estimate based on BGPS
for the entire Gulf region.  Uncertainties,
intrinsic variation of the dust temperature, and different regions of spatial integration
may  account for these  variations in the mass estimates, especially on small scales.

As shown in Figure \ref{fig13}, the largest outflow with the biggest H$_2$-bright
shocks,  the  9\arcmin\  (1.4 pc) long MHO 3417,  emerges from the relatively massive 
`Gulf SW2 main'  clump.  As suggested by \citet{Zhang2011}, the large mass of the 
clump  and the presence of a parsec-scale outflow 
are consistent with the proposal that this clump is in the early stages of massive star
formation. 

\subsection{Fluorescent Edges and Cometary Clouds Point to the Comer{\'o}n  \& Pasquali Star}

Several very long cometary clouds are located southeast of the Gulf in Figure \ref{fig12}
and rendered visible by  their fluorescent limb-brightened cloud edges.  These
clouds are listed in Table~3 with the designation `C' followed by a number (1 through 18)
and marked by cyan circles in the electronic version of this paper.
The longest one (C4, C5, and C6) 
has a length to width ratio of about 20 to 1, and can  be used to locate  the
direction to the primary source of UV irradiation in this part of the NAP complex 
(Figure \ref{fig19a}).   Several  other cometary clouds near the bottom center of 
Figure \ref{fig12} (C1 through C3 and C12 through C17) also
trace the direction of  illumination (Figures \ref{fig19b} and \ref{fig19c}).   
The intersection of lines extended along the orientation  of these cometary  
clouds indicate the location of  the illuminating star.  These lines intersect
near the location of the Comer{\'o}n \& Pasquali star (\# 8), the O5V star located behind the Gulf
(Figure \ref{fig4}). 
Thus, the orientations of cometary clouds with fluorescent edges confirm that
this object is the dominant source of UV radiation in the southeastern portion of 
the NAP complex.

\section{Discussion}

\subsection{The Evolutionary States of Clumps}

Outflow activity,   counts of IR-excess sources,  and H$\alpha$ 
emission line  stars can be used to  constrain the relative evolutionary states of  
nearby 1.1 mm BGPS clumps where the visual extinction is less than about 10 
magnitudes.    Stellar H$\alpha$ emission from active chromospheres and 
accretion flows,  infrared excess emission in the 1.2 to 2.4 $\mu$m and 3.6 to 
8.0 $\mu$m {\it Spitzer}/IRAC images, the presence of  HH objects 
and shock-excited MHOs, and sources visible 
only at wavelengths of 24 $\mu$m or longer 
trace  progressively younger phases in the evolution of young stars.  
Highly embedded Class 0 or Class I  YSOs,  which tend to be younger than 
about $10^5$ years,  are  most likely to  power outflows and jets which excite 
H$_2$  shocks.    Herbig-Haro objects tend to be powered by somewhat less embedded,  
older stars.    As  accretion  onto forming stars abates, so  does the power of 
outflow activity.  Only a few stars older than about  $10^6$ years  have 
associated outflows, and these tend to have low mass loss rates ($\dot M 
< 10^{-8}$ \msol ) which are hard to detect except when externally ionized.    
However,  IR-excess and 
stellar   H$\alpha$ emission can persist  for many millions of years.   
The majority  of {\it Spitzer}-detected IR-excess YSOs have 
spectral energy distributions   (SEDs) indicating that they are in the Class II phase.    
Most of these stars are likely to be older than  $10^6$ years. 
Low-obscuration H$\alpha$  emission-line stars tend to be older still.  Thus,
the BGPS-based mass and the  relative numbers of MHOs, HH objects, IR-excess 
sources, and H$\alpha$ emission-line stars are indicators of dust clump evolution.

BGPS clumps with large mm-wave fluxes and few indicators of  outflow activity, 
few  IR-excess sources, and few H$\alpha$ emission-line 
stars in their vicinity may be relatively young.     On the other hand, clumps with
large numbers of HH objects, H$\alpha$ emission line stars, and MHOs are
likely to be relatively evolved.   Using this metric, the `Gulf core' region with 
the highest concentration of HH objects and  MHOs appears to be the most
evolved region of those studied here.    The number  of H$_2$
shocks divided by the clump mass provides a quantitative estimator of outflow
activity, $I_{out}$.   The total number of catalogued H$_2$ features in Table~3 in the `Gulf core'  
region is  73 and the BGPS mass in this region is about 96 \msol .   Thus,  the
activity index   $I_{out} = N_{MHO} / M_{dense_gas} \approx $ 0.76 for the 
`Gulf core' region.   For the most massive clump, `Gulf SW2 main', 
$I_{out} \approx $ 0.05.  For the `Atlantic main' clump,  $I_{out} \approx $ 0.06 if the 
two non star-forming clumps in Table~4 are excluded.  If these are included, 
$I_{out} \approx $ 0.05.
The index for the Pelican region is $I_{out} \approx $ 0.17, indicating a relatively
advanced evolutionary stage, though not as evolved as `Gulf core'.
For all clumps in the NAP complex excluding `Gulf core', but
including `SW2 main'  $I_{out} \approx $ 0.09.   If `SW2 main' is also excluded,
the remaining clumps have activity index $I_{out} \approx $ 0.13.    

The counting of H$_2$ shocks is highly subjective.   Large and complex objects
are counted once and given the same weight as compact isolated knots are.   
An area-weighted activity index may be a more robust estimator of the momentum 
injection rate and thus more representative of the evolutionary stage.  
However, a simple attempt to include the shock area of several regions in the NAP
shows that the basic results will  be similar.    
Counts of {\it Spitzer}-identified YSOs could also be used as a star-formation rate 
activity and  evolutionary stage indicator.   Since  outflows detectable as
MHOs tend to have about an order of magnitude shorter lifetimes than 
{\it Spitzer}-detected YSOs ($< ~10^5$ years for MHOs; $> ~10^6$ years for YSOs),  
outflows are expected to be a more selective  evolutionary indicator.

On the basis of activity index, the most massive clump, `Gulf SW2'  appears to be
the least evolved, despite containing the largest outflow.  `Gulf core' appears to be
the most evolved (but still actively star-forming) region.    `Atlantic main'  which also has a
very low index poses an interesting situation.  The presence of an IR bubble and
HII region suggests that star formation has stopped in the center of this clump while
the large mass of the dust in the surrounding region  combined with a small number of
outflow and YSOs may indicate a very early phase of star formation in the HII region
surroundings. 

The most massive BGPS clump in the NAP, Gulf SW2,  contains the 
parsec-scale   outflow MHO 3417 and  relatively few  H$_2$ shocks.   Although 
there are a comparable number (many dozens)  of 8 and 24 $\mu$m sources
in both the Gulf core and Gulf SW2 regions, the  BGPS flux 
and derived mass of SW2 is about five times greater than the Gulf core.  Thus, 
the number of mid-IR YSOs and outflows divided by the  1.1 mm  flux and dust mass 
is at least a  factor of four lower in Gulf SW2 than in Gulf core.   
This  also suggests that Gulf SW2 is in an earlier evolutionary stage than the Gulf core
region.  

The large area-covering-factor of HH and MHO shocks in 
the Gulf core  region indicates that much of the volume of this  cloud 
is being  re-processed  by supersonic shocks.   In the northeastern  portion of the Gulf, 
the several dozen outflows traced by HH objects  and MHOs cover about 5 to  
10\% of the projected area of the cloud (Figure \ref{fig15} to \ref{fig18}).    
In the absence of FUV and ionization feedback from massive stars, re-processing 
by outflows may dominate  the self-regulation of star formation and may even 
disrupt the clump.      Such intense outflow activity has been observed in other regions
such as NGC 1333 in the Perseus molecular cloud \citep{Bally1996}.   
Feedback from outflows has been suggested to be the dominant  source of 
energy and momentum injection
in the absence of massive stars \citep{Li2006,Nakamura2011}. 

As star formation abates, outflow activity weakens, and the MHOs and HH 
objects  disappear.   However the aging YSOs can still be traced by IR-excess or 
stellar H$\alpha$ emission.   The G085.06-00.16  HII region, along with most 
of the stars in the  \citet{Rebull2011} `Atlantic cluster'  West of the 
HII region and East of the Pelican
Nebula ionization front may be approaching post-star-forming
dormancy.

\subsection{Triggering?}

The distribution of star formation in the clouds surrounding the NAP complex 
can be used to constrain the modes of star formation.  Three possible modes 
of star formation have  been proposed \citep{Elmegreen1998}. 
 
[1] {\it pre-existing, spontaneous star formation}: 
Star formation can occur spontaneously throughout  the  cloud without
triggering.     In that scenario, YSOs currently located at  cloud edges and at the tips 
of elephant   trunks and pillars would have formed regardless of the expanding 
HII region that has sculpted the local cloud morphology.   
The  velocities of stars and associated cloud cores are 
expected to be uncorrelated with either the HII region velocity or with the 
velocity of  any swept-up shell.   Spontaneous  star formation should also produce 
YSOs deep inside clouds far from  impinging ionization fronts.  The YSOs in the 
interior  of the Pelican cloud, such as the sources of MHO 3400 and MHO 3401  
may be examples. 

[2]  {\it Pressure-triggered gravitational collapse}:  
As an advancing ionization front wraps around  pre-existing clumps,  its  
pressure triggers gravitational collapse and star formation.
Stellar velocities  are  expected to weakly correlate with  the HII region's 
expanding shell.  As ionization exposes a clump,  photo-ablation of its
surface and the resulting Oort-Spitzer rocket effect can accelerate remaining
gas and the embedded YSO away from the ionizing sources.    Young stars
will be located between the center of HII region expansion and cometary wakes
or pillars.   The YSOs at the heads of pillars such as the one which drives 
HH~555 in the Pelican Nebula may be examples of this mode.
 
[3] {\it Collect and collapse:}
In `classic' triggered star formation,  clumps form from gas swept up 
by the expanding HII region.    When enough mass is collected,   
the local gravitational escape speed becomes  larger than the 
spreading velocity due to the shell expansion and gravitational collapse 
occurs.   Forming stars will inherit the expansion velocity of the shell 
when it becomes unstable to its own self-gravity.   For decelerating shells, 
YSOs   will tend to outrun the shell.   
Precision radial velocities and proper motions of stars and gas 
are needed to reliably discriminate between these scenarios.   

There are indications that the `Gulf core'  region is in the foreground relative 
to the rest of the NAP.   The Pelican ionization fronts point toward the 
center of the HII region and various cometary clouds point toward the Comeran and 
Pasquali star \#7, and this star is behind the `Gulf of Mexico' region. 
The redshifted radial velocity of about V$_{LSR}$ = 4 to 6 \kms\  
indicates that the Gulf core region, which supports such a high density of 
YSOs and outflow activity, is plunging inward towards the NAP HII region. 
This is inconsistent with the Ôcollect and collapseÕ scenario.

The Atlantic ridge portion of  the L935 cloud separating the North America and 
Pelican nebulae  is blueshifted by about 3 to 8 \kms\ with respect to the molecular 
gas in the Pelican and Gulf regions of the NAP \citep{BallyScoville1980} .   This
region also obscures the central part of the NAP HII region and must therefore
be located in the foreground.      
\citet{BallyScoville1980} concluded that this portion of the L935 cloud has been 
accelerated by  the expansion of the HII region.   The V$_{LSR}$ = -4 to -6 \kms\
blueshift of the gas associated with the G85.06$-$0.16 HII region, cluster,  and 
outflows in the Atlantic region of L935  may be consistent with the `collect
and collapse' scenario of triggered star formation.
The radial velocities of the clumps  in the Gulf SW1 and
SW2  regions observed  by \citet{Toujima2011} and \citet{Zhang2011} and by the 
Mt. Graham telescope N$_2$H$^+$ and HCO$^+$ measurements presented 
here (V$_{LSR}$ = $-$2.9 to +1.8 \kms ) are consistent with this picture since 
these regions are located near the projected edge of the giant NAP HII region 
(W80; NGC 7000) where the motions induced by HII region expansion are 
expected to be mostly along the plane of the sky.   

The `Gulf core' region,  which supports the highest density of YSOs 
and outflow activity in the NAP complex,  has a redshifted radial velocity of 
about V$_{LSR}$ = 4 to 6 \kms (one faint knot has V$_{LSR}$ = 10 \kms\ in 
Figure \ref{fig2}),  indicating that it may be plunging in towards the  NAP HII region.  
There are indications that the Gulf core region is in the foreground relative to the 
rest of the NAP complex of clouds.   The Pelican ionization fronts point toward
the center of the HII region,  cometary clouds point toward the Comeron and Pasquali star, 
and this star is behind the  `Gulf -of-Mexico' as it is highly obscured.   The redshifted 
radial velocity of about  $V_{LSR}$ = 4 to 6 \kms\   suggests that the Gulf core region, 
which supports  such a high density of YSOs and  outflow activity, is plunging 
in towards the  NAP HII region.  This is inconsistent with the Ôcollect and collapseÕ 
scenario. It is possible that the  back side of the `Gulf core' region is in contact
with the high-pressure HII region, was compressed as a result, and has therefore
recently  experienced a burst of  star formation.  It  may be an example of  
star formation  in a pre-existing cloud  triggered  by  external  pressure.  
On the other hand,  the star formation sites located a parsec or more away 
from the NAP  ionization fronts within the  Pelican Nebula cloud may be 
`spontaneous'  star forming events mostly unaffected by the W80 HII region.

\subsection{Star Formation Efficiency  in the NAP}

The star formation efficiency (SFE) in the NAP can be crudely estimated 
by taking the ratio of the number of YSOs in each region times the median
YSO mass, and dividing  by the masses of the parent clumps  or the
total mass of molecular gas in the NAP. 

The NAP complex is surrounded by a giant molecular cloud complex. 
\citet{BallyScoville1980} found a mass of $3.0 \times 10^4$ \msol\  for the clouds
surrounding and lying in front of the NAP.
\citet{FeldtWendker1993} obtained $^{12}$CO and $^{13}$CO maps of the NAP
finding masses  (scaled from their 500 pc to our 550 pc distance) of $2.8 \times 10^4$ \msol\
and $4.3 \times 10^3$ \msol  ,  respectively.  They argue that the total cloud mass is the
sum of the $^{12}$CO and $^{13}$CO masses since the $^{13}$CO line is only seen 
where  $^{12}$CO is very smooth and likely to be very optically thick.    
 \citet{Dobashi1994} mapped  $^{13}$CO in the 
Cygnus-X region, finding a $^{13}$CO  mass of about $6 \times 10^3$ \msol\  
in the L935 region studied here.   
Thus, the total H$_2$ mass is likely to be about $4 \times 10^4$ \msol\ with at 
least a factor of two uncertainty.  The total mass of H$_2$ detected by BGPS is
about $1.1 \times 10^3$ \msol .   This is likely to be a lower bound due to the
spatial filtering of the BGPS data which loses sensitivity to  structure larger than
approximately 5\arcmin .

\citet{Guieu2009} used 3.6 to 8 $\mu$m {\it Spitzer}/IRAC to identify more than
1,600 candidate YSOs in the NAP.   \citet{Rebull2011} used 24 $\mu$m 
{\it Spitzer}/MIPS to find additional objects increasing  the number of candidates to
over 2,000.  {\it Spitzer} is most sensitive to  young embedded Class I sources, somewhat older  
`flat-spectrum' YSOs, and  more evolved Class II objects with warm circumstellar disks. 
Assuming a median mass of 0.5 \msol , and a total mass for the NAP
of $4 \times 10^4$ \msol\   implies a lower bound on the SFE of about 2\% since
the majority of the more evolved and presumably older Class III YSOs are likely 
to have been missed.    

Restricting attention to the three regions of the NAP containing the largest 
concentration of  Class I sources and all of the BGPS clumps gives somewhat higher
estimates for the SFE.  The Gulf region contains the greatest concentration
of YSOs and BGPS clumps.  \citet{Rebull2011} found 375 YSOs in the  
`Gulf of Mexico'  cluster whose boundaries closely match and enclose the
nearly 800 \msol\  traced by the  BGPS.    For a median stellar mass of
0.5 \msol , the implied SFE of the Gulf  is about 23\% with an uncertainty of at 
least a factor of two.  The Pelican and Atlantic regions in our study are located
at the east and west ends of the `Pelican Cluster'  found by \citet{Rebull2011}.
Our `Pelican'   region contains  two dozen Class I and `flat spectrum' YSOs and 
an additional two dozen Class II objects.  For the same median YSO mass as above, 
the implied SFE is 18\%.   The clusters in our `Atlantic' region only contains four 
Class I and five  `flat spectrum' sources, consistent with the relatively older age of this
region.  Adding approximately two dozen Class II sources implies  a SFE $\sim$
7\%.  However, this may be a severe lower bound as most sources may be
Class III or later objects or ones that have lost most of their circumstellar matter.

\section{Conclusions}

Near-infrared J, H, K$_s$, and H$_2$ images of three regions
of active star formation associated with 1.1 mm dust clumps surrounding 
the North America and Pelican (NAP) nebulae are 
presented.  These images reveal  many new molecular hydrogen objects 
(MHOs) which are shocks powered 
by protostellar  outflows propagating into predominantly molecular 
media.    Alignments and source morphology enables some outflows  
to be identified.  The remaining MHOs can not
be linked into specific flows or associated with potential driving sources
due to indistinct shock orientation or confusion resulting from crowding
and the high density of potential driving sources.  Comparison with 1.1 mm 
dust  continuum emission from the  Bolocam Galactic Plane Survey 
provides constraints on the  evolutionary states of dense molecular clumps.

Summing the masses of the BGPS clumps with radial velocities
near V$_{LSR}$ = 0 \kms , and assuming a dust temperature of 20 K
gives a BGPS-based mass estimate of 1,120 \msol\ for the mass of
dense clumps associated with NGC~7000.
The southwestern portion  of the Gulf of Mexico region of the NAP 
contains the most massive and  brightest BGPS clump, Gulf SW2.

The `Pelican' region contains a  pair of 
outflows  emerging nearly at right angles from an 
infrared source  embedded in a BGPS clump in the Pelican molecular
cloud about  a parsec (in projection) from the Pelican ionization front.
Their S-shaped symmetries indicate that both outflows are powered by 
pulsed, precessing jets, likely from an unresolved binary protostar.
The Pelican region contains an additional two dozen H$_2$ shocks emerging
from a dozen different low-mass (9 to 60 \msol ) BGPS clumps.  Although some
MHOs coincide with previously detected HH objects, most do not have
visual-wavelength counterparts.

The `Atlantic' region contains a region of recent star formation with a 
prominent 8 to 24 $\mu$m bubble and a roughly 2\arcmin\ (0.3 pc) 
diameter HII region, G085.07$-$0.16, with a total radio continuum flux of 
about 0.5 Jy at 5 GHz.  This HII region  contains a small IR cluster 
and a bright source, IRAS 20518+4420, but it 
may be in a mostly post-star forming state.    Only a few
MHOs and an outflow mostly traced by 1.644 $\mu$m [FeII] emission are 
found here.    

The `Gulf of Mexico'  region contains the largest concentration of outflows,
young stars, and the most massive BGPS clumps containing about 
three-quarters of the dense gas and dust in the entire NAP complex.
The region contains well over 300 YSOs in various evolutionary stages,
over 100 H$_2$ shocks grouped into over 50 MHOs,  and a candidate 
massive star forming clump, `Gulf SW2', 
which contains the longest outflow detected so far in the NAP. 
The  northeastern portion of the Gulf (the `Gulf core' region) contains the 
highest concentration of YSOs and outflows in the NAP with  
5\% to 10\% of the clump surface area  covered by H$_2$ shocks
or HH objects.    The density of YSOs and outflows is comparable to
NGC~1333 in Perseus \citep{Bally1996}.  Its ratio of outflow activity to 1.1
mm flux is high.  The cloud may be in the process 
of being disrupted by the energy and momentum deposited by outflow 
activity.  

\begin{figure}
\epsscale{1.0}
\center{\includegraphics[width=1.0\textwidth,angle=0]
   {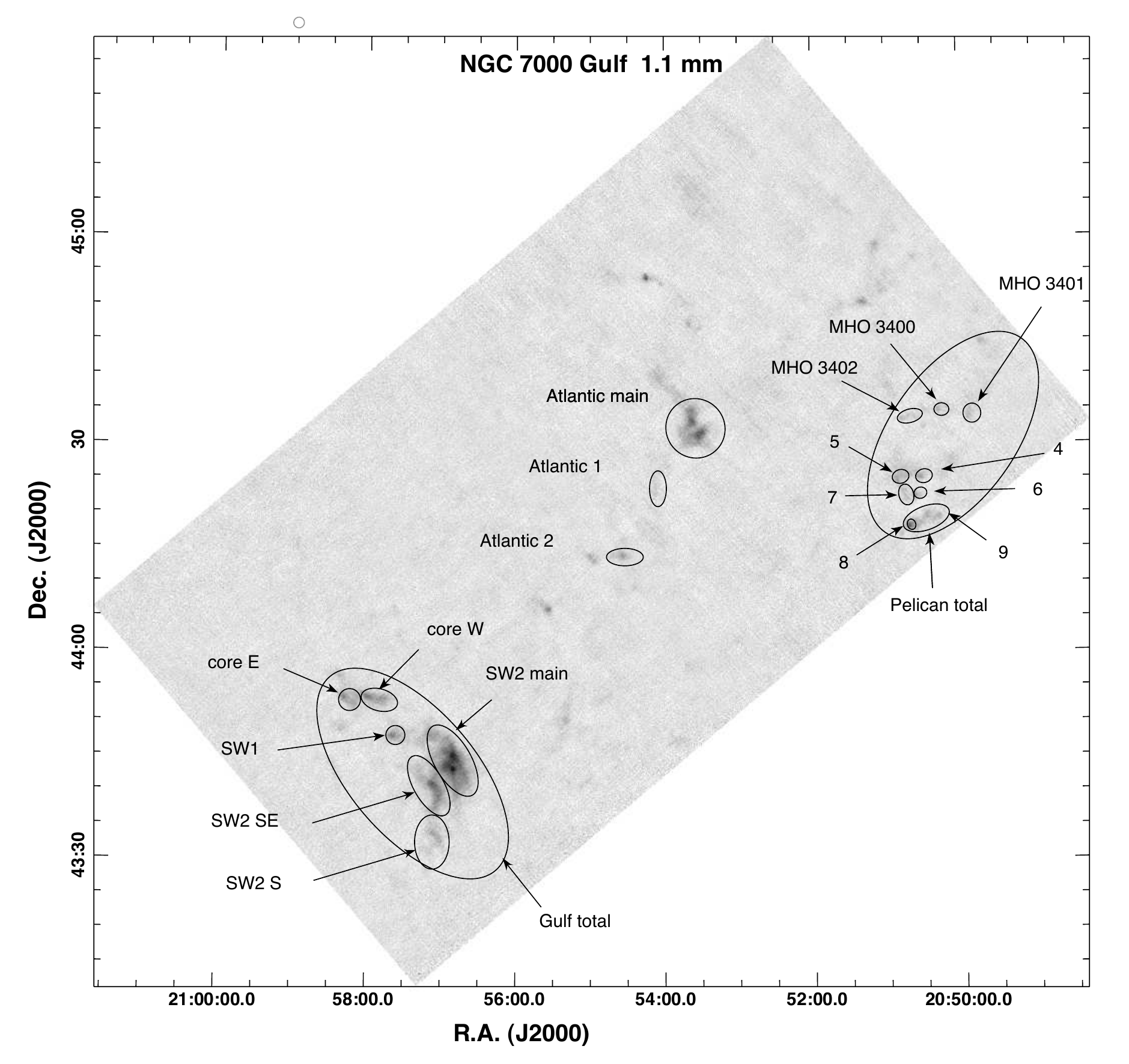}}
\caption{The 1.1 mm BGPS image showing the regions inside which 
                masses listed in Table 4 were evaluated.  Masses were 
                calculated only for objects having radial velocities indicating 
                an association with NGC~7000 (the NAP) and thought to lie
                at a distance of about 550 pc.  
}
\label{fig3}
\end{figure}

\begin{figure}
\epsscale{1.0}
\center{\includegraphics[width=1.0\textwidth,angle=0]  {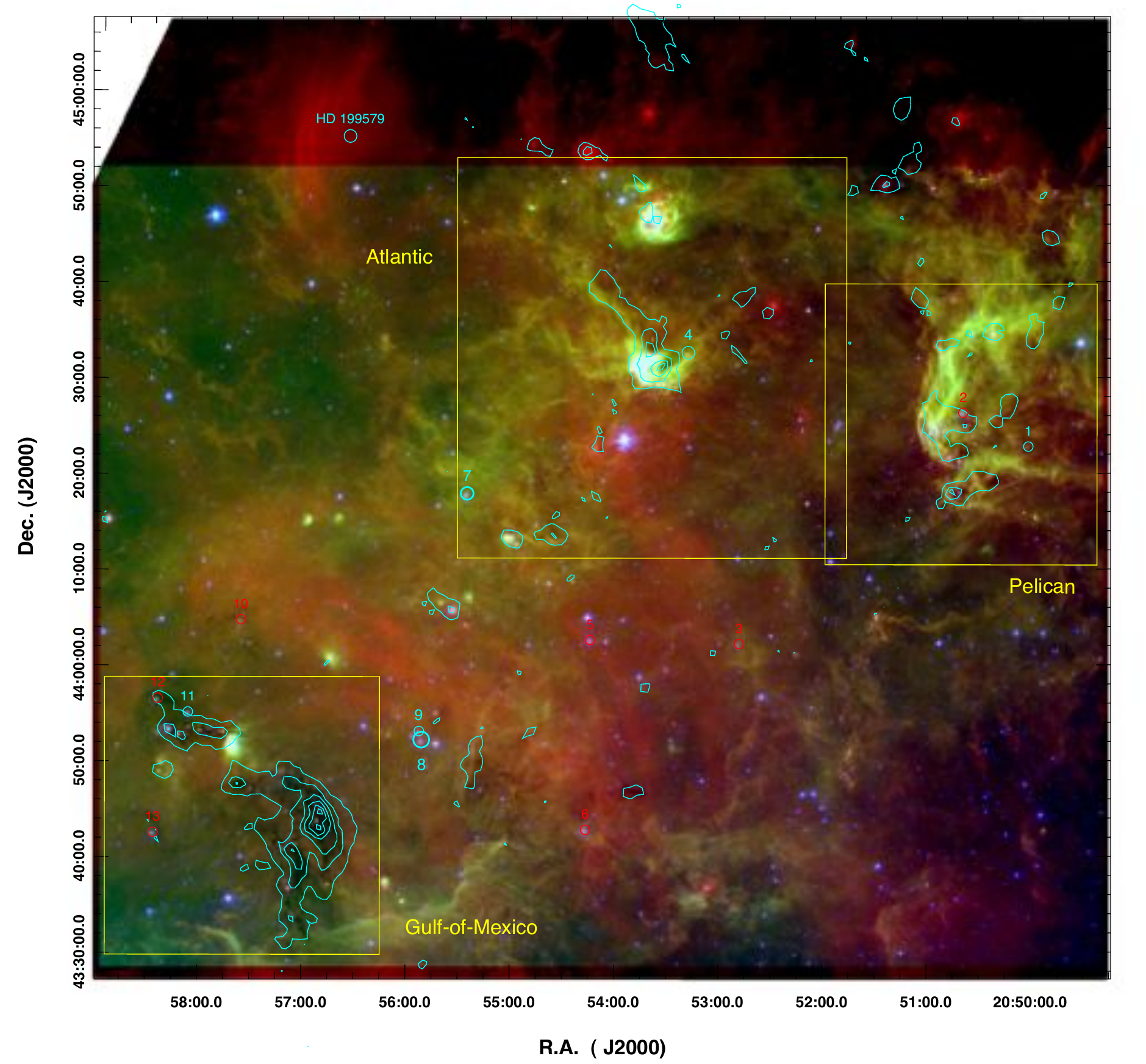}}
\caption{A Spitzer Space Telescope image showing emission at 
24 $\mu$m (red), 8.0 $\mu$m (green), and 4.5 $\mu$m (blue).  BGPS
contours (Figure 1) are superimposed.    Stars from Straizys \& Laugalys (2008) 
are indicated along with the NEWFIRM fields of view.  Those marked in cyan 
are OB stars.  Those marked in red are post main-sequence massive stars.
}
\label{fig4}
\end{figure}

\begin{figure}
\epsscale{1.0}
\center{\includegraphics[width=1.0\textwidth,angle=0]  {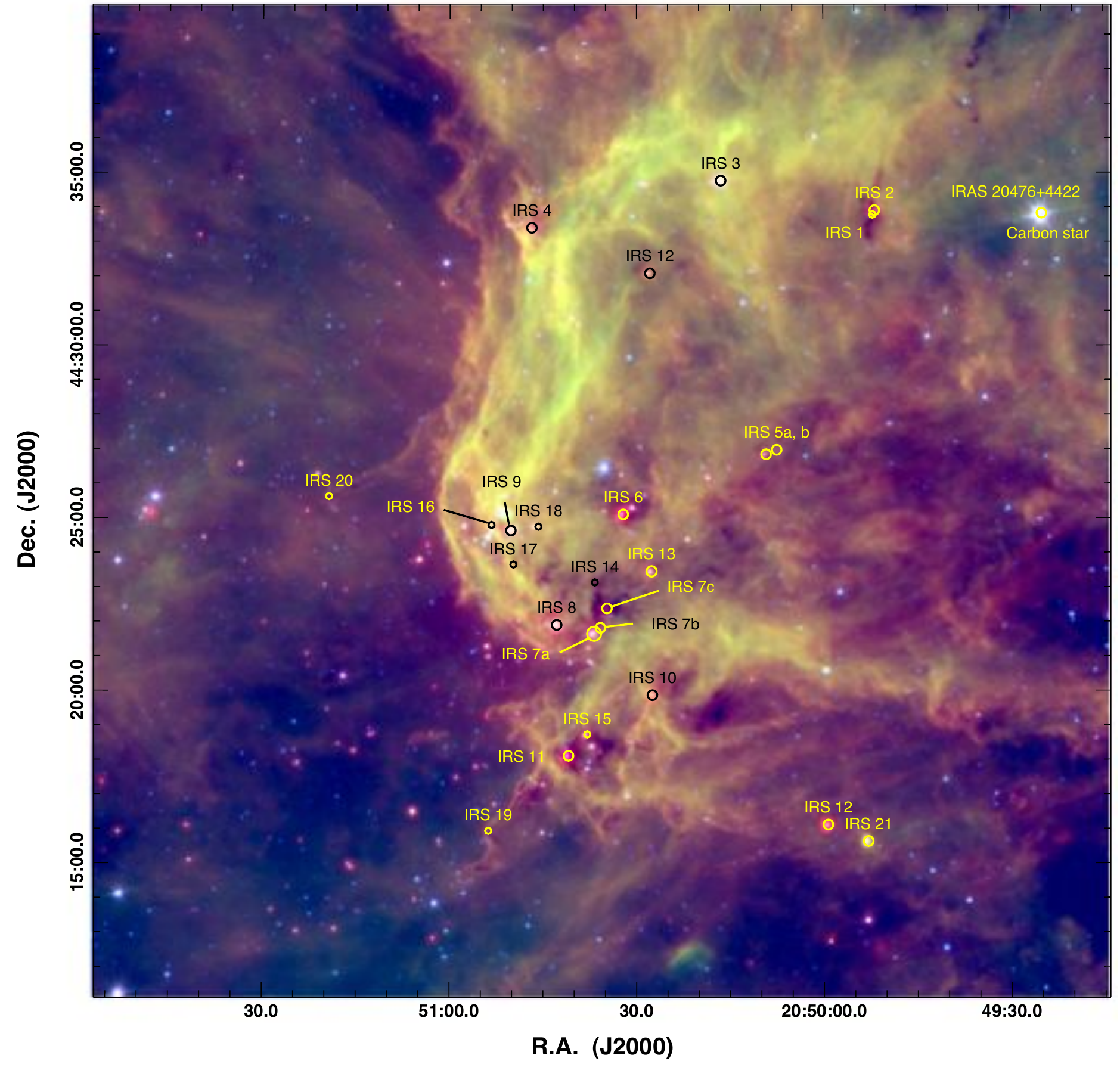}}
\caption{A color composite image showing 4.5 (blue),  8 (green), and 24 $\mu$m (red)
emission in the Pelican Region  with the locations of 70 $\mu$m sources shown
with yellow circles (or black where the background is bright). 
Numbers correspond to the entries in Table~1 and the text.}
\label{fig5}
\end{figure}

\begin{figure}
\epsscale{1.0}
\center{\includegraphics[width=1.0\textwidth,angle=0]  {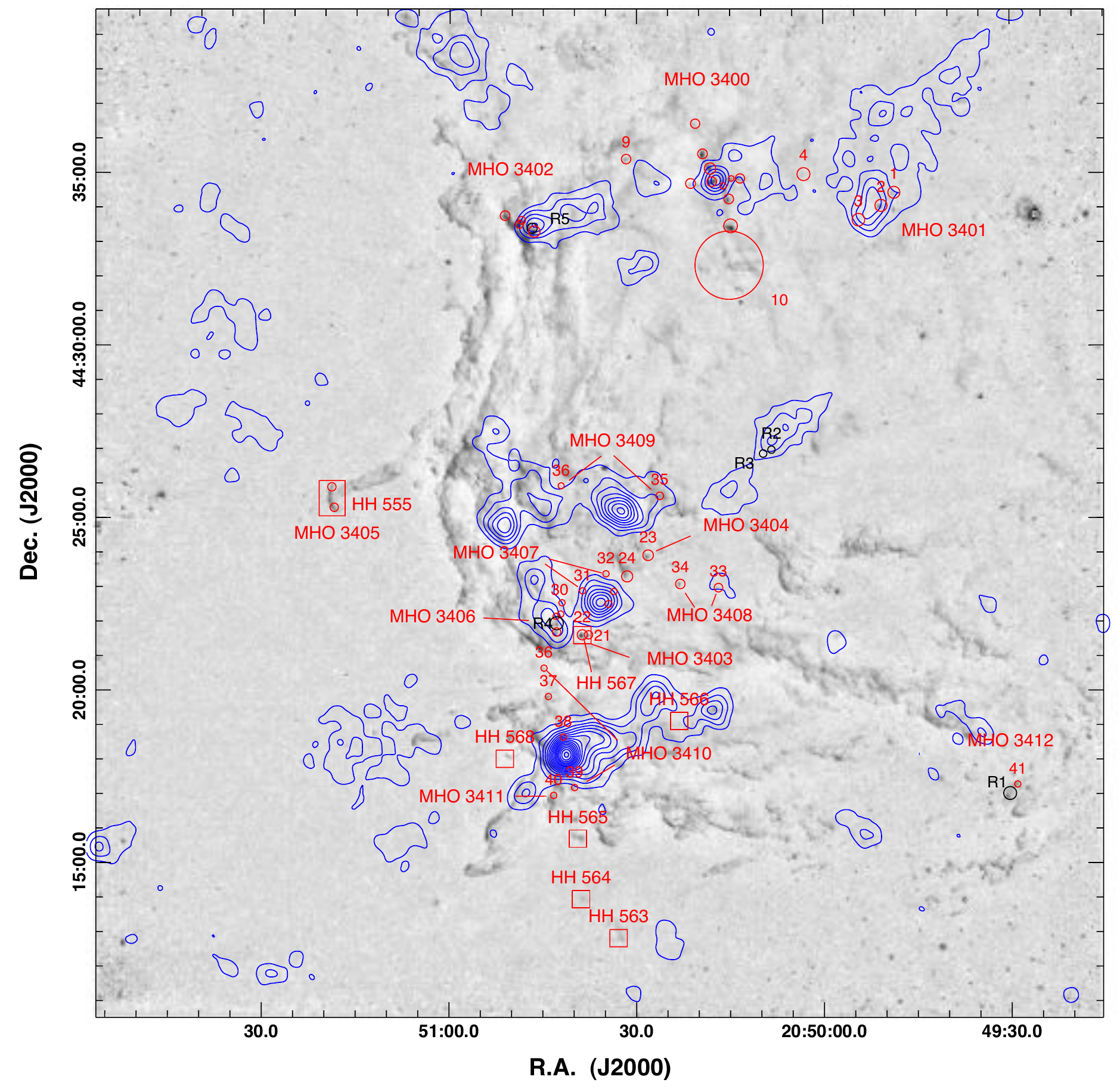}}
\caption{A continuum-subtracted 2.12 $\mu$m H$_2$ image
showing the Pelican Nebula region with   contours showing 
the 1.1 mm dust continuum emission contours from  the Bolocam 
Galactic Plane Survey (BGPS).  Contour levels are shown at 
intervals of 0.05 Jy from 0.05 Jy to 1.0 Jy
per 33\arcsec\ beam.   Circles  show the locations of shock-excited molecular  
hydrogen objects (MHOs).    Small red numbers above some red circles 
correspond to the Table~1 entries for the Pelican region.  
Black circles mark  near-IR reflection nebulae.   The locations of previously 
discovered Herbig-Haro objects \citep{BallyReipurth2003} are indicated by 
squares. }
\label{fig6}
\end{figure}

\begin{figure}
\epsscale{1.0}
\center{\includegraphics[width=1.0\textwidth,angle=0]
  {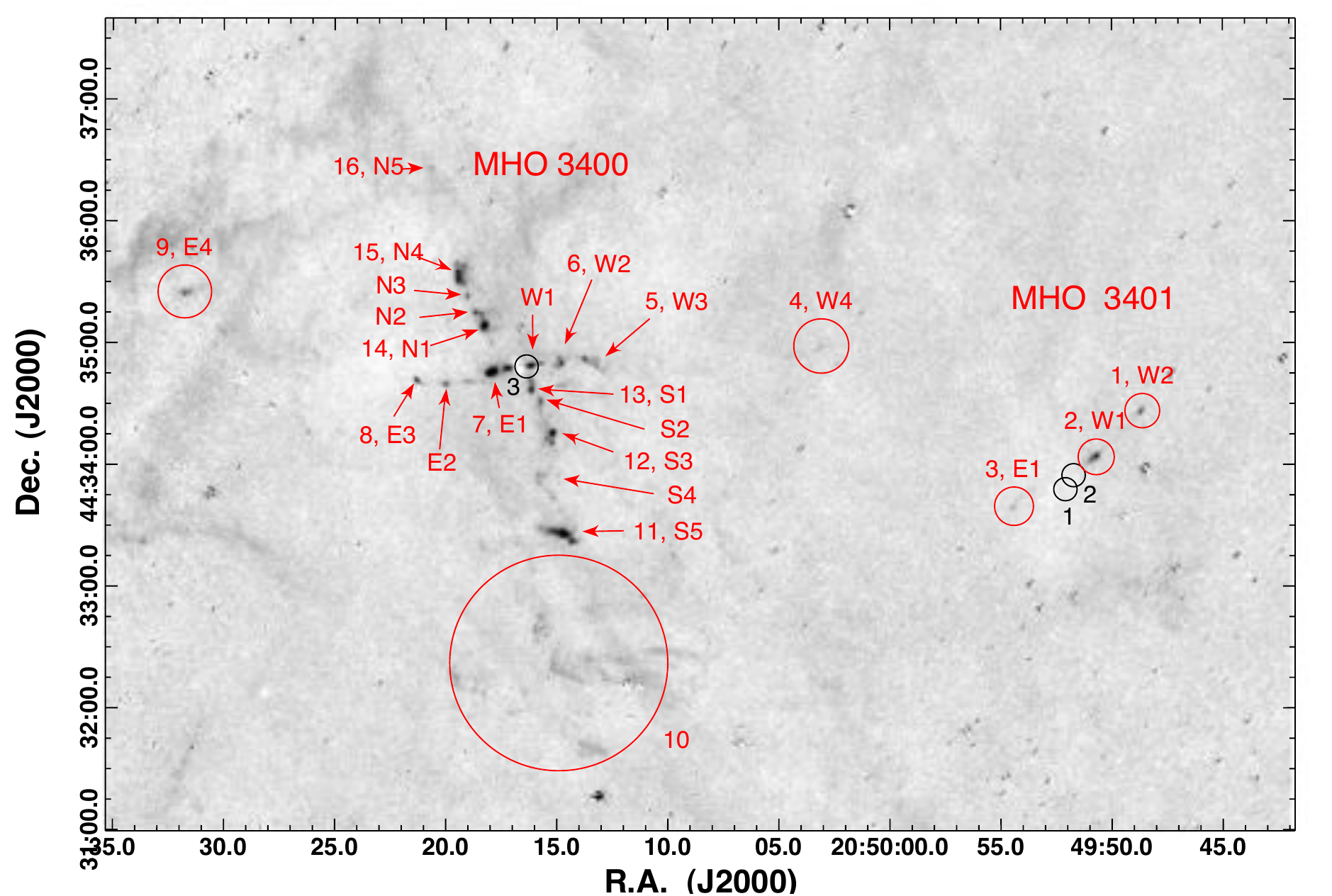}}
\caption{A continuum subtracted H$_2$ image showing the 
quadrupolar outflow MHO 3400  embedded in the northwest clump
in the Pelican Nebula (left) and the outflow MHO 3401  emerging from
a protostar embedded in a BGPS clump.  The large circle encloses
a feature  which may be a shock along the MHO 3401 axis or a fluorescent 
edge.  Black circles  mark locations of the Spitzer-detected
IR sources discussed in the text.  The source labeled as 2 (IRS 2 in Table 1) is more
embedded and likely the source of MHO 3401.  
Source 3 (listed as IRS 3 in Table 1) at the center
of MHO 3400 is  IRAS 20485+4423.  }
\label{fig7}
\end{figure}

\begin{figure}
\epsscale{1.0}
\center{\includegraphics[width=1.0\textwidth,angle=0]
  {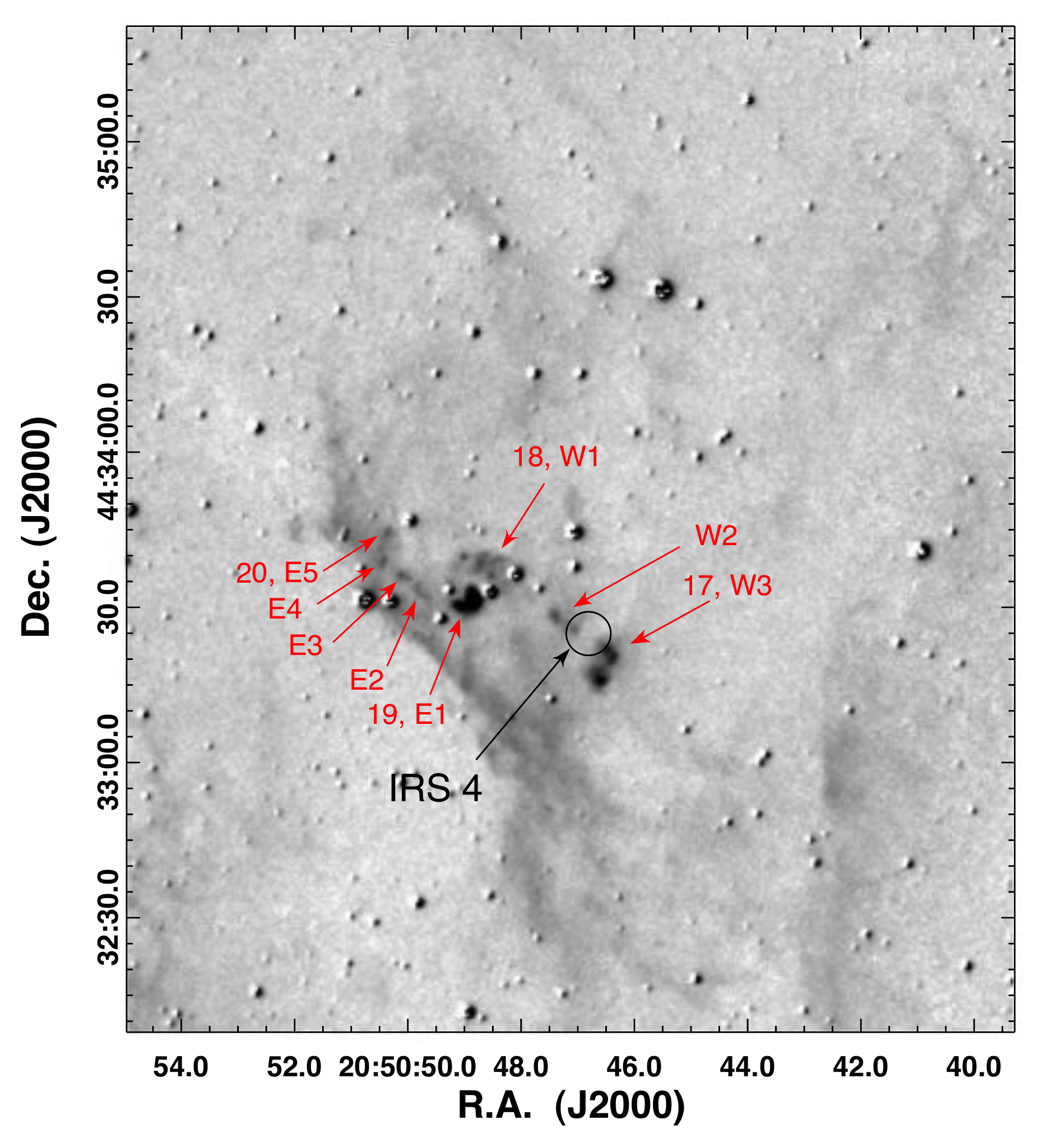}}
\caption{A continuum subtracted 2.12 $\mu$m H$_2$  image showing 
the field around MHO 3402 at the ionization front in the Pelican region.  
As discussed in the text the knots  preceded by `W' (west) and `E' (east) may trace 
at least two distinct outflows from YSOs in this small aggregate.  The source
IRS 4 corresponds to the 8 to 70 $\mu$m object listed in Table~1.}
\label{fig8}
\end{figure}

\begin{figure}
\epsscale{1.0}
\center{\includegraphics[width=1.0\textwidth,angle=0]
  {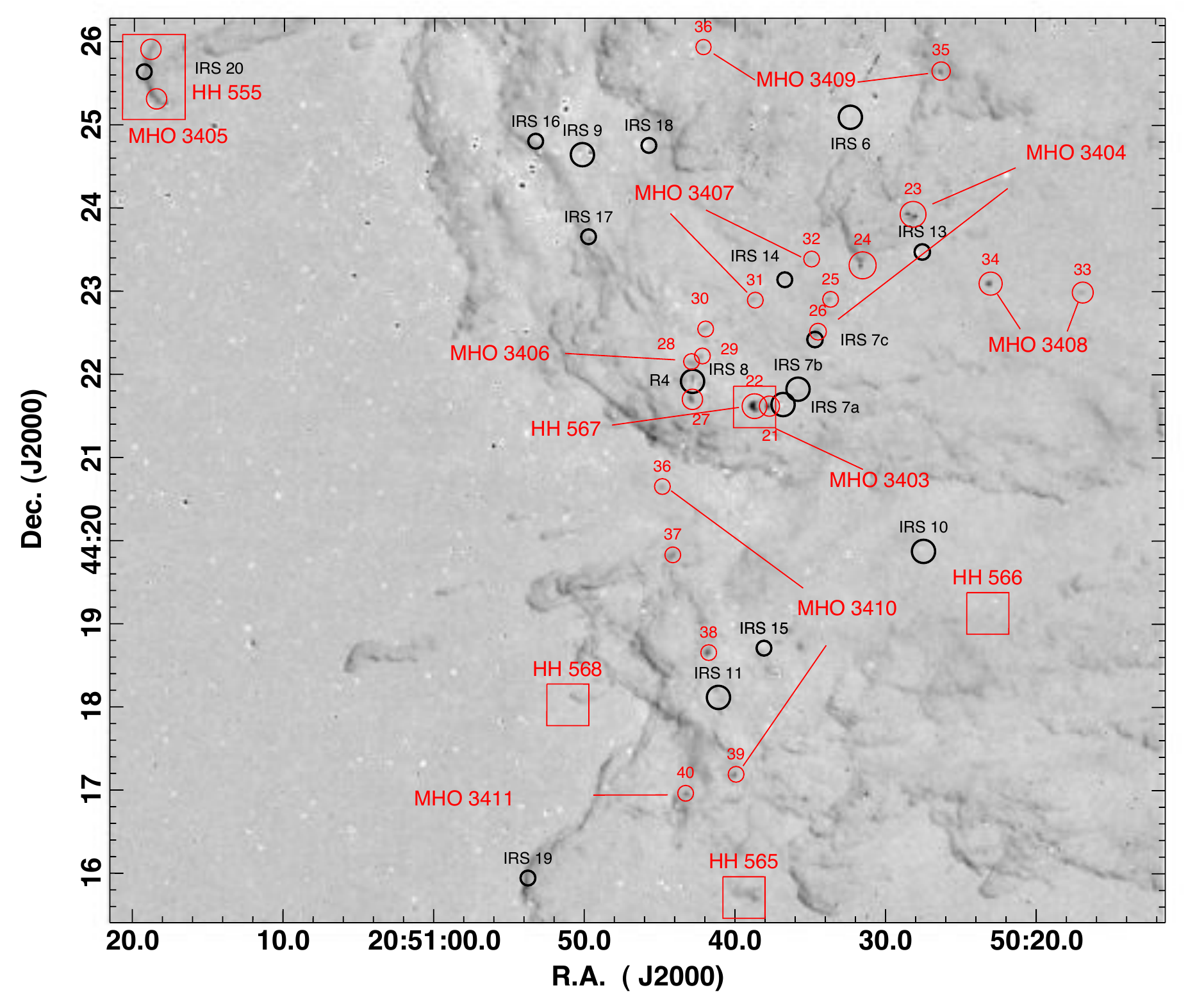}} 
\caption{A 2.12 $\mu$m continuum subtracted H$_2$ 
image showing  the cluster of MHOs and highly embedded YSOs in 
the southeast portion of the Pelican region.   
The brightest, highly embedded 
Spitzer detected YSOs are shown:  black 
circles are 70 $\mu$m sources.}
\label{fig9}
\end{figure}

\begin{figure}
\epsscale{1.0}
\center{\includegraphics[width=1.0\textwidth,angle=0]
  {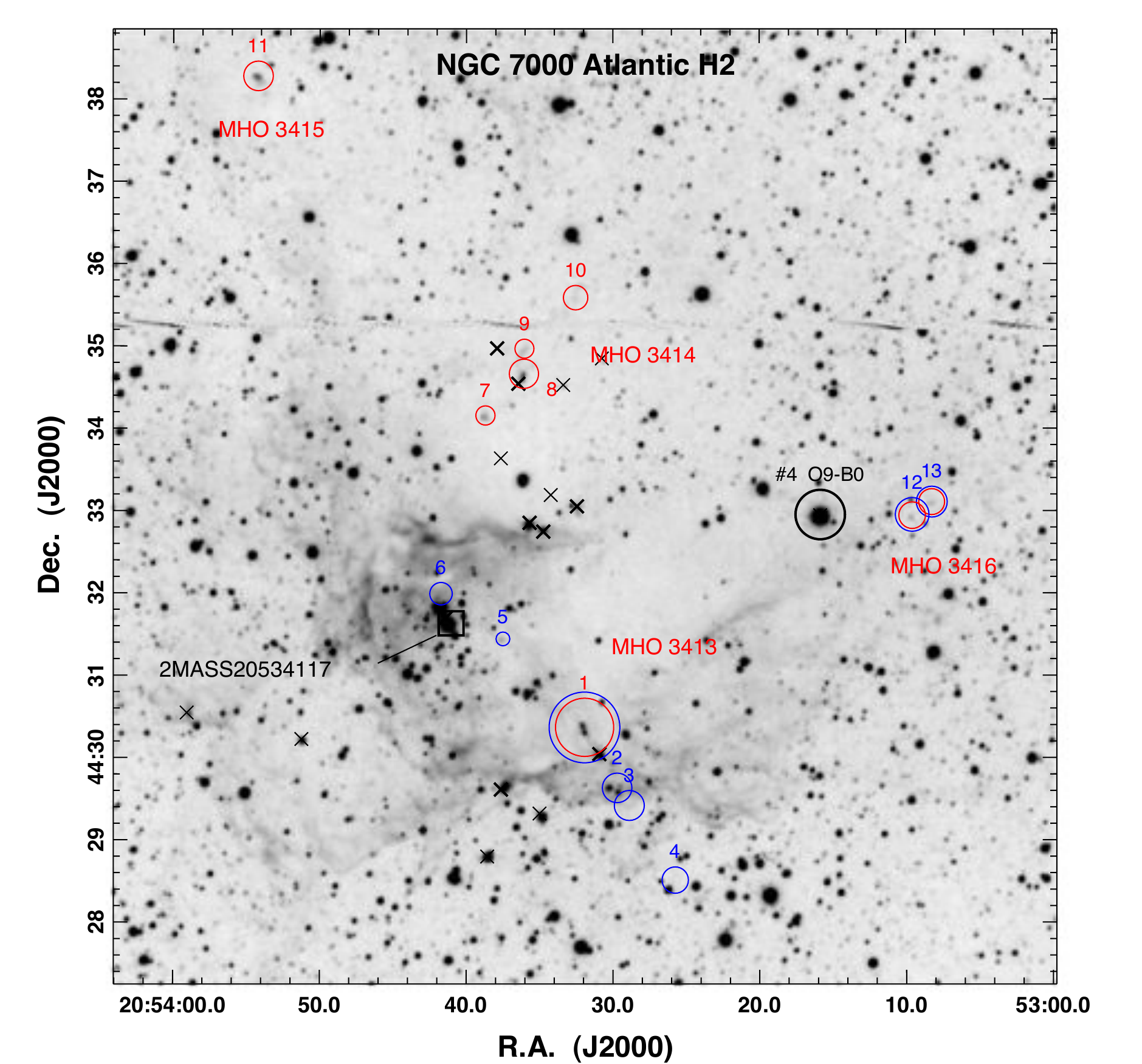}}
\caption{A 2.12 $\mu$m H$_2$  image showing 
the field around  cluster and HII region G085.06$-$0.16 
in the ``Atlantic core" region.  
The brightest 2 $\mu$m source, 2MASS 20534117+44311378 
discussed in the text (only the digits before the `+' are given),
is marked.
Red circles show candidate H$_2$ shocks; blue circles are the
[FeII] knots shown in Figure 11. The corresponding 
K$_s$ image was plagued by clouds and could not be used
for continuum subtraction. The black circle shows Straizys and
Laugalys (2008) star \#4, suspected to be an O9 or B0 massive
star.   The black crosses mark 24 $\mu$m sources.  The brighter ones
have darker marks.  The horizontal band near the top is a cosmetic defect.}
\label{fig10}
\end{figure}

\begin{figure}
\epsscale{1.0}
\center{\includegraphics[width=1.0\textwidth,angle=0]
  {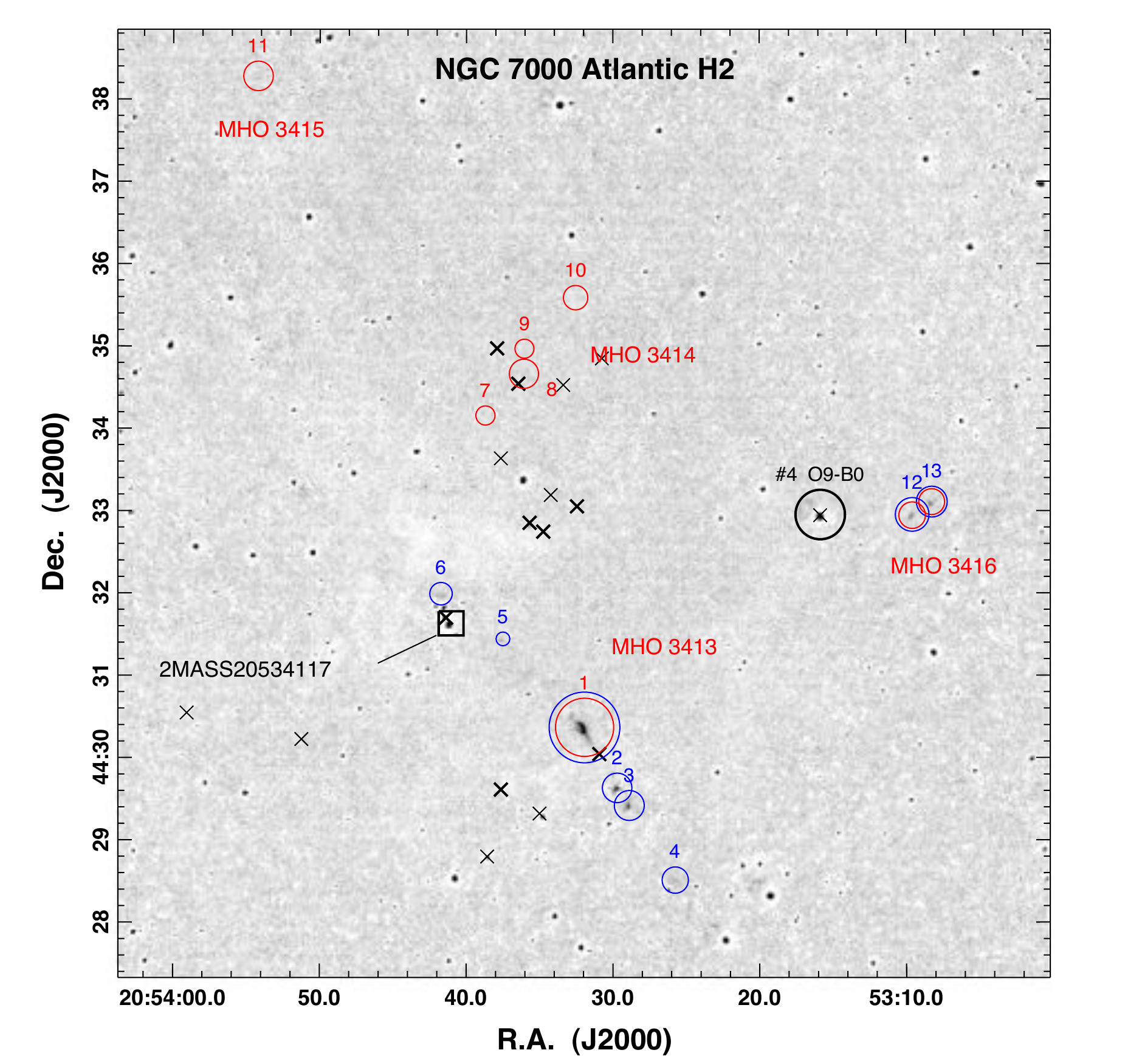}}
\caption{A continuum subtracted 1.64 $\mu$m [FeII]  image showing 
the jet near  the 2MASS source (black square) in the ``Atlantic gulf " region. 
Red circles show candidate H$_2$ shocks; blue circles show the
[FeII] knots. The black circle shows Straizys and
Laugalys (2008) star \#4, suspected to be an O9 or B0 massive
star.   The black crosses mark 24 $\mu$m sources.  The brighter ones
have darker marks.}
\label{fig11}
\end{figure}

\begin{figure}
\epsscale{1.0}
\center{\includegraphics[width=1.0\textwidth,angle=0]
{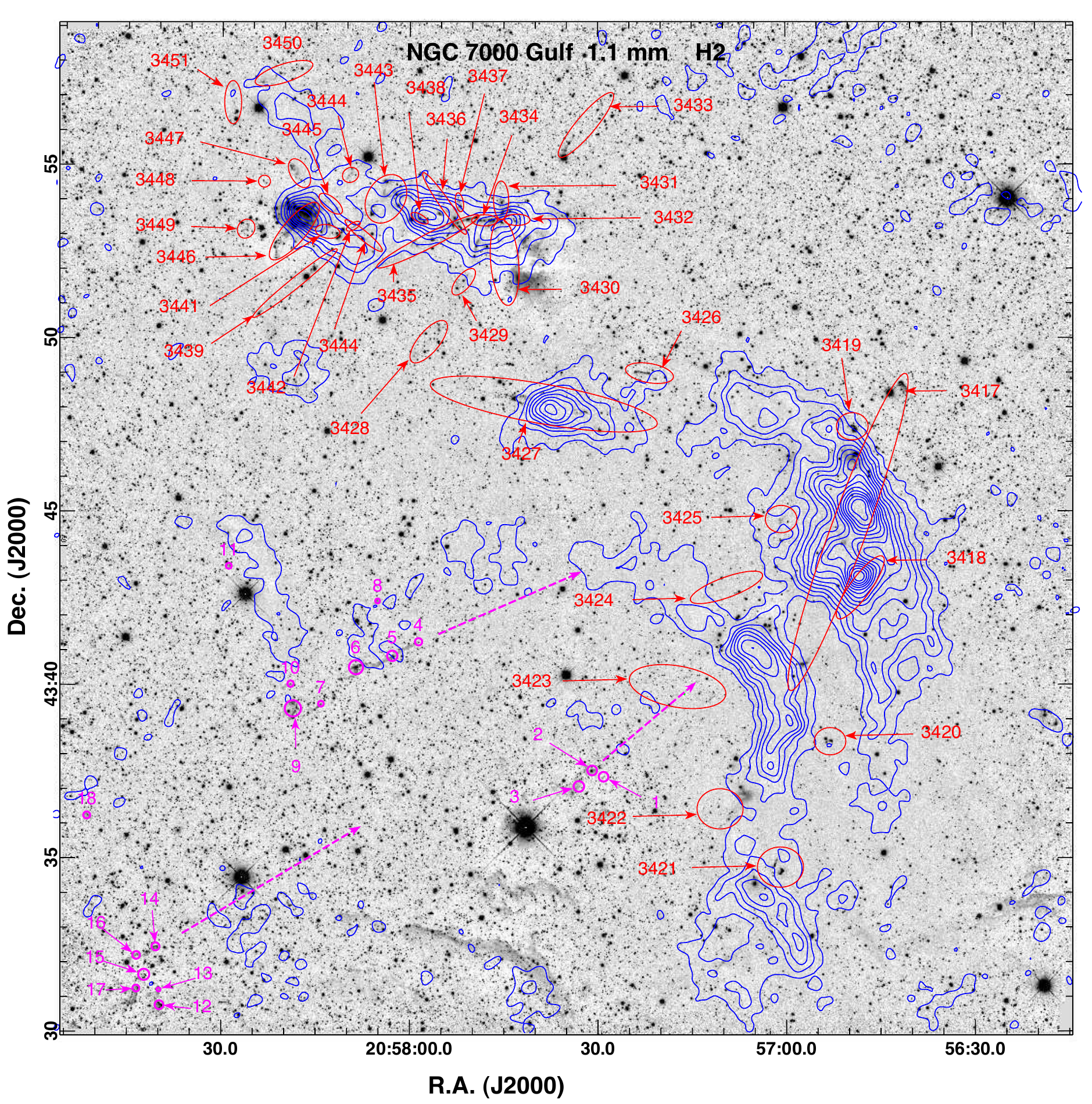}}
\caption{A 2.12 $\mu$m H$_2$  image showing the Gulf of Mexico field.
Red ovals mark MHOs listed in Table 3.    
The cyan circles in the lower left mark cometary clouds
traced by fluorescent H$_2$ emission from cloud edges listed in Table~3
with a letter `C' in the Comments column.  The mean orientations of the
three groups of cometary clouds are indicated by the dashed arrows.
Figures \ref{fig19a}, \ref{fig19b},  and \ref{fig19c} show close-up views of these
objects 
These elongated 
clouds point to one of the exciting stars of the NAP,  star \#8 (the spectral type O5
Comer{\'o}n \&  Pasquali star) .   Contours show $\lambda$ = 1.1 mm BGPS
dust continuum emission with levels at 
0.19,  0.26, 0.34, 0.41, 0.48, 0.55, 0.63, 0.70, 0.78,  0.85, 0, 93, 1.0,  and 1.07 Jy/beam.}  
\label{fig12}
\end{figure}

\begin{figure}
\epsscale{1.0}
\center{\includegraphics[width=0.7\textwidth,angle=0]
{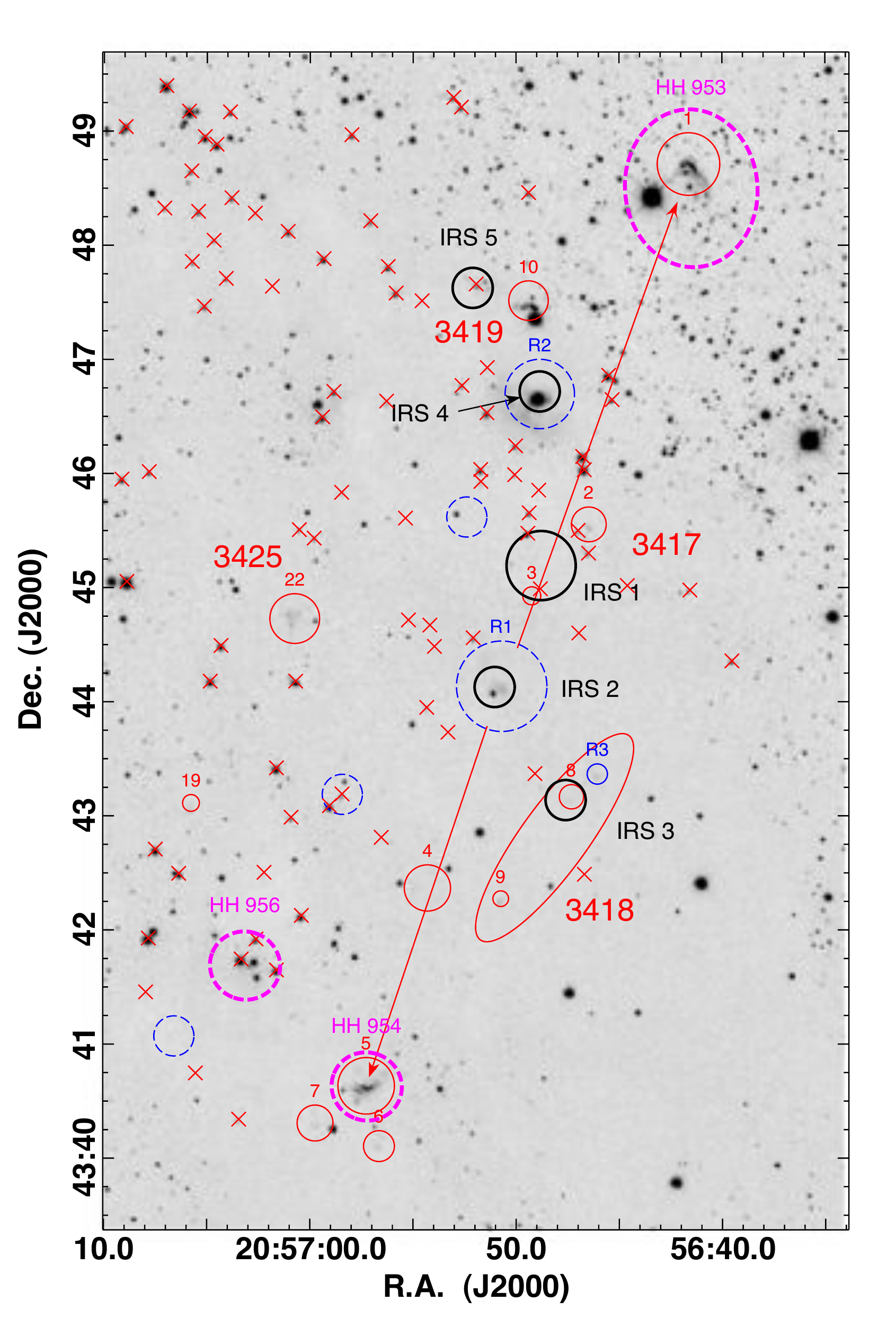}}
\caption{An H$_2$ image showing details in the Gulf SW2 region containing the
large molecular hydrogen outflow MHO 3417.   Thin circles and ovals  (red) 
mark MHOs with numbers listed in Table~3.  
Crosses (red 'x' symbols in the electronic version) 
mark locations of YSOs identified in Guieu et al. (2009).   Thin dashed circles 
(blue) marked with the letter `R'  followed by a number
mark 2 $\mu$m reflection nebulae indicated in Table~3.  Thick black solid circles
mark 70 $\mu$m sources.  Thick dotted ovals and circles mark locations of
HH objects. 
}
\label{fig13}
\end{figure}

\begin{figure}
\epsscale{1.0}
\center{\includegraphics[width=1.0\textwidth,angle=0]
{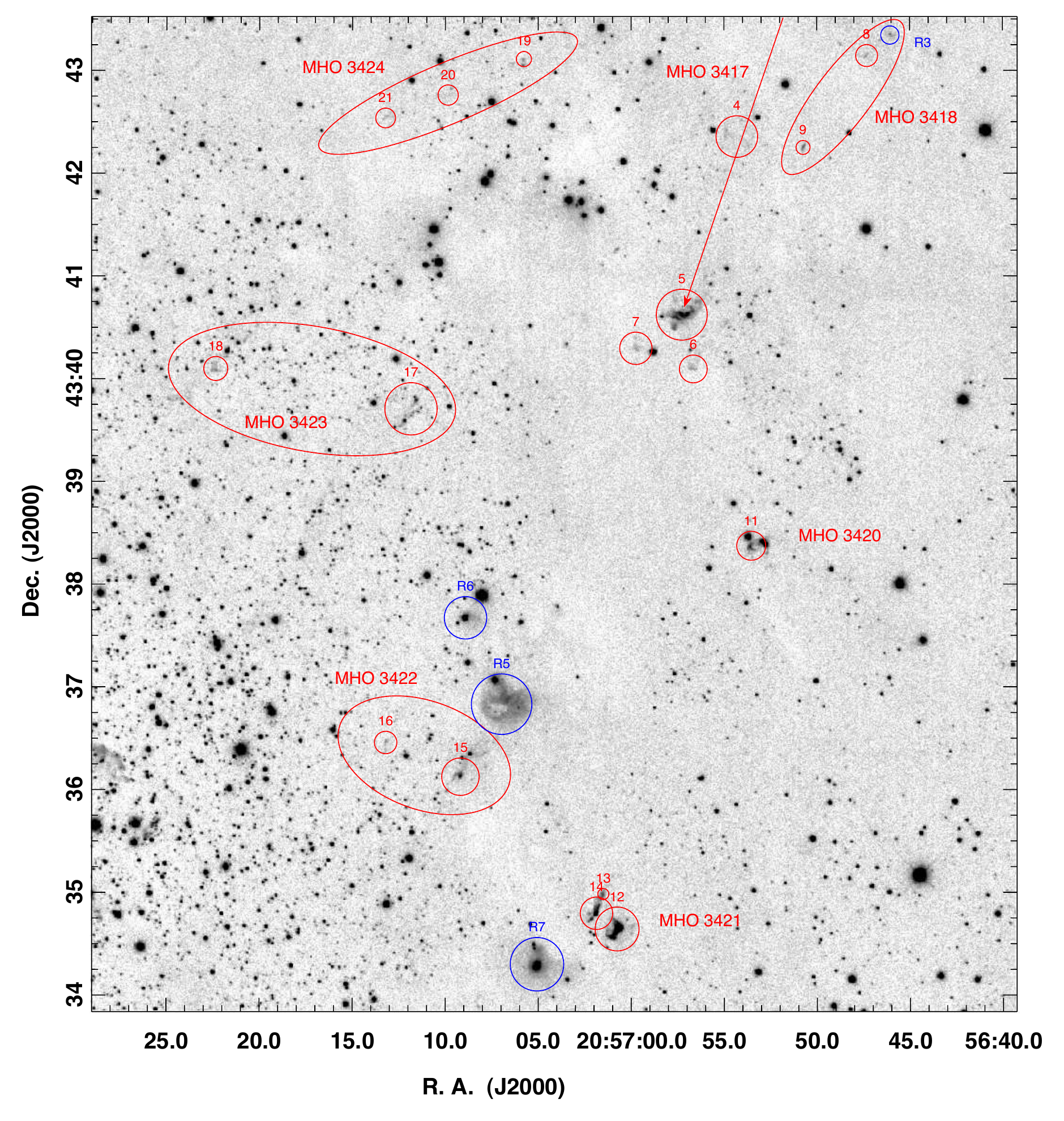}}
\caption{An H$_2$ image showing the field containing `Gulf SW2 SE' and `SW2 S' 
and MHO 3420 through  MHO 3424. 
}
\label{fig13b}
\end{figure}

\begin{figure}
\epsscale{1.0}
\center{\includegraphics[width=1.00\textwidth,angle=90]
    {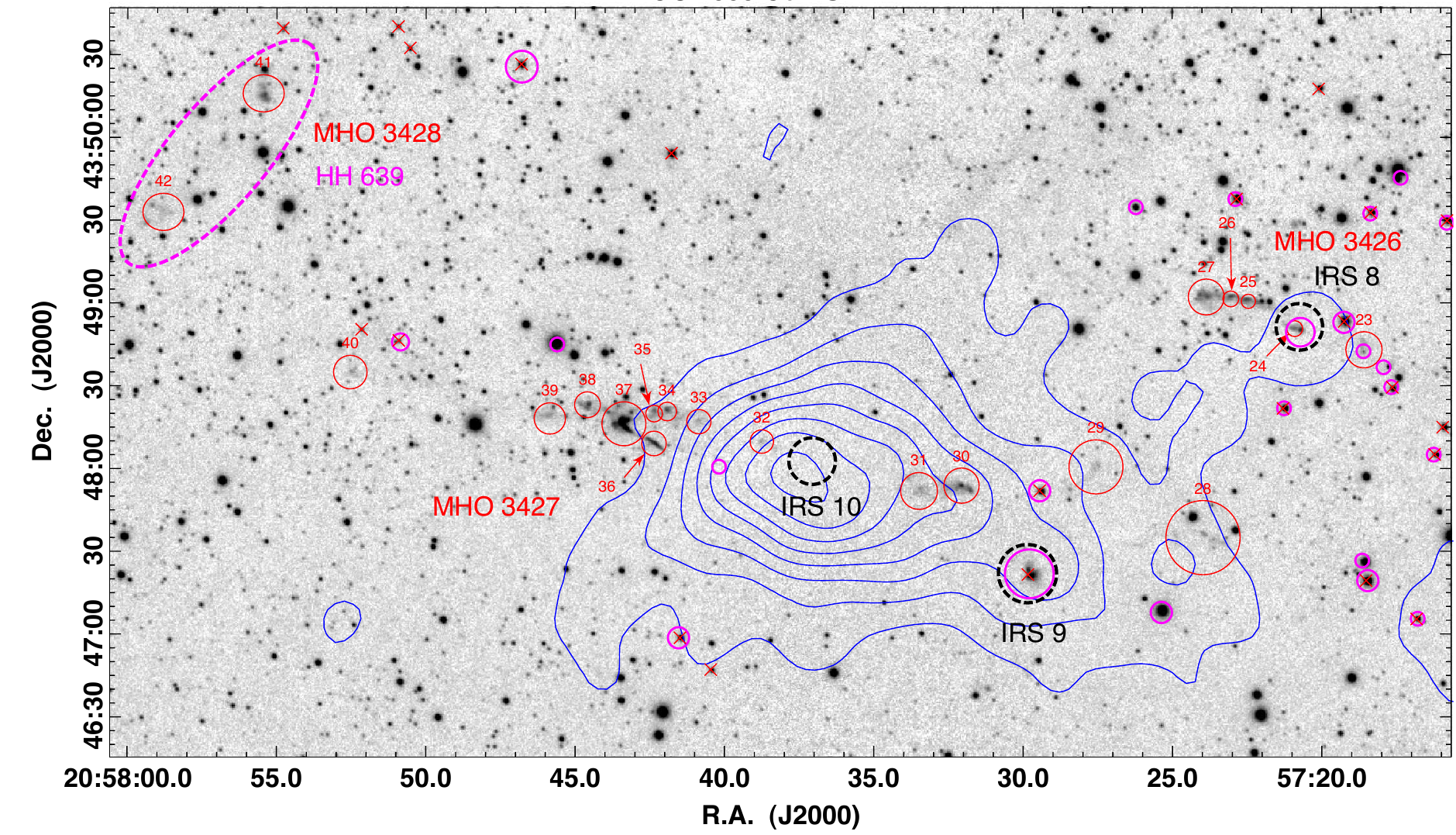}}
\caption{Molecular hydrogen outflows emerging from the 
Gulf SW 1 core (IRDC G84.963-1.117) in NGC 7000.  
Thin red circles with numbers mark MHOs 
located in the Gulf of Mexico region (Table~3).
Bold dashed  circles mark Spitzer 70 $\mu$m sources.
Bold solid  circles (magenta in electronic version) 
mark Spitzer 24 $\mu$m sources;
'crosses ('x's)  mark YSOs identified by Guieu et al. (2009).
The source of the large east-west H$_2$ outflow, MHO 3427, 
may be a Class 0 protostar visible only at a wavelength of
70 $\mu$m and marked by the black circle above IRS 10 in 
the image.   Contours (blue) show  the 1.1 mm dust continuum 
emission detected by the Bolocam  Galactic Plane Survey.  
Contour levels are shown at  intervals of 0.074 Jy from 0.04 Jy to 1.7 Jy
per 33\arcsec\ beam.   }
\label{fig14}
\end{figure}

\begin{figure}
\epsscale{1.0}
\center{\includegraphics[width=1.1\textwidth,angle=0]
   {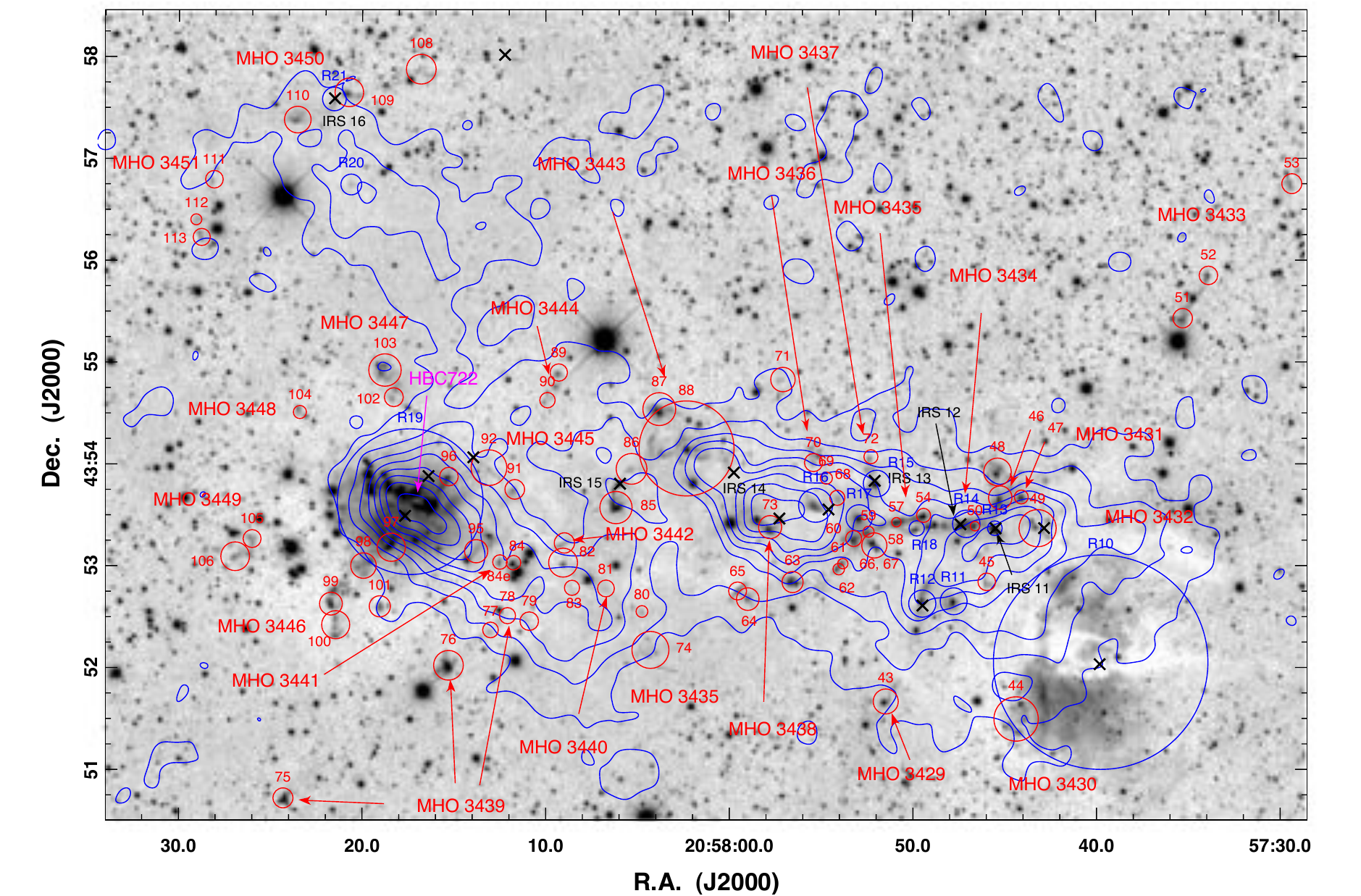}}
\caption{Molecular hydrogen objects (red in the electronic version of the paper) 
and reflection nebulae (blue circles in the 
electronic version of the paper) located in the Gulf `core E' and `core W'  regions.
The intensity is displayed with a log stretch.
The numbers correspond to the entries in Table~3.  
Reflection nebulae are preceded by a letter `R'.
70 $\mu$m sources are marked with an `x'.  Contours trace the 1.1 mm dust
continuum emission. 
Circles indicate the locations of  MHOs and reflection nebulae as listed in 
Table~3.  Some  objects listed in Table~3 such as knots 55 and 56 that
lie between 54 and 57 are not marked to avoid confusion from overlapping 
symbols.   Knot 82n in Table~3 is marked as MHO 3442.
}
\label{fig15}
\end{figure}

\begin{figure}
\epsscale{1.0}
\center{\includegraphics[width=1.1\textwidth,angle=0]
   {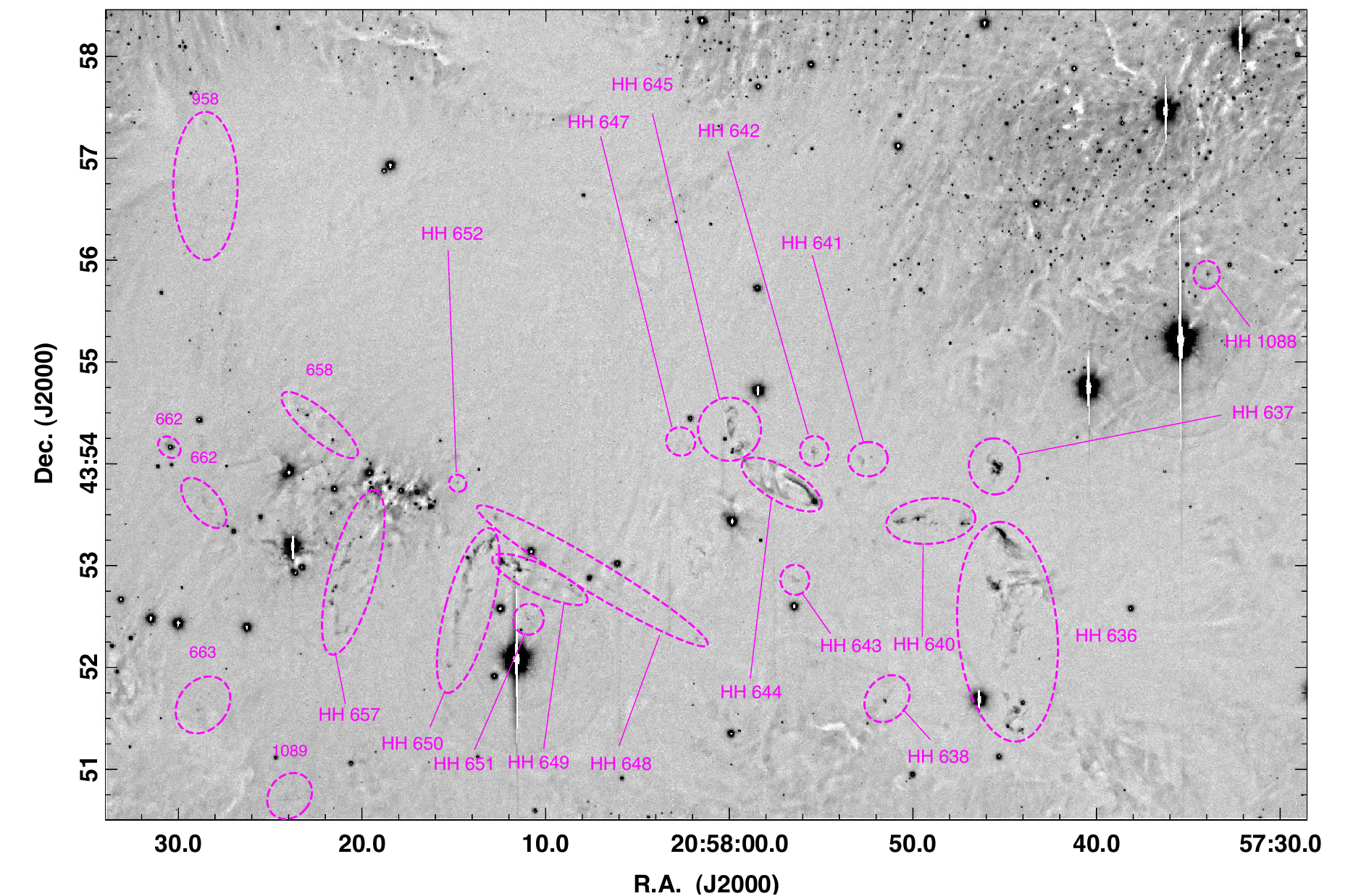}}
\caption{A greyscale image showing $\lambda$ 0.6717/0.6731 $\mu$m 
[SII] emission.  The image was processed with a filtering algorithm to
remove large-scale, large-amplitude background variations.  First, the image
is convolved with a 21-pixel top-hat function.   Second, the convolved image
is scaled by a numerical factor of 0.9 to create a mask image.  Third, the mask
image is subtracted from the original image.  
The region shown is the same as Figure \ref{fig15}. }
\label{fig16}
\end{figure}

\begin{figure}
\epsscale{1.0}
\center{\includegraphics[width=1.1\textwidth,angle=0]
   {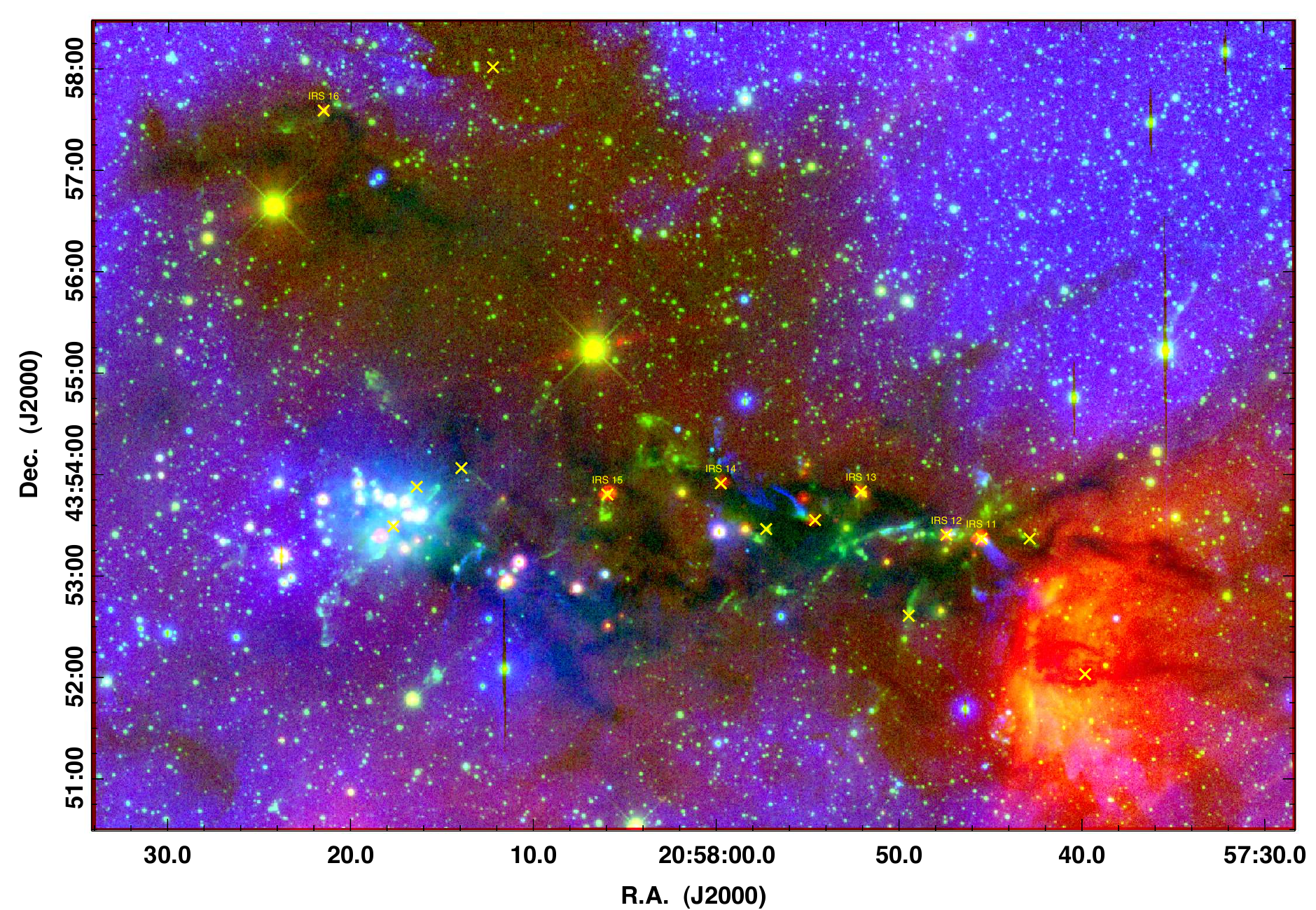}}
\caption{(Electronic version only)
A color image combining  NEWFIRM $\lambda$ 2.12 $\mu$m H$_2$ 
emission (green),  Subaru  0.67 $\mu$m [SII] (blue),  and  
Spitzer 8 $\mu$m (red).     Yellow `X' symbols mark locations of highly
embedded 70 $\mu$m sources.  The two bright yellow sources
are red giant stars unrelated to the NAP.
The region shown is the same as Figure \ref{fig15}. 
}
\label{fig17}
\end{figure}

\begin{figure}
\epsscale{1.0}
\center{\includegraphics[width=1.1\textwidth,angle=0]
   {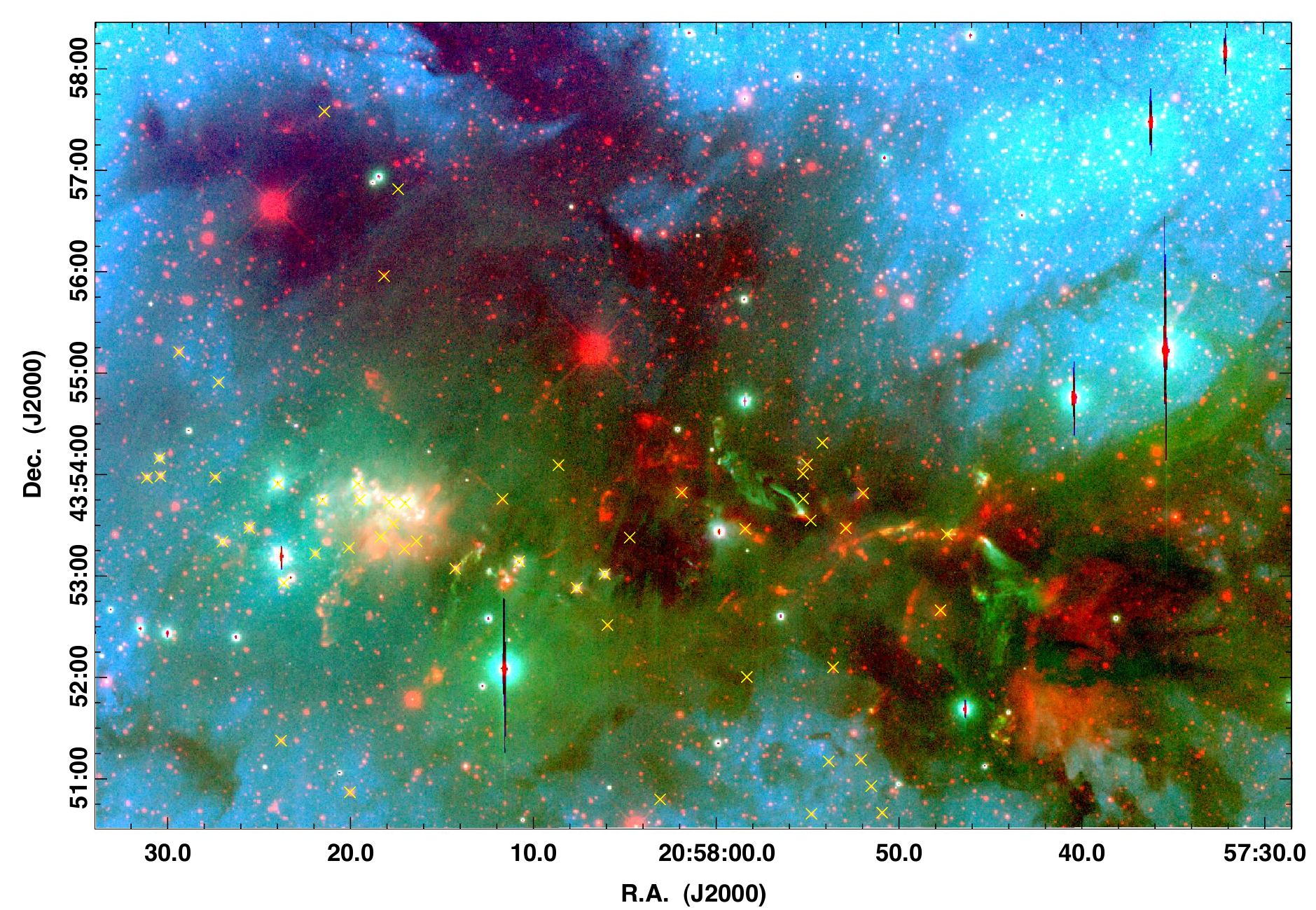}}
\caption{(Electronic version only)
A color image showing 0.6563 $\mu$m H$\alpha$ (blue), 
$\lambda \lambda $ 0.6717/0.6731 $\mu$m [SII] (green), and
$\lambda$ 2.12 $\mu$m H$_2$ 
emission (red).   Crosses mark YSOs identified by Guieu et al. (2009).
The region shown is the same as Figure \ref{fig15}. 
}
\label{fig18}
\end{figure}

\begin{figure}
\epsscale{1.0}
\center{\includegraphics[width=1.1\textwidth,angle=0]
   {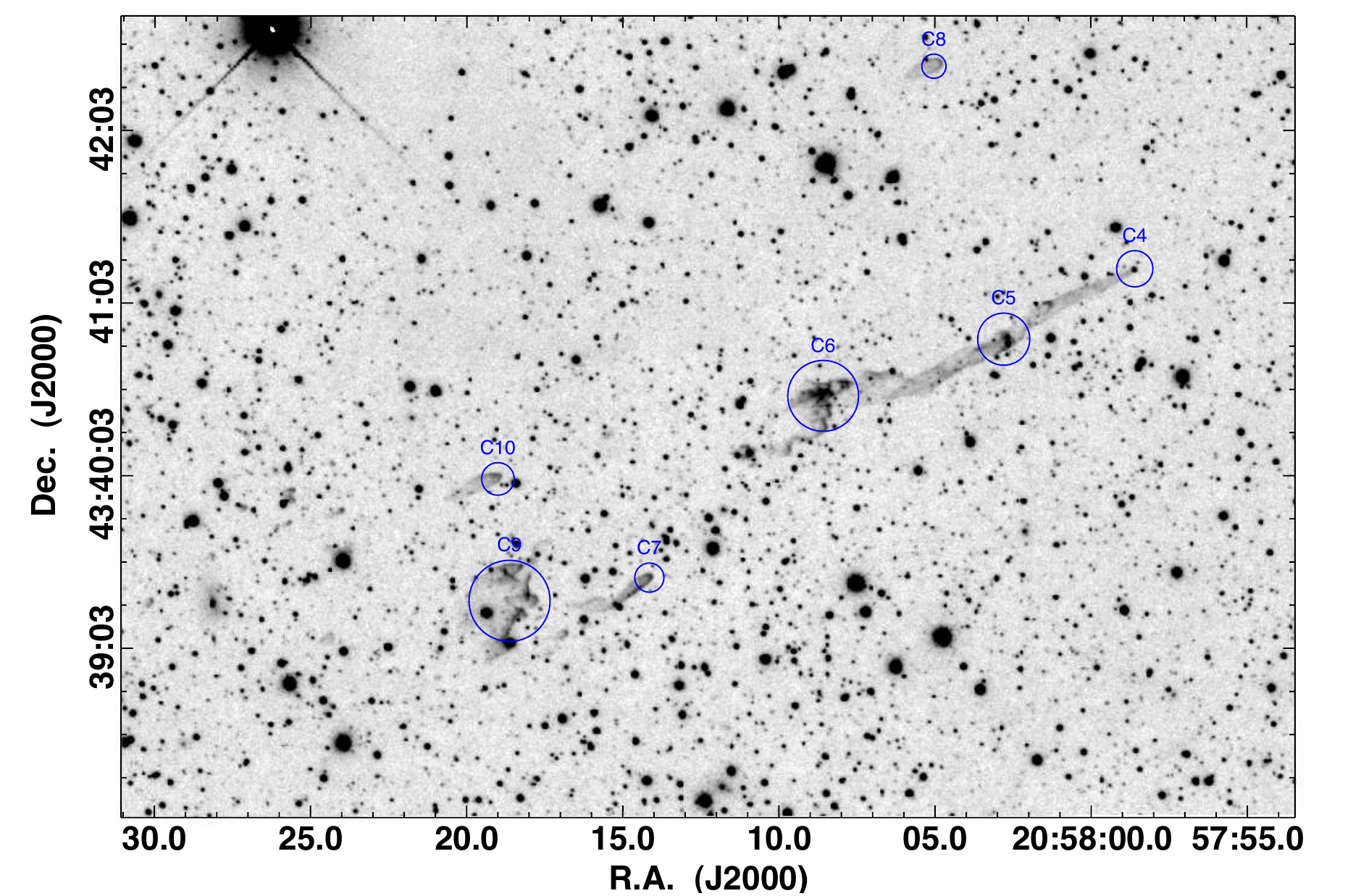}}
\caption{
2.12 $\mu$m image showing several cometary clouds (C4 through C10) southeast of the Gulf of Mexico.
}
\label{fig19a}
\end{figure}

\begin{figure}
\epsscale{1.0}
\center{\includegraphics[width=1.1\textwidth,angle=0]
   {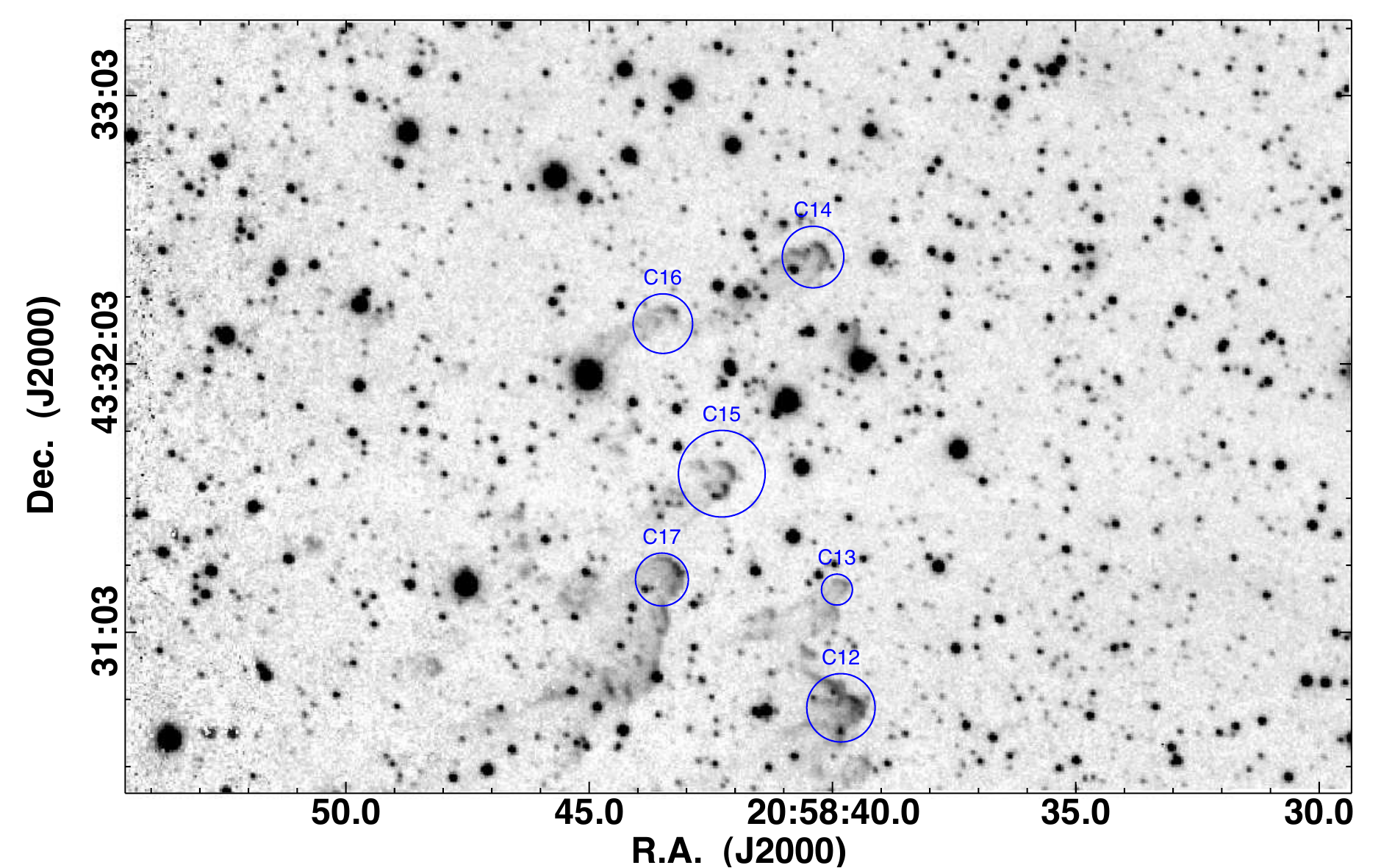}}
\caption{
2.12 $\mu$m image showing several cometary clouds (C12 through C17) southeast of the Gulf of Mexico.
}
\label{fig19b}
\end{figure}

\begin{figure}
\epsscale{1.0}
\center{\includegraphics[width=1.1\textwidth,angle=0]
   {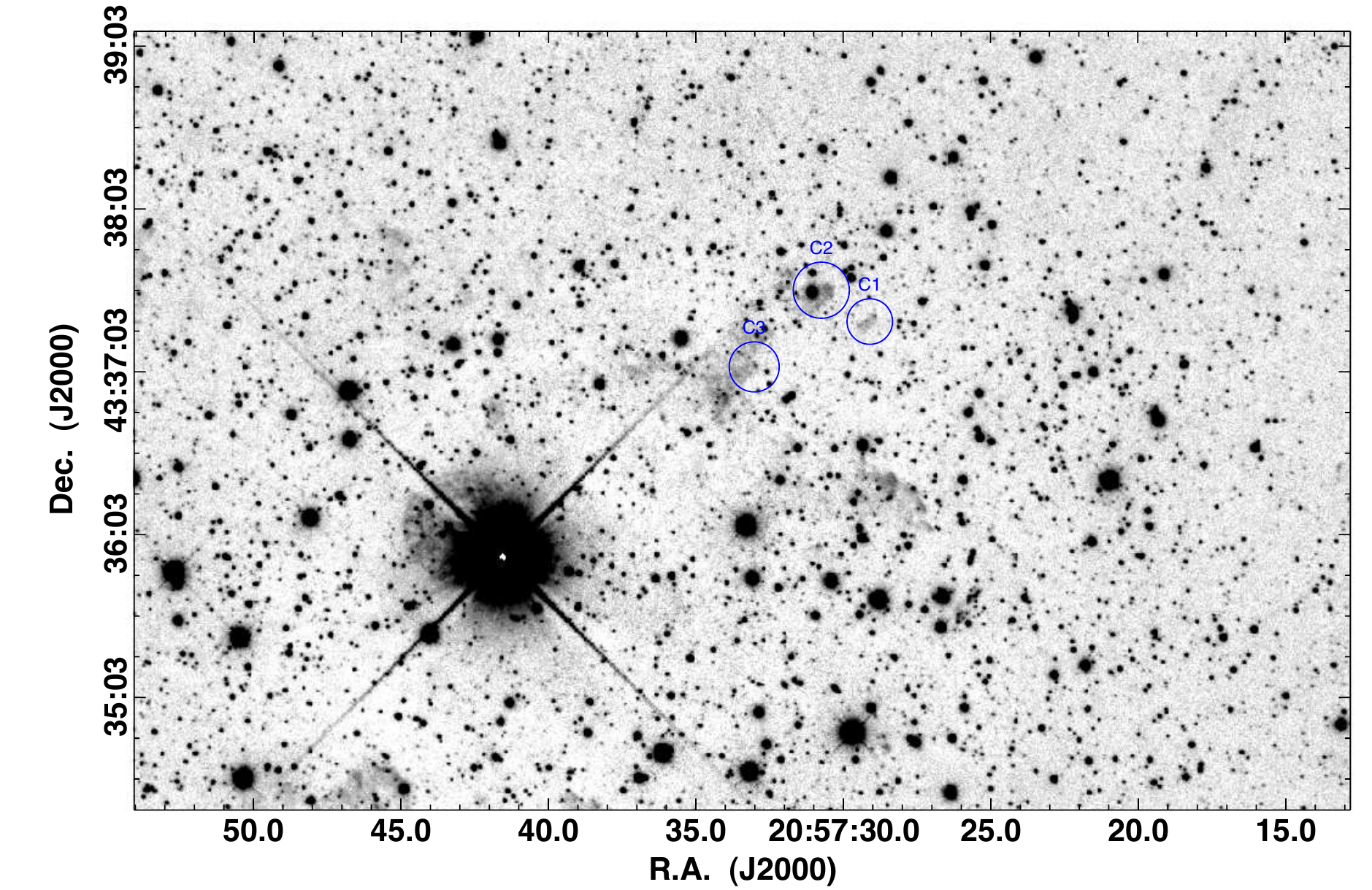}}
\caption{
2.12 $\mu$m image showing several cometary clouds (C1 through C3) southeast of the Gulf of Mexico.
}
\label{fig19c}
\end{figure}

The Gulf SW2 clump has a mass of about 500 \msol .  However, it only
contains a relatively small number of MHOs, including the parsec-scale 
outflow MHO 3417, which may be powered by a protostar that could be 
evolving into a massive star.      The ratio of YSOs or MHOs to BGPS mass
is relatively low.  Thus, the Gulf SW2 clump may be in a relatively earlier
evolutionary stage than `Gulf core' (core E and core W in Figure \ref{fig3}).  
An activity index, formed from the 
ratio of number of shocks divided by clump mass is proposed as
an indicator of the relative evolutionary stage of a clump.

\acknowledgments{This research was supported by NSF grants 
\# AST 0708403 (The Bolocam Millimeter Wavelength Survey of the Northern Galactic Plane) 
and
\# AST 1009847 (Formation of Galactic massive Stars and Star Clusters).   
BR acknowledges support by the National
Aeronautics and Space Administration through the NASA Astrobiology
Institute under Cooperative Agreement No. NNA09DA77A issued through
the Office of Space Science. 
GSS is supported through grants received from NASA.
This project made use of ds9 (http://hea-www.harvard.edu/RD/ds9/site/Home.html).  
This research has made use of the SIMBAD database, operated at CDS, 
Strasbourg, France.
Observations  reported here were obtained at the Kitt Peak National 
Observatory which is operated by  the National Optical Astronomy Observatories a
nd AURA Inc. with support from the National Science foundation.  
We thank an anonymous referee for a very thorough reading of the manuscript
and useful suggestions which improved the paper and figures.}

\clearpage

\bibliography{ms}

\clearpage

%

%

%

%

%

%

%

%

%

%

%

%

%

%

%

%

%

%


\clearpage


%


%

%

%

\clearpage


\begin{deluxetable}{cclll}
  
\centering

\tablecaption{ 
      MHOs and Reflection Nebulae in the Pelican 
      \label{table1}}
    
\tablehead{
    \colhead{$\alpha$} & 
    \colhead{$\delta$}  & 
    \colhead{\#$^1$}    &
    \colhead{MHO$^2$}       &
     \colhead{Comments}
       \\

    \colhead{J2000} & 
    \colhead{J2000} & 
    \colhead{ }           &
    \colhead{ }           &
    \colhead{} 
   
    }
      
\startdata

  20 50 03.1   &  +44 34 58   & 4  &  3400 & W4  west-end of east-west flow \\
  20 50 13.4   &  +44 34 51   & 5  &  3400 & W3 \\
  20 50 14.8   &  +44 34 51   & 6  &  3400 & W2 \\
  20 50 16.1   &  +44 34 50   &     &  3400 & W1  jet at PA $\sim$ 275$\deg$  \\
  20 50 17.9   &  +44 34 46   & 7  &  3400 & E1   jet at PA $\sim$ 95$\deg$ \\
  20 50 20.0   &  +44 34 40   &     &  3400 & E2 \\
  20 50 21.3   &  +44 34 42   & 8  &  3400 & E3 \\
  20 50 31.7   &  +44 35 25   & 9  &  3400 & E4  east end of east-west flow \\

  20 50 15.1   &  +44 32 20   & 10 &           & bow from MHO 3401 or fluorescent edge? \\
  20 50 14.9   &  +44 33 25   & 11 & 3400 &  S5  south-end of pair of north-south flow \\
  20 50 15.8   &  +44 33 51   &       & 3400 &  S4 dim complex of knots \\
  20 50 15.2   &  +44 34 15   & 12 & 3400 &  S3 bright knot \\
  20 50 15.7   &  +44 34 32   &       & 3400 &  S2 bright knot  \\
  20 50 16.1   &  +44 34 38   & 13 & 3400 &  S1 jet at PA $\sim$ 185$\deg$ \\
  20 50 18.1   &  +44 35 08   & 14 & 3400 &  N1 \\
  20 50 18.6   &  +44 35 15   &       & 3400 &  N2 \\
  20 50 19.0   &  +44 35 24   &       & 3400 &  N3 \\
  20 50 19.4   &  +44 35 34   & 15 &  3400 & N4 north-end of 4 knot chain \\
  20 50 20.6   &  +44 36 26   & 16 &  3400 & N5 north-end of north-south flow? \\

  20 49 48.6   &  +44 34 26   & 1  & 3401  & W2  Flow from isolated clump \\
  20 49 50.7   &  +44 34 04   & 2  & 3401  & W1 \\
  20 49 54.3   &  +44 33 39   & 3  & 3401  & E1 \\

  20 50 46.4   &  +44 33 19   & 17 & 3402 & W3  Cluster of knots; 2 flows? \\
  20 50 47.3   &  +44 33 29   &       & 3402 & W2 \\
  20 50 48.5   &  +44 33 38   & 18 &  3402 & W1 \\
  20 50 48.8   &  +44 33 31   & 19 & 3402  & E1 \\
  20 50 49.8   &  +44 33 33   &       & 3402 &  E2 \\
  20 50 50.0   &  +44 33 36   &       & 3402 &  E3 \\
  20 50 50.4   &  +44 33 38   &       & 3402 &  E4 \\
  20 50 50.3   &  +44 33 44   & 20  & 3402 & E5 \\
   
  20 50 37.7   &  +44 21 38   & 21  & 3403 & HH 567  \\
  20 50 38.7   &  +44 21 38   & 22  & 3403 & HH 569 bow \\
  
  20 50 28.2   &  +44 23 56   & 23   & 3404 & part of SE-NW chain \\
  20 50 31.5   &  +44 23 19   & 24   & 3404 &      " \\
  20 50 33.6   &  +44 22 55   & 25   & 3404 &      " \\
  20 50 34.5   &  +44 22 31   & 26   & 3404 &      "  \\
  
  20 51 18.9   &  +44 25 54   & -      & 3405 &    HH 555 north \\
  20 51 18.5   &  +44 25 19   & -      & 3405 &    HH 555 south \\

  20 50 42.8   &  +44 21 42   & 27 & 3406 &  south of IRAS 2489+4410  \\
  20 50 42.9   &  +44 22 10   & 28 & 3406 &  north of  IRAS 2489+4410 \\
  20 50 42.2   &  +44 22 14   & 29 & 3406 &    "  \\
  
  20 50 42.0   &  +44 22 33   & 30 & 3407 &    "   \\
  20 50 38.7   &  +44 22 54   & 31 & 3407 &  northwest of  IRAS 2489+4410  \\
  20 50 34.9   &  +44 23 24   & 32 & 3407 &     \\
 
  20 50 16.9   &  +44 22 59   & 33  & 3408 & fainter of two knots \\
  20 50 23.0   &  +44 23 06   & 34  & 3408 & brighter knot \\
  
  20 50 26.3   &  +44 25 39   & 35  & 3409 & compact knot west of IRS 6  \\
  20 50 42.1   &  +44 25 57   & 35  & 3409 & compact knot northeast of IRS 6  \\
  
  20 50 44.8   &   +44 20 40  & 36  & 3410  & north of IRAS 20489+4406 \\
  20 50 44.1   &   +44 19 50  & 37  & 3410  &   "   \\
  20 50 41.7   &   +44 18 40  & 38  & 3410  &   "   \\
  20 50 39.9   &   +44 17 12  & 39  & 3410  & south of IRAS 20489+4406  \\
  20 50 39.4   &   +44 15 43  &   -    & 3410  &  HH 565   \\ 
  20 50 38.9   &   +44 13 58  &   -    & 3410  &  HH 564   \\ 
  20 50 32.9   &   +44 12 50  &    -   & 3410  &  HH 563   \\ 
       
  20 50 43.3   &   +44 16 58  & 40  & 3411  &    east of HH 3410  \\
    
  20 49 29.0   &  +44 17 17   & 41 & 3412  &   compact bow at west edge \\
    
  20 49 30.2   &  +44 17 01   &       &          &  R1 \\
  20 50 08.4   &  +44 26 59   &       &         &   R2 \\
  20 50 09.7   &  +44 26 53   &       &         &   R3 \\
  20 50 42.8   &  +44 21 56   &       &         &  R4 \\
  20 50 46.8   &  +44 33 24   &       &         &  R5 \\
  
  20 49 52.1	 & +44 33 48      & & & Pelican IRS 1 \\
  20 49 51.8   	 & +44 33 55 	 & & & Pelican IRS 2; source of MHO 3401 \\
  20 50 16.6   	 & +44 34 49  	 & & & Pelican IRS 3; source of MHO 3400, IRAS  IRAS 20485+4423  \\
  20 50 46.8  	 &  +44 33 25   	  & & &Pelican IRS4; MHO 3402 \\
  20 50 09.2	 & +44 26 51	 & & & Pelican IRS5a \\ 
  20 50 07.5	 & +44 26 59	 & & & Pelican IRS5b \\
  20 50 32.3	 & +44 25 06	 & & & Pelican IRS6 \\
  20 50 36.8	 & +44 21 40	 & & & Pelican IRS7a \\
  20 50 35.8	 & +44 21 50	 & & & Pelican IRS7b \\
  20 50 34.8	 & +44 22 23	 & & & Pelican IRS7c \\
  20 50 42.8	 & +44 21 55	 & & & Pelican IRS8 \\
  20 50 50.1	 & +44 24 39	 & & & Pelican IRS 9 \\
  20 50 27.4	 & +44 19 53	 & & & Pelican IRS 10 \\
  20 50 40.8	 & +44 18 07	 & & & Pelican IRS 11 \\
  20 50 27.8	 & +44 32 06	 & & & Pelican IRS 12 \\
  20 50 27.5	 & +44 23 28	 & & & Pelican IRS 13 \\
  20 50 36.7	 & +44 23 09	 & & & Pelican IRS 14 \\
  20 50 37.9	 & +44 18 45	 & & & Pelican IRS 15 \\
  20 50 53.3	 & +44 24 49	 & & & Pelican IRS 16 \\
  20 50 49.8	 & +44 23 40	 & & & Pelican IRS 17 \\
  20 50 45.7	 & +44 24 46	 & & & Pelican IRS 18 \\
  20 50 53.7	 & +44 15 57	 & & & Pelican IRS 19 \\
  20 51 19.3 	 & +44 25 38	 & & & Pelican IRS 20 \\
  20 49 52.8	 & +44 15 38	 & & & Pelican IRS 21 \\

\enddata

\tablecomments{
 {\bf[1]}:  The numbers refer to those marked in Figure \ref{fig5}.
  {\bf[2]}:  MHO (Molecular Hydrogen Object) numbers refer to the catalog entries in the MHO catalog (http://www.astro.ljmu.ac.uk/MHCat/). 
 {\bf[3]}:  Reflection nebulae are indicated by a letter `R' followed by
a number which corresponds to the entries in the Figures.
               }
\end{deluxetable}

\clearpage

\begin{deluxetable}{cclll}
  
\centering

\tablecaption{ 
      MHOs and Reflection Nebulae in the Atlantic Region 
      \label{table2}}
    
\tablehead{
    \colhead{$\alpha$} & 
    \colhead{$\delta$} & 
     \colhead{$\#$}& 
      \colhead{MHO} & 
    \colhead{Comments}
       \\

    \colhead{J2000} & 
    \colhead{J2000} & 
    \colhead{}  &
    \colhead{}  &
    \colhead{} 
   
    }
      
\startdata

 20 53 31.9 & +44 30 22 & 1 & MHO 3413   &  [FeII] and H$_2$ \\
 20 53 29.7 & +44 29 38 & 2 &      "                &  [FeII] only \\
 20 53 28.9 & +44 29 25 & 3 &     "                 &  [FeII] only \\ 
 20 53 25.8 & +44 28 31 & 4 &     "                 &  [FeII] only \\
 20 53 37.5 & +44 31 26 & 5 &     "                 &   counterjet [FeII] only \\
 20 53 41.7 & +44 31 59 & 6 &    "                  &   [FeII] only; very faint \\
 20 53 30.9 & +44 30 03 &    &  IR source    & Source of MHO 3413 \& 3415 ? \\
 20 53 38.7 & +44 34 09 & 7 & MHO 3414   &  H$_2$ only; north of  IRAS20518+4420 \\
 20 53 36.1 & +44 34 39 & 8 &    "                  &    H$_2$   \\
 20 53 36.0 & +44 34 58 & 9 &    "                  &    H$_2$   \\
 20 53 32.5 & +44 35 35 & 10 &  "                  &    H$_2$  \\
 20 53 54.2 & +44 38 16 & 11 & MHO 3415 & H$_2$ knot   \\
 20 53 09.6 & +44 32 57 & 12 &  MHO 3416 & [FeII] and H$_2$; west of  IRAS20518+4420 \\ 
 20 53 08.3 & +44 33 07 & 13 &      "               &  [FeII] and H$_2$ \\
 
\enddata

\tablecomments{
 {\bf[1]}:  Labeling same as  in  Table 1. 
               }
\end{deluxetable}

\clearpage

\begin{deluxetable}{cclll}
  
\centering

\tablecaption{ 
      MHOs, Reflection Nebulae, and IR Sources in the Gulf-of-Mexico Region 
      \label{table2}}
    
\tablehead{
    \colhead{$\alpha$} & 
    \colhead{$\delta$} & 
    \colhead{\#$^1$} & 
    \colhead{MHO$^2$} &
    \colhead{Comments}
       \\

    \colhead{J2000} & 
    \colhead{J2000} & 
    \colhead{ }           &
    \colhead{ }           &
    \colhead{} 
   
    }
      
\startdata
                        & Gulf       & SW2 & 	&  \\
                        \hline
    
  20:56:41.6  &  43:48:41  &   1         & 3417   	&    HH 953 large bow, north end\\
  20:56:46.4  &  43:45:31  &   2         & "          	&   \\
  20:56:49.2  &  43:44:54  &   3         & "               	& \\
  20:56:54.3  &  43:42:21  &   4         & "               	& \\
  
  20:56:57.2  &  43:40:37  &   5         &  3417      	&  HH 954 large bow, south end \\
  20:56:56.6  &  43:40:05  &   6         & "               	&  faint  knot \\
  20:56:59.7  &  43:40:17  &   7         & "              	&  faint knot  \\
  20:56:49.0  &  43:45:13  &   IRS 1 &                 	&  S$_{70}$ = 6.1 Jy    \\
  20:56:51.1  &  43:44:06  &   IRS 2 &               	&   S$_{70}$ = 2.9 Jy    \\
  
  20:56:47.3  &  43:43:08  &   8         &   3418	 &  west of 3417 \\
  20:56:50.7  &  43:42:15  &   9  		& "  		&   \\
  20:56:47.7  &  43:46:43  &   IRS 3  & 		&  S$_{70}$ = 3.4 Jy    \\
  
  20:56:49.3  &  43:47:29  &   10 	&  3419 	& bright bow, east of 3417   \\
  20:56:48.9  &  43:44:06  &   IRS 4	&		&  S$_{70}$ = 1.9 Jy    \\
  20:56:52.1  &  43:47:38  &   IRS 5	&		&  S$_{70}$ = 1.9 Jy    \\

  20:56:53.5  &  43:38:22  &   11  	& 3420	& isolated compact bow \\
  
  20:57:00.7  &  43:34:38  &   12  	& 3421	& irregular bright H$_2$ knots ; HH 1087 \\
  20:57:01.4  &  43:34:59  &   13     	& "  		& near R7 \\
  20:57:01.8  &  43:34:47  &   14      	& " 		& \\
  20:57:05.1  &  43:34:28  &   IRS 6	& 		&   S$_{70}$ = 0.4 Jy    \\

  20:57:09.1  &  43:36:07  &   15   	&  3422	&  southeast of R5 \\
  20:57:13.2  &  43:36:27  &   16      	& " 		&   \\
  20:57:07.1  &  43:37:04  &   IRS 7 	& 		&  S$_{70}$ = 2.7 Jy    \\
  
  20:57:11.8  &  43:39:42  &   17  	&  3423	&  east of knot 5 \\
  20:57:22.3  &  43:40:05  &   18    	&  " 		& \\
  
  20:57:05.7  &  43:43:06  &   19   	&  3424	&  east of MHO 3418 \\
  20:57:09.8  &  43:42:45  &   20     	&  "  		& \\
  20:57:13.2  &  43:42:32  &   21  	&   "   	& \\
  
  20:57:00.8  &  43:44:43  &   22   	&  3425	&  south facing bow \\
\hline
                          & Gulf       & SW1         	&		&   \\
\hline

  20:57:18.5  &  43:48:42  &   23  	&   3426  	& west  \\
  20:57:20.7  &  43:48:50  &   24  	&   "    	&  west \\
  20:57:22.4  &  43:49:00  &   25 	&   "     	& east \\
  20:57:23.0  &  43:49:01  &   26   	&   "    	&   east   \\
  20:57:23.8  &  43:49:02  &   27   	&  "    	&  east  \\
 20:57:20.7  &  43:48:51  &   IRS 8	&	"	&  S$_{70}$ = 0.4 Jy    \\
  
  20:57:23.9  &  43:47:35  &   28 	&  3427   	& west  \\
  20:57:27.5  &  43:48:00  &   29 	&   " 		& \\
  20:57:32.0  &  43:47:53  &   30 	&   "  		& \\
  20:57:33.4  &  43:47:52  &   31  	&  "  		&  \\
 20:57:29.9  &  43:47:22  &   IRS 9	& 		&   S$_{70}$ = 0.6 Jy    \\
 20:57:37.1  &  43:48:03  &   IRS 10	& 		&   S$_{70}$ = 0.1 Jy    \\
  
  20:57:38.7  &  43:48:09  &   32  	& 3427 	& east   \\
  20:57:40.8  &  43:48:17  &   33 	&   "  		& \\
  20:57:41.9  &  43:48:20  &   34 	&   "  		& \\
  20:57:42.3  &  43:48:20  &   35 	&  "		&  \\
  20:57:42.3  &  43:48:09  &   36  	&  " 		& \\
  20:57:43.3  &  43:48:16  &   37 	&   "  		& \\
  20:57:44.5  &  43:48:23  &   38  	&  " 		& \\
  20:57:45.8  &  43:48:18  &   39  	&  " 		&  \\
  20:57:52.5  &  43:48:35  &   40 	&  "  		& \\
  \hline 
                             & Gulf          	&  core    	&    		&      \\
  \hline              
  20:57:55.4  &  43:50:16  &   41 	&  3428	&  HH 639 \\
  20:57:58.8  &  43:49:33  &   42 	&  "  		& \\
  
  20:57:51.4  &  43:51:40  &   43   	&  3429	&   HH 638 \\
  
  20:57:44.3  &  43:51:29  &   44   	&  3430	&   HH 636 \\
  20:57:45.9  &  43:52:50  &   45   	&    "   	&  \\
  20:57:45.5  &  43:53:22  &    IRS 11 &		& \\
    
  20:57:45.2  &  43:53:40  &   46   	&  3431	& HH 637 \\
  20:57:44.0  &  43:53:40  &   47 	& "  		& \\
  20:57:45.3  &  43:53:55  &   48  	& "  		& \\

  20:57:43.1  &  43:53:22  &   49  	&  3432	& HH 640 counterflow?  \\
  20:57:46.6  &  43:53:23  &   50  	&  3434	& HH 640 \\
  20:57:47.4  &  43:53:25  &    IRS 12 &		& \\

  20:57:35.2  &  43:55:25  &   51  	&  3433	&  \\
  20:57:33.8  &  43:55:51  &   52  	&   "   	& HH 1088 \\
  20:57:29.3  &  43:56:45  &   53   	&   "  		&  \\
  
  20:57:49.4  &  43:53:29  &   54  	&  3435	&  HH 640 \\
  20:57:49.8  &  43:53:28  &   55 	&    " 		& between 54 \& 57 \\
  20:57:50.5  &  43:53:27  &   56  	&   " 		& between 54 \& 57 \\
  20:57:50.8  &  43:53:25  &   57  	&   "  		& \\
  20:57:51.6  &  43:53:23  &   58  	&   " 		& \\
  20:57:52.4  &  43:53:20  &   59 	&  "  		& \\
  20:57:53.2  &  43:53:16  &   60  	& "  		& \\
  20:57:53.8  &  43:53:01  &   61  	& 	"	& continuation of MHO 3435 ?  \\
  20:57:54.0  &  43:52:58  &   62   	& 	"	&   \\
  20:57:56.5  &  43:52:50  &   63 	& 		& HH 643  \\
  20:57:58.9  &  43:52:40  &   64  	& 		& \\
  20:57:59.5  &  43:52:45  &   65  	& 		& \\
  20:58:04.3  &  43:52:10  &   74  	& 		& \\

  20:57:51.8  &  43:53:10  &   66  	&  3436 	& \\
  20:57:52.1  &  43:53:13  &   67 	&		& \\
  20:57:54.1  &  43:53:40  &   68 	& 		&  \\
  20:57:54.7  &  43:53:51  &   69  	& 		& \\
  20:57:55.4  &  43:54:01  &   70  	& HH 642	& \\
  20:57:57.0  &  43:54:50  &   71 	& 		&  \\

  20:57:52.2  &  43:54:04  &   72  	&  3437 	& HH 641 R15 \\
  20:57:52.1  &  43:53:50  &  IRS 13 	& 		& \\

  20:57:57.7  &  43:53:22  &   73  	&  3438 	& \\
 
  20:58:24.3  &  43:50:43  &   75  	&  3439 	& HH 1089 \\
  20:58:15.2  &  43:52:01  &   76  	&   " 		&  \\
  20:58:13.0  &  43:52:22  &   77  	&   "  		& \\
  20:58:12.0  &  43:52:30  &   78   	&  "  		& \\
  20:58:10.8  &  43:52:27  &   79   	& 		& HH  651  \\
  
  20:58:04.7  &  43:52:33  &   80  	&  3440  	& HH 648 \\
  20:58:06.7  &  43:52:46  &   81  	&    "  	& \\
  20:58:09.0  &  43:53:02  &   82  	&   "   	& \\
  
  20:58:08.5  &  43:52:47  &   83 	&  3441 	&  HH 649  \\
  20:58:11.8  &  43:53:02  &   84 	&  "  		& \\
  20:58:12.5  &  43:53:02  &   84e  	&  "  		& \\  

  20:58:08.9  &  43:53:13  &   82n 	& 3442	& filament north of 82      \\

  20:58:06.1  &  43:53:34  &   85  	& 3443	&   \\  
  20:58:05.3  &  43:53:57  &   86  	&  "  		& \\
  20:58:03.8  &  43:54:32  &   87  	&  "  		& \\
  20:58:02.3  &  43:54:09  &   88  	&  "  		&  HH 647  \\
  20:57:59.7  &  43:53:55  &   IRS 14 	&		&  \\
  20:58:06.0  &  43:53:49  &   IRS 15 	& 		& \\

  20:58:09.3  &  43:54:54  &   89  	& 3444	&   \\
  20:58:09.9  &  43:54:37  &   90  	& " 		&  \\

  20:58:11.6  &  43:53:45  &   91 	& 3445 	&    \\
  20:58:13.0  &  43:53:57  &   92   	&   " 		& \\
  
  20:58:13.7  &  43:53:08  &   95  	& 		& HH 650 \\
  
  20:58:15.3  &  43:53:52  &   96  	&  3446 	& HH 657 \\
  20:58:18.4  &  43:53:11  &   97  	&    " 		& \\
  20:58:19.9  &  43:53:00  &   98  	&    "   	& \\
  20:58:21.7  &  43:52:37  &   99  	&    "  	& \\
  20:58:21.4  &  43:52:25  & 100   	&  "  		& \\
  20:58:19.0  &  43:52:36  &  101  	&  " 		& \\
  
  20:58:18.2  &  43:54:39  &   102  	&  3447	&  \\
  20:58:18.8  &  43:54:55  &   103   	&    " 		& \\
  
  20:58:23.4  &  43:54:30  &   104 	&  3448 	& HH 658 \\
  
  20:58:26.0  &  43:53:16  &   105  	&  3449 	& \\
  20:58:26.9  &  43:53:05  &   106   	&   " 		& \\
  
  20:58:16.8  &  43:57:52  &   108  	&  3450 	& \\
  20:58:20.7  &  43:57:39  &   109 	&   " 		&  \\
  20:58:23.5  &  43:57:22  &   110  	&  "  		&  \\
  20:58:21.5  &  43:57:35  &    IRS 16	& 		&  \\
  
  20:58:28.0  &  43:56:47  &   111  	&  3451 	& \\
  20:58:29.0  &  43:56:24  &   112  	&   " 		&  \\
  20:58:28.7  &  43:56:13  &   113   	&  " 		& \\
\hline                      
 		&  Reflection 	&Nebulae	&	&  \\
 \hline                       
  20:56:50.7  &  43:44:06  &   R1  	&		& IRS 2 MHO 3417 \\
  20:56:48.8  &  43:46:40  &   R2  	&		& IRS 4 MHO 3419 \\
  20:56:46.0  &  43:43:20  &   R3  	& 		& MHO 3418  \\
  20:56:17.0  &  43:38:51  &   R4 	& 		&  \\
  20:57:06.9  &  43:36:50  &   R5  	& 		& IRS 7 MHO 3422 \\
  20:57:08.9  &  43:37:40  &   R6   	&     		&    \\
  20:57:05.0  &  43:34:18  &   R7 	& 		&  IRS 6  MHO 3421 \\
  20:57:10.1  &  43:32:00  &   R8  	& 		& \\
  20:57:29.7  &  43:47:22  &   R9 	& 		&  IRS 9 Gulf Core SW1  \\
  20:57:39.7  &  43:52:03  &   R10 	& 		& west-end of Gulf core W \\
  20:57:47.7  &  43:52:39  &   R11 	& 		&  \\
  20:57:49.4  &  43:52:37  &   R12 	& 		&  MHO 3429 \\
  20:57:45.5  &  43:53:22  &   R13 	& 		&  IRS 11 \\
  20:57:47.0  &  43:53:24  &   R14 	& 		&  IRS 12  \\
  20:57:51.9  &  43:53:48  &   R15 	& 		&  IRS 13 \\
  20:57:55.2  &  43:53:35  &   R16 	& 		&  \\
  20:57:52.9  &  43:53:28  &   R17 	& 		&  \\
  20:57:49.7  &  43:53:22  &   R18  	& 		& \\
  20:58:18.2  &  43:53:28  &   R19 	& 		& Herbig cluster  \\
  20:58:20.6  &  43:56:44  &   R20  	& 		& south of R21  \\
  20:58:21.5  &  43:57:35  &   R21  	& 		& MHO 3450 \\
\hline  
                            		& Cometary 	& Clouds 	& 		&  \\
 \hline                           
  20:57:29.0  &  43:37:21  &   C1  	& 		& \\
  20:57:30.7  &  43:37:33  &   C2  	& 		& \\
  20:57:33.0  &  43:37:05  &   C3  	& 		& \\
  20:57:58.5  &  43:41:14  &   C4  	& 		& \\
  20:58:02.7  &  43:40:50  &   C5  	& 		& \\
  20:58:08.5  &  43:40:30  &   C6  	& 		& \\
  20:58:14.1  &  43:39:27  &   C7  	& 		& \\
  20:58:05.0  &  43:42:25  &   C8  	& 		& \\
  20:58:18.6  &  43:39:19  &   C9  	& 		& \\
  20:58:19.0  &  43:40:01  &   C10  	& 		& \\
  20:58:28.8  &  43:43:26  &   C11 	& 		&  \\
  20:58:39.8  &  43:30:46  &   C12  	& 		& \\
  20:58:39.9  &  43:31:12  &   C13  	& 		& \\
  20:58:40.4  &  43:32:26  &   C14  	& 		& \\
  20:58:42.2  &  43:31:38  &   C15  	& 		& \\
  20:58:43.4  &  43:32:12  &   C16  	& 		& \\
  20:58:43.5  &  43:31:14  &   C17  	& 		& \\
  20:58:51.4  &  43:36:14  &   C18  	& 		& \\

\enddata

\tablecomments{
 {\bf[1]:}  Numbers  as in Table 1.   Numbers following a letter `C'
refer to cometary clouds whose edges are lit up with 
fluorescent H$_2$ emission.  Numbers following a letter `R' refer
to infrared reflection nebula detected in the K$_s$ images. 
Numbers preceded by `IRS' refer to infrared sources in the
Gulf region.
{\bf[2]:} MHO numbers.
               }
\end{deluxetable}

\clearpage

\begin{deluxetable}{lccccl}
  
\centering

\tablecaption{ 
      Properties of Bolocam Clumps and Cores
      \label{table4}}
    
\tablehead{
     
    \colhead{$Object$}  & 
    \colhead{$\alpha$}   & 
    \colhead{$\delta$}    & 
    \colhead{$S_{\nu}$} &
    \colhead{M}                &
    \colhead{Comments}
       \\
    \colhead{}            &
    \colhead{J2000} & 
    \colhead{J2000} &
     \colhead{(Jy)}    &
    \colhead{(M$_{\odot}$)} &   
    \colhead{} 
   
    }
      
\startdata

Pelican MHO 3401   &  20 49 50.6 & +44 34 11 & 1.2 	&    9		& r = 77\arcsec\  by 83\arcsec\  oval \\
Pelican MHO 3400	&  20 50 15.3 & +44 34 46 & 1.2	&    9		& r = 65\arcsec\ by 55\arcsec\  oval \\
Pelican MHO 3402   	&  20 50 40.9 & +44 33 51 & 1.6	&  12		& r = 58\arcsec\ by 111\arcsec\  oval \\
Pelican 4            	&  20 50 30.1 & +44 25 11 & 1.4	&  10		& r = 74\arcsec\ by 60\arcsec\ oval\\
Pelican 5            	&  20 50 49.0 & +44 25 07 & 2.2	&  16		& r = 74\arcsec\ by 60\arcsec\ oval\\
Pelican 6            	&  20 50 33.2 & +44 22 44 & 1.5	&  11		& r = 56 \arcsec\ by 49\arcsec\ oval\\
Pelican 7            	&  20 50 44.6 & +44 22 30 & 1.9	&  14		& r = 64\arcsec\ by 92\arcsec\ oval \\
Pelican 8            	&  20 50 40.7 & +44 18 12 & 2.6	&  19 	&  r = 36\arcsec\ by 46\arcsec\ oval \\
Pelican 9            	&  20 50 28.7 & +44 19 05 & 7.8	&  56		& contains 8; 209\arcsec\ by 103\arcsec\ oval \\
Pelican total       	& 20 50 06.1  & +44 30 59 & 17.9  	& 129  	& 550\arcsec\ by 1030\arcsec\ oval \\
                                     &            		&		&		&  		& \\
Atlantic main       	& 20 53 34.8 & +44 32 24 & 30.1	&  216	& r = 257\arcsec\  \\
Atlantic 1           		& 20 54 05.3 & +44 23 36 &  2.0	&  14		& 155\arcsec\ by 73\arcsec\ oval   \\
Atlantic 2           		& 20 54 32.1 & +44 13 45 &  2.8	&  20		& 160\arcsec\ by 76\arcsec\ oval  \\
                                     &            		&		&		&  		& \\
Gulf core E          	& 20 58 13.0 & +43 53 00 &   4.4 	&   32	& r = 95\arcsec\   \\
Gulf core W          	& 20 57 49.1 & +43 53 00	&   8.9 	&   64	& r = 162\arcsec\ by 95\arcsec\ oval  \\
Gulf SW1             	& 20 57 36.0 & +43 47 56 &   4.2	&   30	& r = 82\arcsec\  \\
Gulf SW2 main       	& 20 56 45.7 & +43 43 41	&  66.3    	& 477   	& r = 345\arcsec\ by 159\arcsec\ oval \\
Gulf SW2 SE         	& 20 57 09.5 & +43 40:39	&  16.6	&  119	& r = 292\arcsec\ by 130\arcsec\ oval \\
Gulf SW2 S         	& 20 57 06.3 & +43 32 31	&    3.6 	&   26	& r = 235\arcsec\ by 150\arcsec\ oval \\
Gulf total         		& 20 57 22.0 & +43 42 24 &108.0 	& 776	& r = 1105\arcsec\ by 548\arcsec\ oval \\

\enddata

\tablecomments{
 {\bf[1]}:  Masses are computed from the area-integrated flux S$_{\nu}$ 
 in units of Janskys
  using   $M = 14.26 (e^{(13.01/T_K)} -1) D^2(kpc) S_{\nu}(Jy)$ M$_{\odot}$
  where $D(kpc)$ is the distance in kpc.  For 
  an adopted distance $D(kpc)$ = 0.55 kpc and  $T_K$ = 20 Kelvin, 
  $ M = 7.19 S_{\nu}(Jy)$ M$_{\odot}$.  
 {\bf[2]}: 
  The entries for Pelican total and Gulf total are based on the integrated fluxes measured in the large ovals and not derived by summing the masses of the individual clumps.  
                 }
                 
\end{deluxetable}

\clearpage

\end{document}